\documentclass[12pt,a4paper,final]{iopart}

\expandafter\let\csname equation*\endcsname\relax
\expandafter\let\csname endequation*\endcsname\relax
\usepackage{amsmath}
\usepackage{iopams}
\usepackage{graphicx}
\usepackage[breaklinks=true,colorlinks=true,linkcolor=blue,urlcolor=blue,citecolor=blue]{hyperref}

\usepackage{array}
\newcolumntype{L}[1]{>{\raggedright\let\newline\\\arraybackslash\hspace{0pt}}m{#1}}
\newcolumntype{C}[1]{>{\centering\let\newline\\\arraybackslash\hspace{0pt}}m{#1}}
\newcolumntype{R}[1]{>{\raggedleft\let\newline\\\arraybackslash\hspace{0pt}}m{#1}}
\usepackage{subfigure}
\usepackage{graphicx}
\usepackage{dcolumn}
\usepackage{bm}
\usepackage{epstopdf}
\usepackage{epsfig}
\usepackage{url}
\usepackage{textcomp}
\usepackage{stmaryrd}
\usepackage{wasysym}
\usepackage{lipsum,lineno}
\usepackage{enumitem} 
\usepackage{xcolor}

\begin{document}

\title[Snakes in square, honeycomb, and triangular lattices]{Snakes in square, honeycomb and triangular  lattices}

\author{R. Kusdiantara$^{1,2}$, H. Susanto$^{2}$}
\address{$^1$Industrial and Financial Mathematics Research Group, Department of Mathematics, Institut Teknologi Bandung, Jl.\ Ganesha 10, Bandung, 40132, Indonesia}
\address{$^2$Department of Mathematical Sciences, University of Essex, Wivenhoe Park, Colchester CO4 3SQ, United Kingdom}

\ead{rudy@math.itb.ac.id,hsusanto@essex.ac.uk}


%

\begin{abstract}
We present a study of time-independent solutions of the two-dimensional discrete Allen-Cahn equation with cubic and quintic nonlinearity. 
Three different types of lattices are considered, i.e., square, honeycomb, and triangular lattices. 
The equation admits uniform and localised states. We can obtain localised solutions by combining two different states of uniform solutions, which can develop a snaking structure in the bifurcation diagrams. We find that the complexity and width of the snaking diagrams depend on the number of ``patch interfaces'' admitted by the lattice systems. We introduce an active-cell approximation to analyse the saddle-node bifurcation and stabilities of the corresponding solutions along the snaking curves. Numerical simulations show that the active-cell approximation gives good agreement for all of the lattice types when the coupling is weak. {We also consider planar fronts that support our hypothesis on the relation between the complexity of a bifurcation diagram and the number of interface of its corresponding solutions.} 
\end{abstract}


\section{Introduction}

In recent years, a great deal of interest has been focussed on the study of  homoclinic snaking \cite{Woods1999} appearing in pattern formations in nonlinear systems, such as those in the Swift-Hohenberg equations \cite{Burke2006,Burke2007,Burke2007a}, cellular buckling \cite{Hunt2000}, neuronal model \cite{Laing2001,Avitabile2015}, and optical systems \cite{Firth2007,Yulin2008,Yulin2010,Yulin2011}, leading to a rather complete understanding of their properties and mechanism of formation in the lower dimension. Homoclinic snaking has also been observed in different experiments, e.g., in magnetic fluids \cite{Lloyd2015}, liquid crystals \cite{Bortolozzo2009,Haudin2011}, shell bucklings \cite{Thompson2015}, optical cavities \cite{Tlidi2012}, and semiconductor optical systems \cite{Barbay2008}. {Generally, localised solutions can be present due to the existence of bistability regimes between two states, i.e., they can both be homogeneous or patterned states, or even a mix of the two. We can obtain a localised state when we combine them back to back, 
which are connected by fronts \cite{Pomeau1986}. }

Higher dimensional snaking has been studied as well \cite{Lloyd2008,Uecker2014,Avitabile2010}. Using the planar Swift-Hohenberg equation, several numerical observations show exotic solutions, such as stripes or fronts, localised spots and hexagon patches \cite{Lloyd2008,Coullet2000,Hilali1995,Sakaguchi1996,kozy13} and localised radial solutions \cite{McCalla2010,Lloyd2009}. Localised square patterns have also been observed in the same equation with an additional nonlinear gradient term \cite{Sakaguchi1997}. Planar neuronal models also exhibit similar exotic solutions \cite{Rankin2013}. Patchwork quilt state, i.e., regular triangles, \cite{Golubitsky1984} is foreseen in bistable systems with the symmetry $u\rightarrow-u$. Snaking involving various superpatterns \cite{Dionne1997,Judd2000} is also anticipated. Snaking of localised structures called convectons in three-dimensional doubly diffusive convections has also been studied \cite{Beaume2011,Beaume2013}. 

All in all, details of the snaking behaviour are rather more involved in higher dimensional case, such as overlapping pinning regions, complexity involving Maxwell points, and growing patterns by nucleating individual structure which break and recover the basic symmetry of the state \cite{Lloyd2008}. Even a good qualitative picture of those behaviours is still an open problem \cite{Knobloch2008a}. 

This paper provides a further step towards understanding the problem. However, rather than considering spatially continuous systems, we study a discrete one as it may provide a better control over, e.g., the patterns of localised states that may appear by determining the lattice types. Our idea exploits homoclinic snaking that is also observed in spatially discrete systems \cite{Chong2009,Chong2011,Kusdiantara2017,Taylor2010,Susanto2011,Susanto2018}. While in continuous systems snaking is caused by pinning between fronts and the underlying oscillatory states, in discrete setups it is due to the pinning of fronts and the imposed lattices, i.e., in continous equations homoclinic snaking occurs when localised states add ``rolls'' at the fronts, in discrete systems they add ``cell''. In two-dimensional discrete systems, homoclinic snaking has been numerically studied in \cite{Chong2009,Taylor2010}, where it was shown that bifurcation diagrams of localised solutions can exhibit a complex behaviour, which is not clearly well understood yet. 

In this work, we consider square, honeycomb, and triangular lattices. Our main finding is that the complexity and width of the snaking diagrams depend on the number of ``patch interfaces'' admitted by the lattice patterns. Here, we consider a two-dimensional discrete Allen-Cahn equation with cubic and quintic nonlinearity  \cite{Taylor2010}. The equation can be considered to come from the two-dimensional discrete nonlinear Schr\"odinger equation
\begin{equation}
i\dot{\psi}_n+c\Delta \psi_n+2|\psi_n|^2\psi_n-|\psi_n|^4\psi_n=0,
\label{eq:dnls35}
\end{equation}
where $\psi_n(t)$ is an array of complex field and
$\Delta$ is a Laplacian operator of the nearest neighbour differences. 
By substituting $\psi_n(t)=u_n e^{-i\mu t}$, where $u_n$ is real stationary field into equation \eqref{eq:dnls35}, we obtain 
\begin{equation}
	\mu u_n+c\Delta u_n + 2u_n^3-u_n^5=0,
\end{equation}
which is the time-independent discrete Allen-Cahn equation. 
In this report, we study homoclinic snaking of localised solutions (patches and planar fronts) admitted by the system when the coupling between lattices is weak. In particular, we consider three different types of lattices, i.e., square, honeycomb, and triangular, which to our best knowledge have not been studied in the context of homoclinic snaking. Another main result is that we classify all the relevant structures causing saddle-node bifurcations that form the boundaries of the pinning regions. 

The paper is constructed as follows.
The two-dimensional discrete Allen-Cahn equation is discussed in Section \ref{sec2d:math_mod_uni}.
We also discuss uniform states and their stability in the section.
In Section \ref{sec2d:loc_snake}, localised states in the form of patches and their homoclinic snaking for square, honeycomb, and triangular lattices are being discussed and calculated.  
 In Section \ref{sec2d:saddle_anal}, we discuss active-cell approximations of the saddle-node bifurcations and details of the snaking structures reported in Section \ref{sec2d:loc_snake}. In Section \ref{secadd} we study planar fronts, which are (quasi)1D solutions and hence in a way are simpler than patches, and their snaking. Our results in the section support our hypothesis on the relation between the complexity of a bifurcation diagram and the number of interfaces of its corresponding solutions. Section \ref{sec2d:conclusions_2d} is our conclusions.

\section{Mathematical model and uniform state}\label{sec2d:math_mod_uni}
\begin{figure}[h!]
	\centering
	\subfigure[]{\includegraphics[scale=0.45]{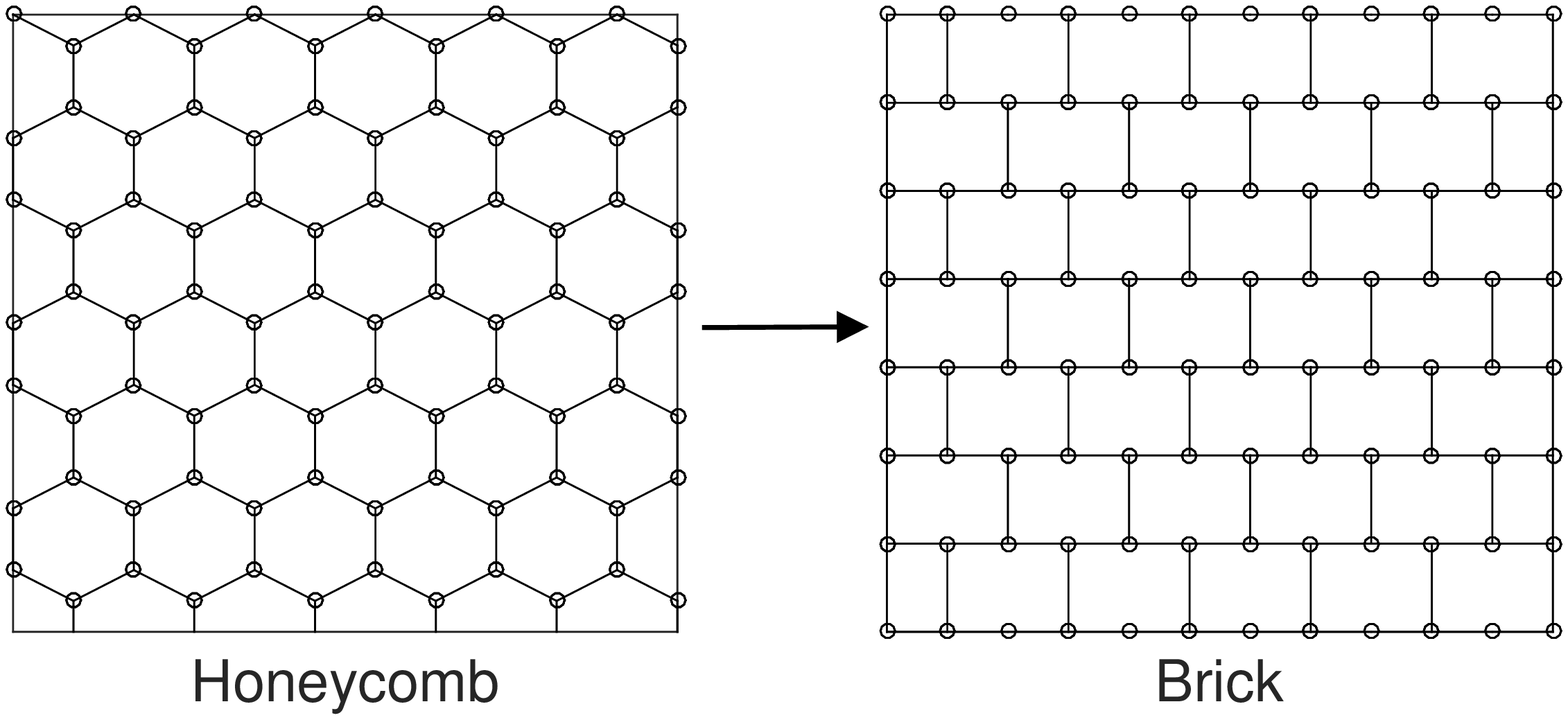}\label{subfig:honeycomb2brick}}
	\subfigure[]{\includegraphics[scale=0.45]{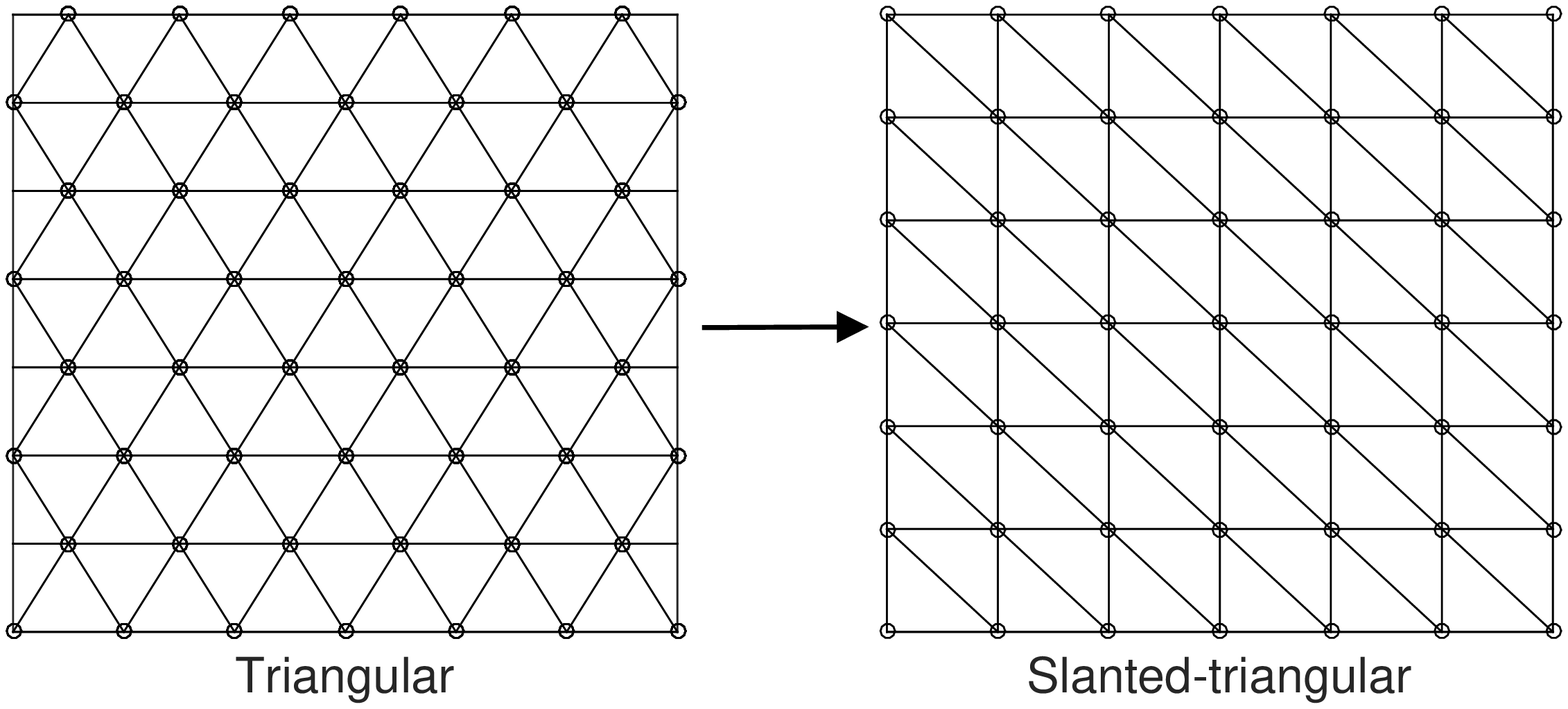}\label{subfig:triangular2slanted}}
	\caption{
		Honeycomb and triangular lattices.
		We transform the lattices into brick and slanted-triangular ones for the sake of computations and plotting in this report.
	}
	\label{fig:tranform_lattice}
\end{figure}

In this study, we consider the two-dimensional (2D) discrete Allen-Cahn equation, which is given by
\begin{equation}
\dot{u}_{n,m} 
=\mu u_{n,m} + 2u_{n,m}^3 - u_{n,m}^5+c^{\square}\Delta^\square u_{n,m},
\label{eq:dnls_all}
\end{equation}
where $u_{n,m}$ is a real stationary field defined on 2D integer lattice,  $\mu$ is a real bifurcation parameter, $c^\square$ is the coupling strength of the nearest-cell, and $\Delta^\square$ is a discrete Laplacian operator on the 2D integer lattices $\mathbb{Z}^2$. 
We consider three lattice types, namely,
\begin{itemize}[leftmargin=*]
	\item Square lattice : 
	\begin{eqnarray}
	\begin{array}{ccl}
	c^{\square}\Delta^\square u_{n,m}&=& c^+\Delta^+ u_{n,m}\\
	&=&c^+\left(u_{n+1,m}+u_{n-1,m}+u_{n,m+1}+u_{n,m-1}-4u_{n,m}\right),
	\end{array}
	\end{eqnarray}	
	\item Honeycomb lattice :
	\begin{eqnarray}
	\begin{array}{ccl}
	c^{\square}\Delta^\square u_{n,m}&=& c^\Yup\Delta^{\Yup_\pm} u_{n,m}\\
	&=&c^\Yup\left(u_{n+1,m}+u_{n-1,m}+u_{n,m\pm1}-3u_{n,m}\right),
	\end{array}
	\end{eqnarray}
	where $\Delta^{\Yup_+}$ and $\Delta^{\Yup_-}$ correspond to the case when $n+m$ is even or odd, respectively, 
	\item Triangular lattice :
	\begin{equation}
	\begin{array}{ccl}
	c^{\square}\Delta^\square u_{n,m}&=&c^{\varhexstar}\Delta^{\varhexstar} u_{n,m}\\
	&=&c^{\varhexstar} \left(u_{n+1,m}+u_{n-1,m}+u_{n,m+1}+u_{n,m-1}+u_{n-1,m+1}+u_{n+1,m-1}\right.\\
	&&\left.-6u_{n,m}\right).
	\end{array}
	\end{equation}	
\end{itemize}
A sketch of honeycomb and triangular lattices is shown in figure\ \ref{fig:tranform_lattice}. For the purpose of computations and plotting, in the following we transform the lattices into a square domain, i.e., brick and slanted-triangular lattices, as sketched also in figure\ \ref{fig:tranform_lattice}. 

{We choose the discrete Allen-Cahn as our toy model here because of its simplicity. Yet, at the same time it shares similar qualitative pictures with other complicated systems, such as the discrete Swift-Hohenberg equation (see \cite{Kusdiantara2017}). The Allen-Cahn equation also resembles the discrete nonlinear Schr\"odinger equation that models many physical systems, see \cite{Chong2009,Chong2011} for the motivational details.}

In particular, we study the time-independent solution of equation \eqref{eq:dnls_all}, i.e.,
\begin{equation}
\mu u_{n,m} + 2u_{n,m}^3 - u_{n,m}^5+c^{\square}\Delta^\square u_{n,m}=0.
\label{eq:dnls_ti}
\end{equation}
To determine the linear stability of a solution $\tilde{u}_{n,m}$, we write
\begin{equation}
u_{n,m}=\tilde{u}_{n,m}+\epsilon e^{\lambda t}\hat{u}_{n,m}.
\label{eq:anz_2dstab}
\end{equation}
By substituting \eqref{eq:anz_2dstab} into \eqref{eq:dnls_all} and linearising around $\epsilon=0$, we obtain the linear equation
\begin{equation}
\lambda \hat{u}_{n,m}=\mathcal{L}\hat{u}_{n,m},
\end{equation}
where

\begin{equation}
\mathcal{L}=\mu + 6\tilde{u}^2_{n,m}-5\tilde{u}^4_{n,m}+c^\square \Delta^\square.
\end{equation}
A uniform solution is said to be stable when all {$\lambda\leq 0$} and unstable when $\exists \lambda>0$.
\begin{figure}[htbp]
	\centering
	\includegraphics[scale=0.42]{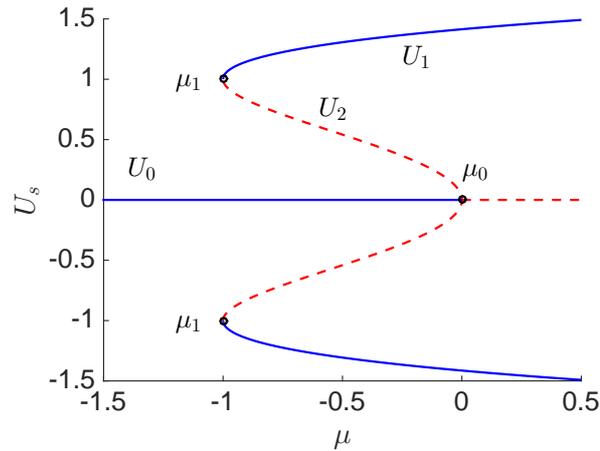}
	\caption{Uniform solution of the discrete Allen-Cahn equation.
		The blue solid and red dashed lines indicate stable and unstable solutions, respectively.
	}
	\label{fig:uniform_dnls}
\end{figure}

The 2D discrete Allen-Cahn equation \eqref{eq:dnls_all} exhibits the same uniform solution as the 1D case that has been studied by Taylor and Dawes in \cite{Taylor2010}, which is given by
\begin{equation}
0 = \mu U_s + 2U_s^3 - U_s^5,
\end{equation}
that can be solved to yield
\begin{equation}
U_0=0\quad \text{and}\quad U_{1,2}^2=1\pm\sqrt{1+\mu}.
\end{equation}
We plot the solutions for varying $\mu$ in figure\ \ref{fig:uniform_dnls}.
To determine the linear stability of the uniform solutions $\tilde{u}_{n,m} = U_s$, where $s=0,1,2$, 
one has $\hat{u}_{n,m}=e^{i\left(kn+lm\right)}$, where $k$ and $l$ are the wave number of the perturbations in the $n$ and $m$ directions, from which we obtain for the square and triangular lattices the dispersion relation
\begin{equation}
\lambda(k,l)=\mu +6 U_s^2-5U_s^4+\gamma^\square(k,l), \quad \square=+,\varhexstar,
\label{eq:eig_fun}
\end{equation}
where
\begin{equation}
\begin{array}{ccl}
\gamma^+(k,l)&=&2c^+\left(\cos(k)+\cos(l)-2\right),\\
\gamma^{\varhexstar}(k,l)&=&2c^{\varhexstar}\left(\cos(k-l)+\cos(k)+\cos(l)-3\right),
\end{array}
\end{equation}
respectively.
As for the honeycomb lattice, we need to rewrite equation \eqref{eq:dnls_all} into
\begin{equation}
\begin{array}{ccl}
\dot{\psi}_{n,m} &=& \mu \psi_{n,m} + 2\psi_{n,m}^3 - \psi_{n,m}^5+c^{\Yup}\left(\varphi_{n,m}+\varphi_{n,m-1}+\varphi_{n-1,m}-3\psi_{n,m}\right),\\
\dot{\varphi}_{n,m} &=& \mu \varphi_{n,m} + 2\varphi_{n,m}^3 - \varphi_{n,m}^5+c^{\Yup}\left(\psi_{n,m}+\psi_{n,m+1}+\psi_{n+1,m}-3\varphi_{n,m}\right).
\end{array}
\label{eq:honeycomb_v2}
\end{equation}
The perturbation ansatz in this case would be
\begin{equation}
\left(
\begin{array}{c}
\psi_{n,m}\\
\varphi_{n,m}
\end{array}\right)=
U_s+
\left(
\begin{array}{c}
\hat{\psi}_{n,m}\\
\hat{\varphi}_{n,m}
\end{array}\right)\epsilon e^{\lambda t}.
\label{eq:anz_honeycomb}
\end{equation}
By substituting \eqref{eq:anz_honeycomb} into \eqref{eq:honeycomb_v2} and linearising around $\epsilon=0$, we obtain the eigenvalue problem
\begin{equation}
\lambda\left(
\begin{array}{c}
\hat{\psi}_{n,m}\\
\hat{\varphi}_{n,m}
\end{array}\right)=
\left(
\begin{array}{cc}
\mu +6 U_s^2-5U_s^4-3c^{\Yup}&c^{\Yup}\,\xi^-(k,l)\\
c^{\Yup}\,\xi^+(k,l)&\mu +6 U_s^2-5U_s^4-3c^{\Yup}
\end{array}\right)
\left(
\begin{array}{c}
\hat{\psi}_{n,m}\\
\hat{\varphi}_{n,m}
\end{array}\right),
\end{equation}
where
\begin{equation}
\xi^\pm(k,l)=1+\cos(k)+\cos(l)\pm i\left(\sin(k)+\sin(l)\right).
\end{equation}
Hence, we have the dispersion relation for the honeycomb lattice, i.e., 
\begin{equation}
\lambda(k,l)=\mu +6 U_s^2-5U_s^4-3c^{\Yup}\pm c^{\Yup}\sqrt{2\left(\cos(k-l)+\cos(k)+\cos(l)\right)+3}.
\label{eq:disper_honey}
\end{equation}

The points $\mu_{s}$, $s=0,1$ is figure\ \ref{fig:uniform_dnls} denote the stability change of $U_s$. 
They correspond to a condition when the maximum of the dispersion relation \eqref{eq:eig_fun} and \eqref{eq:disper_honey} touch the $k,l$ plane, which is attained at $k=l=0$ for all of the lattice types. 
One can note that we have bistability interval $\mu\in\left[\mu_{1},\mu_0\right]$ for the uniform solutions, see figure\ \ref{fig:uniform_dnls}.
Furthermore, the bifurcation diagram and the stability of the uniform solution in figure \ref{fig:uniform_dnls} is the same as those in the 1D model \cite{Taylor2010,Chong2009}.

\begin{figure*}[t!]
	\centering
	\subfigure[Site-centred solution of square lattice]{\includegraphics[scale=0.5]{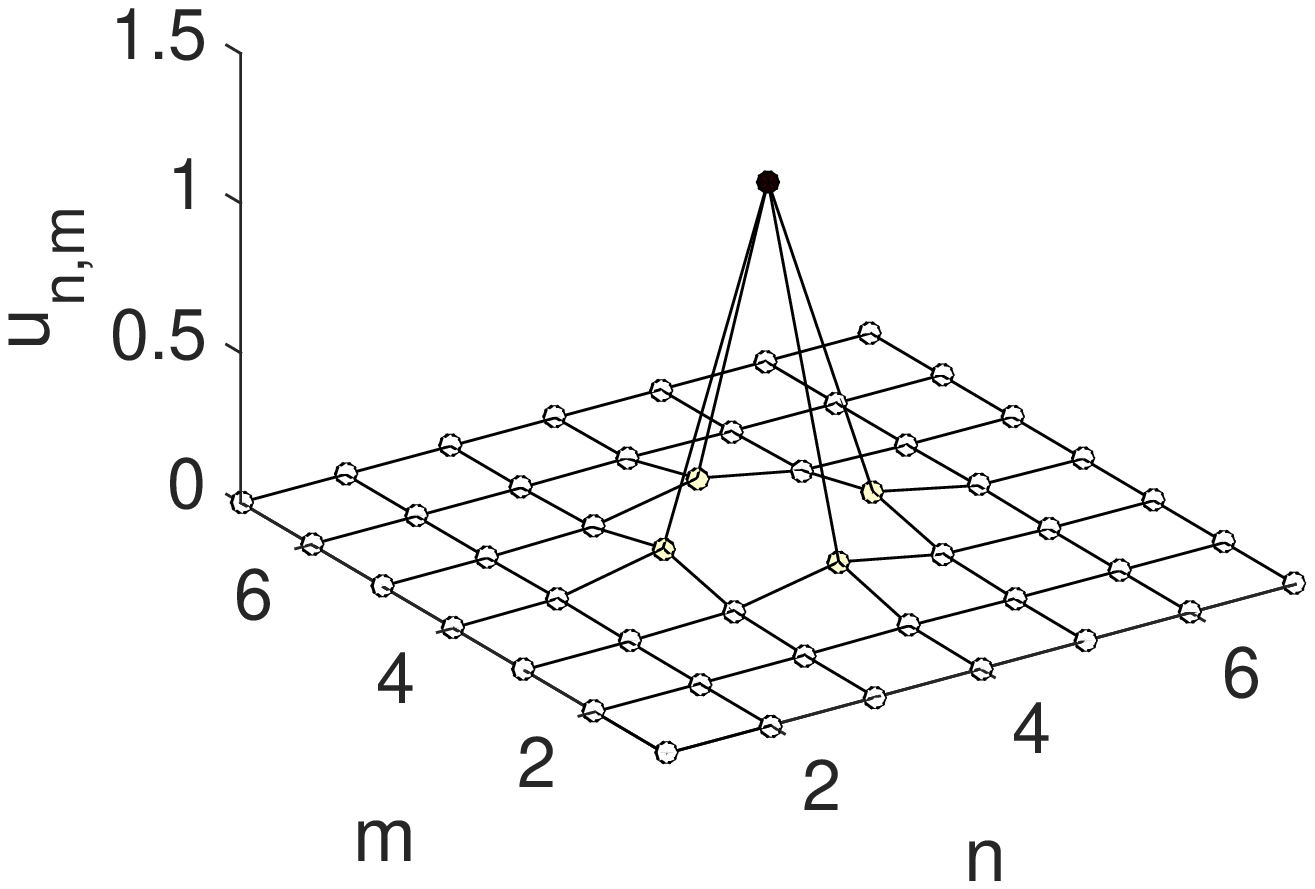}\label{subfig:site_square}}
	\subfigure[Bond-centred solution of square lattice]{\includegraphics[scale=0.5]{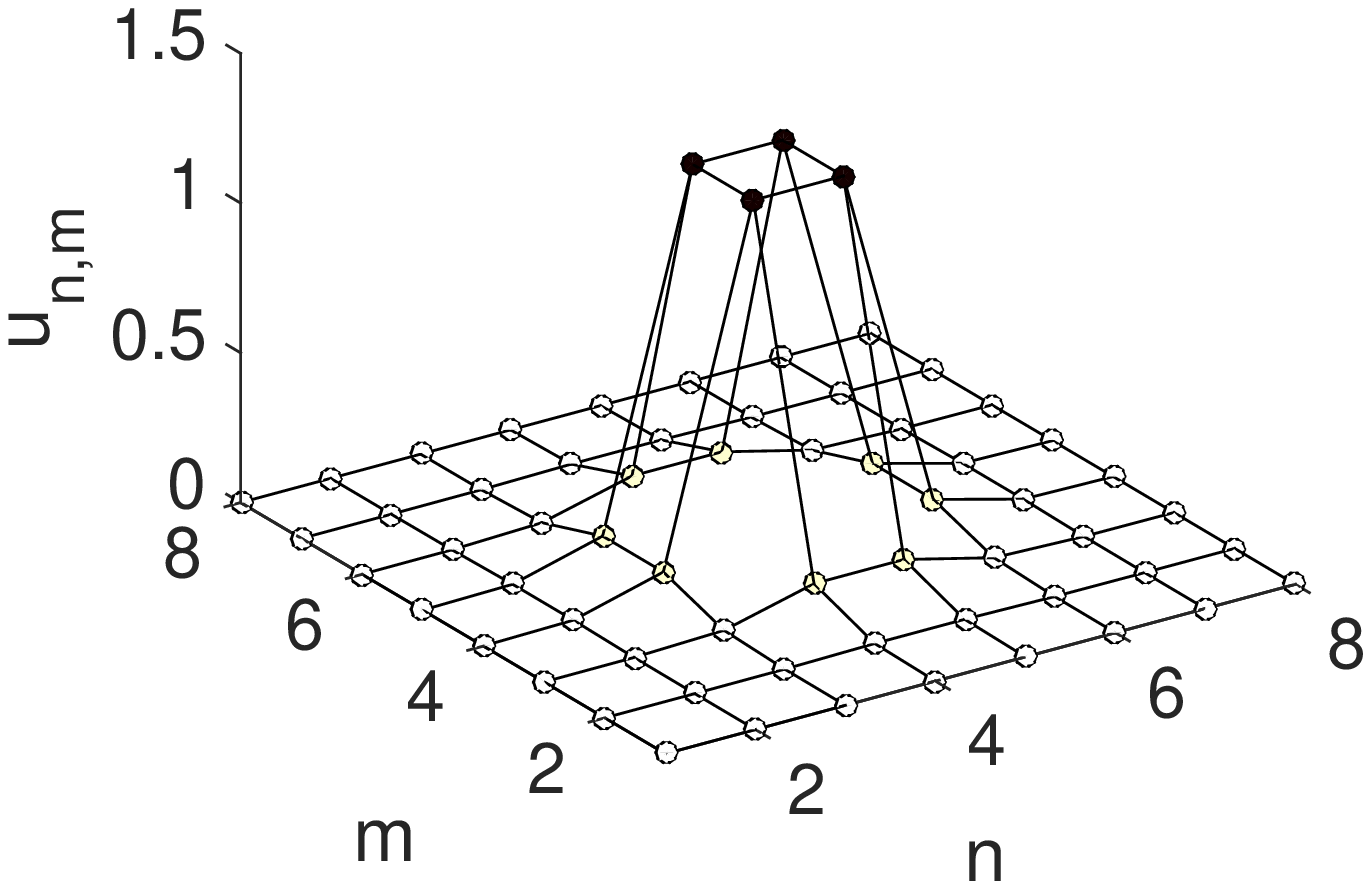}\label{subfig:bond_square}}
	\subfigure[Site-centred solution of honeycomb lattice]{\includegraphics[scale=0.5]{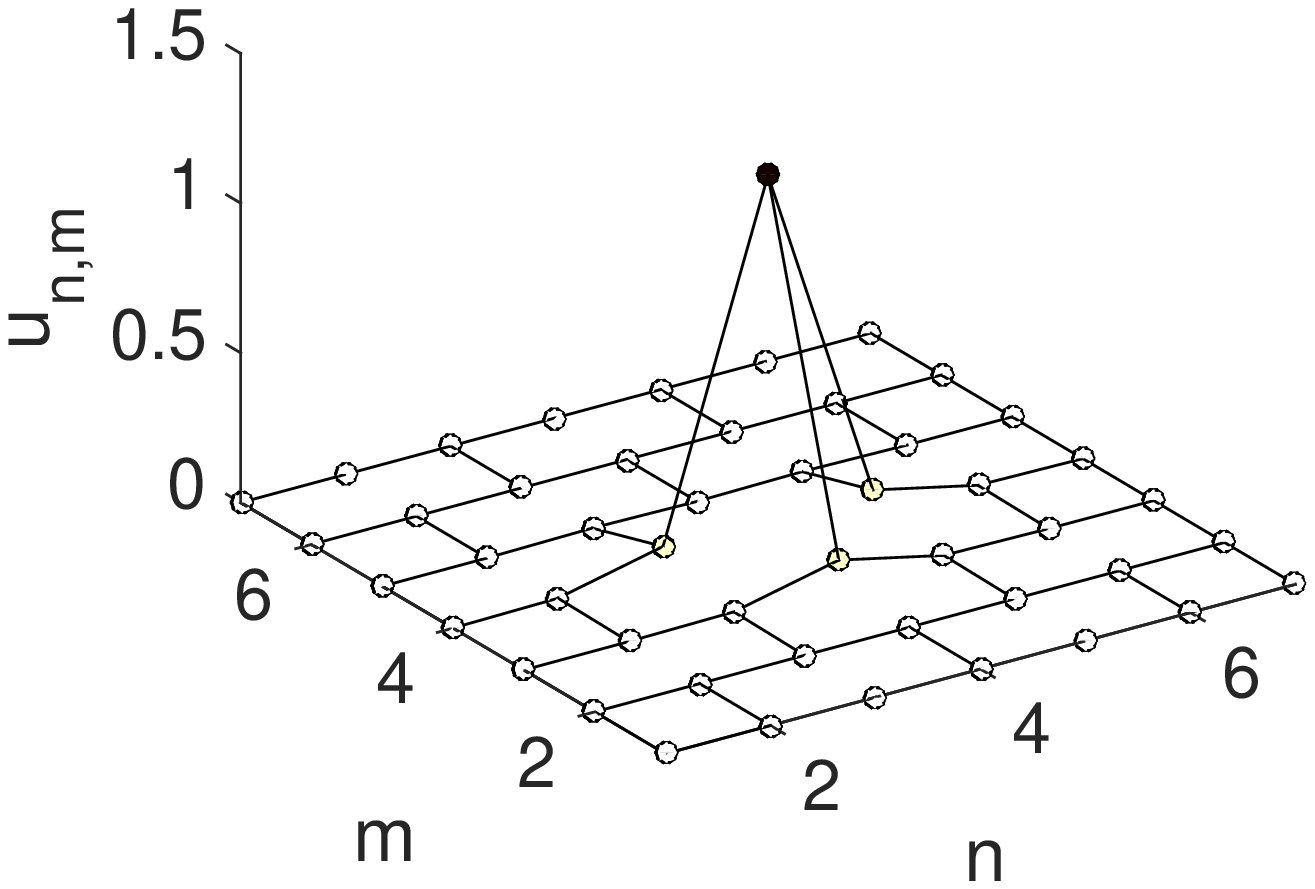}\label{subfig:site_brick}}
	\subfigure[Bond-centred solution of honeycomb lattice]{\includegraphics[scale=0.5]{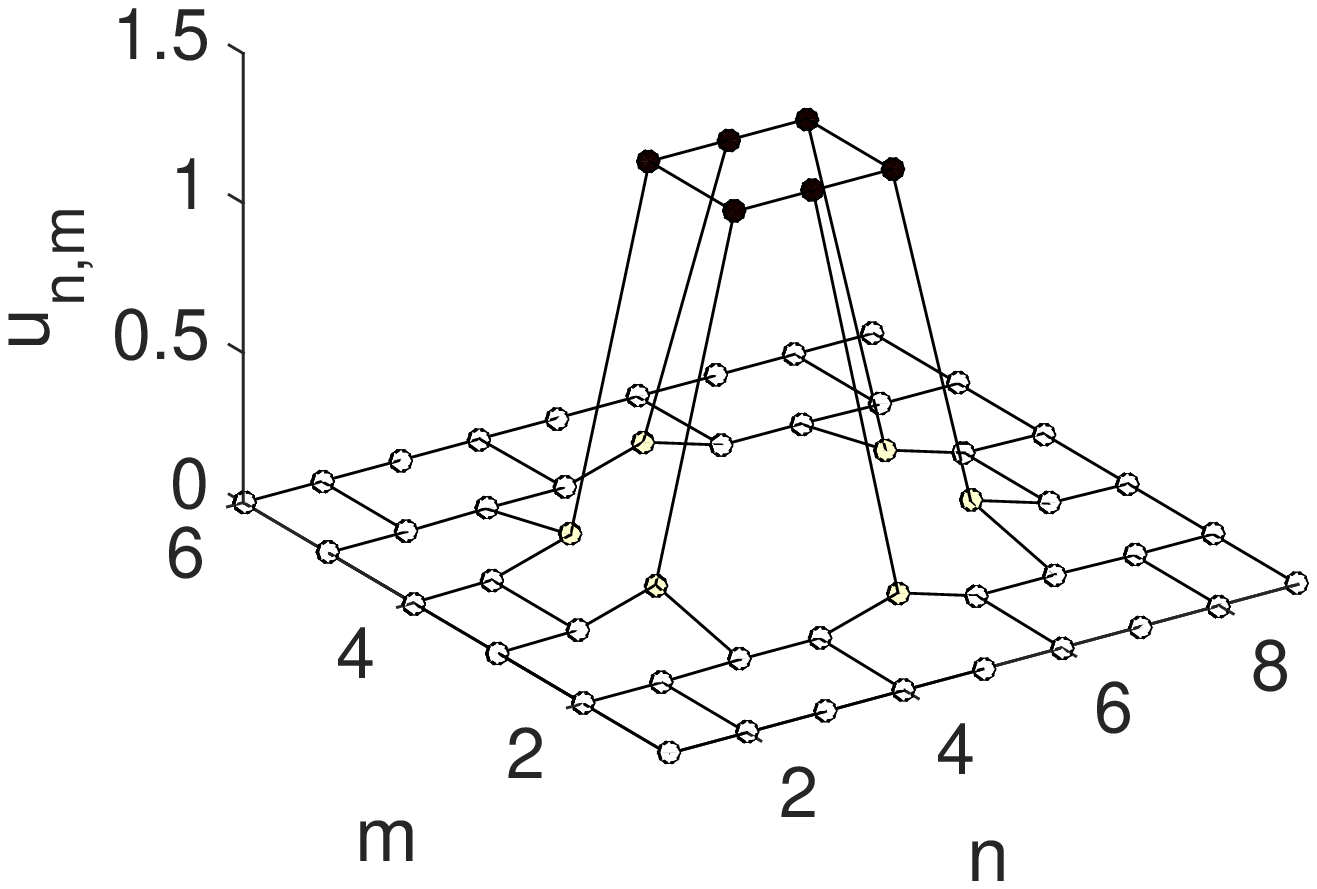}\label{subfig:bond_brick}}
	\subfigure[Site-centred solution of triangular lattice]{\includegraphics[scale=0.5]{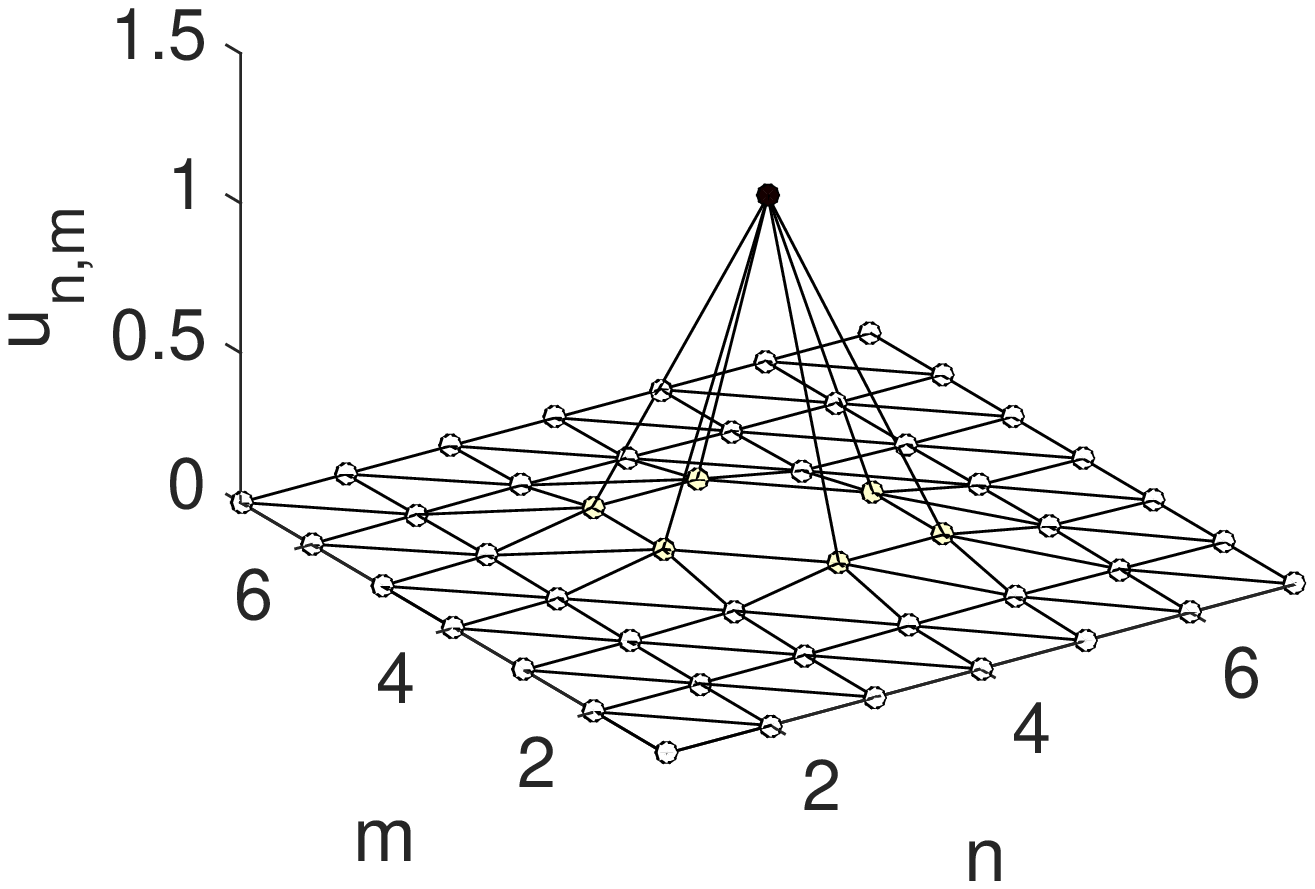}\label{subfig:site_triangular}}
	\subfigure[Bond-centred solution of triangular lattice]{\includegraphics[scale=0.5]{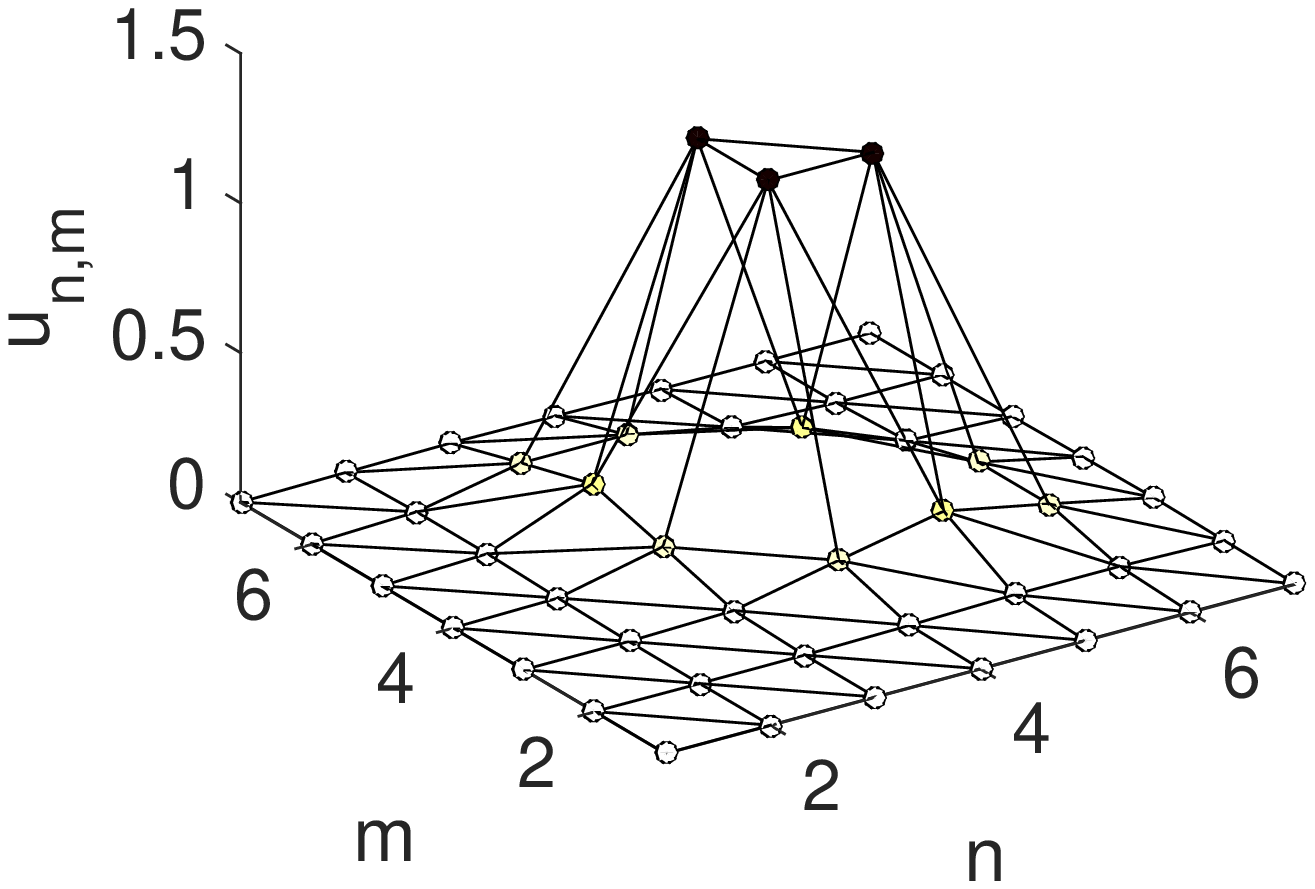}\label{subfig:bond_triangular}}
	\caption{Structures of fundamental localised solutions for $c^+=c^\Yup=c^{\varhexstar}=0.05$ and $\mu=-0.6$.
	}
	\label{fig:site_bond_example}
\end{figure*}

\section{Localised solution and snaking}\label{sec2d:loc_snake}
The discrete Allen-Cahn equation \eqref{eq:dnls_all} admits solutions that are localised in both planar directions, i.e., patches, and bifurcate from the uniform solution $U_0$ at point $\mu_0$. 
We are particularly interested in fundamental localised solutions, i.e., site-centred and bond-centred solutions, which are the counter-part of onsite and intersite solutions in the 1D case. 
They are formed by two bistable states from the uniform solutions, i.e., the non-zero state $U_1$ as the ``upper'' state and zero state $U_0$ as the ``background" state.

{To compute them, we solve the time-independent equation of (\ref{eq:dnls_all}) numerically using a Newton-Raphson method with periodic boundary conditions for all of the lattice types. 
Herein, we use $20\times20$ lattice domain. When performing numerical continuations, we use a pseudo-arclength method to continue the computations past turning points \cite{kell87}. The bifurcation diagrams are then presented in the $(M,\mu)$-plane, where $M$ is a scaled version of the $\mathbb{L}_2$ norm or ``mass'' norm \cite{Taylor2010}
\begin{equation}
M=\left(\sum_{n,m}\frac{u_{n,m}^2}{1+\sqrt{1+\mu}}\right)^\frac{1}{2}.
\end{equation}
Spectrum of the solutions is also calculated numerically using a standard matrix eigenvalue solver.}

\begin{figure*}[htbp!]
	\centering
	\subfigure[$c^+=0.05$]{\includegraphics[scale=0.425]{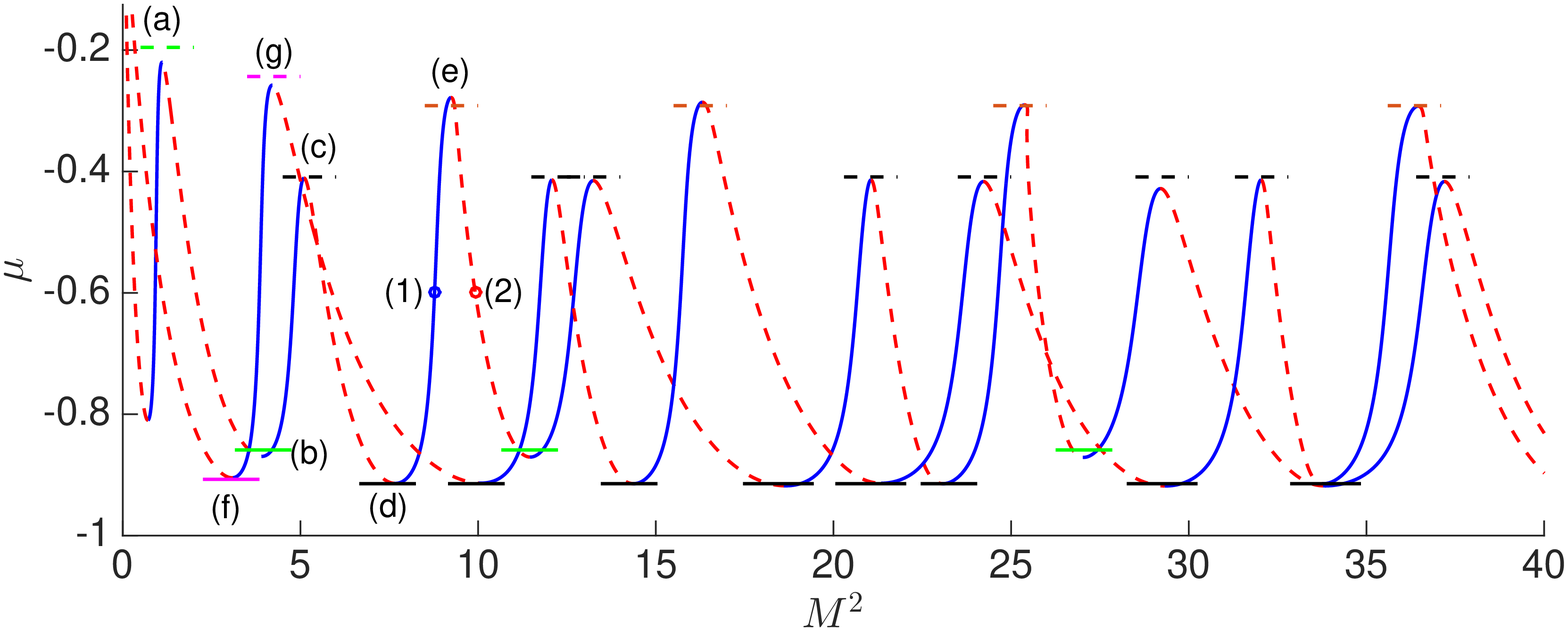}\label{subfig:bifur_cp_0_05}}
	\subfigure[$c^+=0.15$]{\includegraphics[scale=0.425]{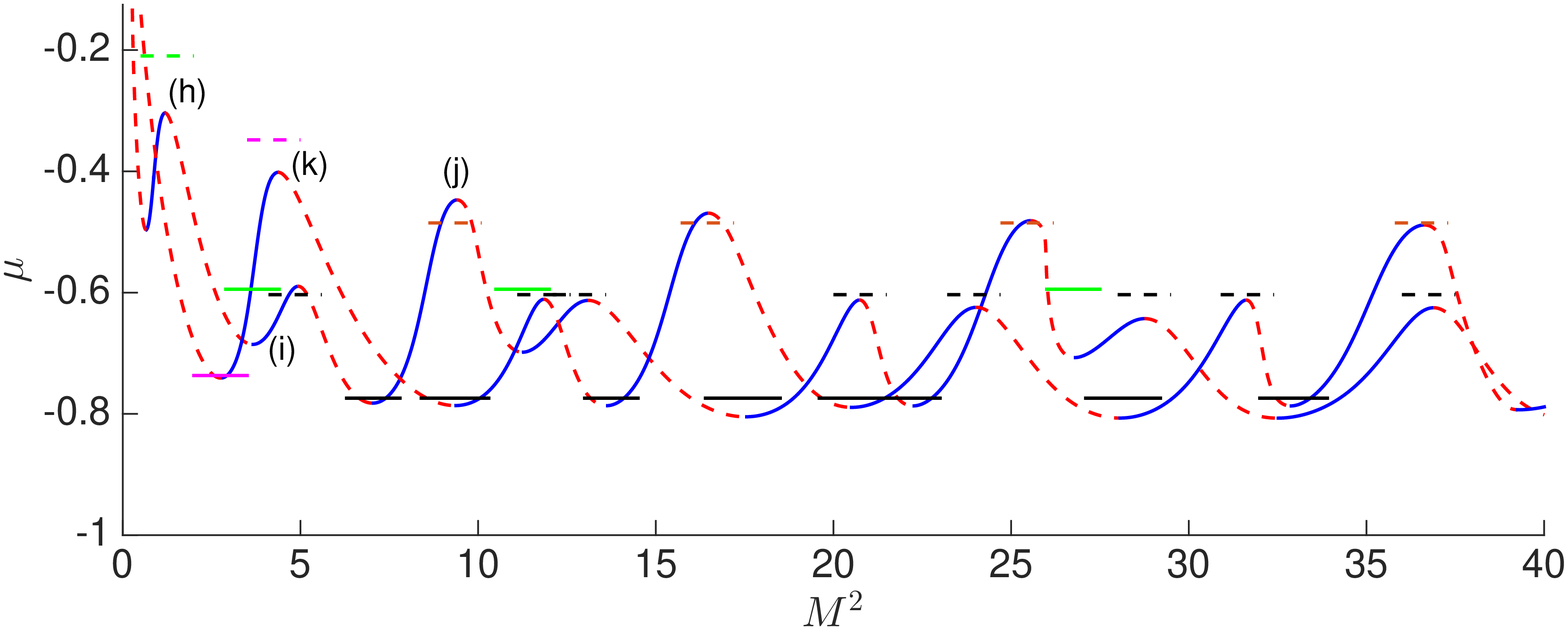}\label{subfig:bifur_cp_0_15}}
	\caption{Panels (a) and (b) show bifurcation diagrams for square lattice. 
		The solid and dashed lines around turning points of the snaking curves are the approximation of ``lower'' and ``upper'' saddle-node bifurcations, respectively.
		The green, black, magenta, and brown line colors correspond to saddle-node bifurcations from our active-cell approximations of type 1, 2, 3, and 4, {see Section \ref{sec2d:saddle_anal}}. Solution profiles at the turning points labelled as (a)-(e) in the top panel are shown in Fig.\ \ref{fig:prof_cp}. Points (1)-(2) will be used to describe the solution stability in Fig.\ \ref{fig:active_cell_compare}. Points (h)-(k) are discussed in the text. 
	}
	\label{fig:bifur_cp}
\end{figure*}

\begin{figure*}[htbp!]
	\centering
	\subfigure[]{\includegraphics[scale=0.19]{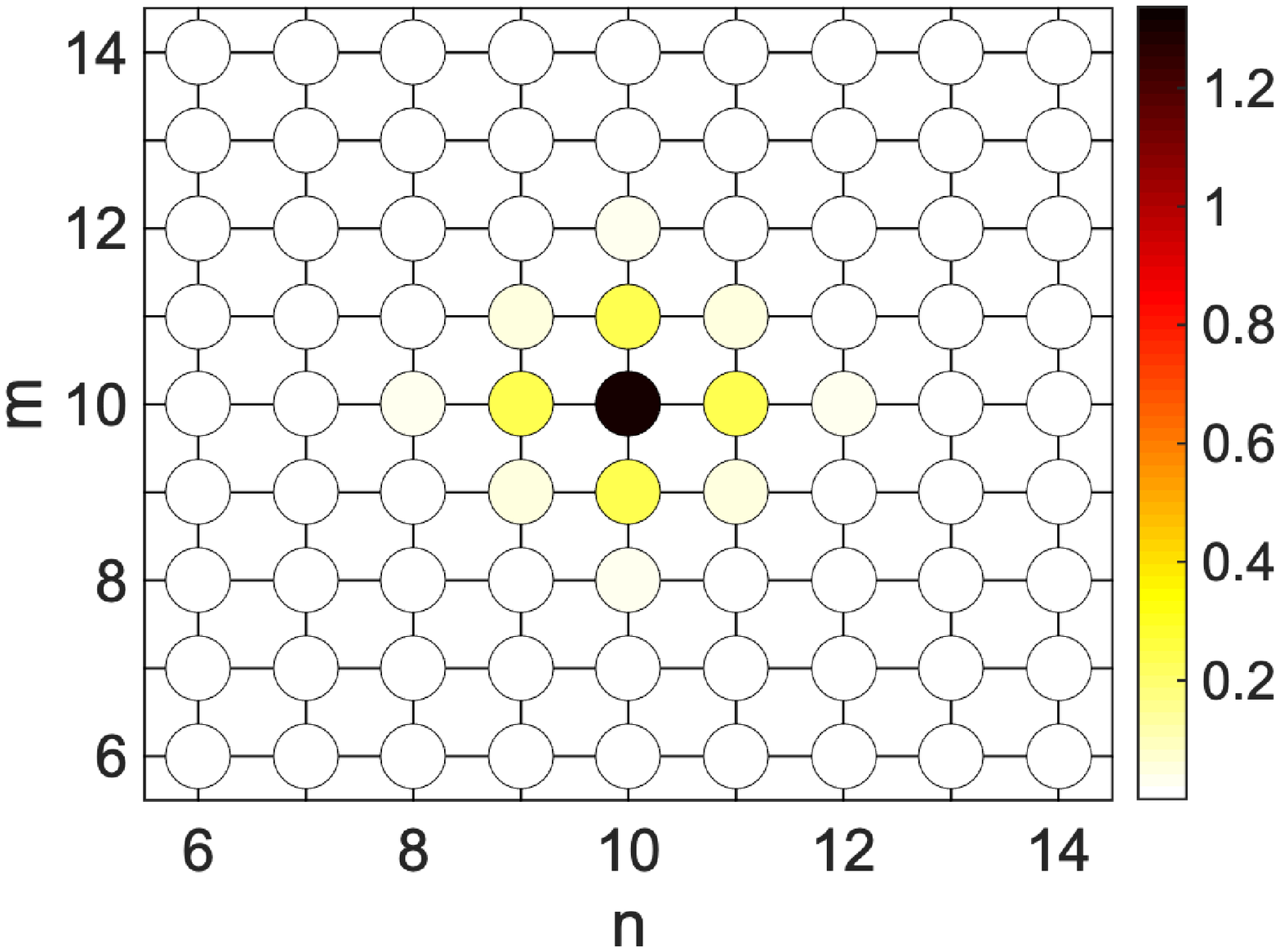}\label{subfig:prof_cp_0_05_a}}
	\subfigure[]{\includegraphics[scale=0.19]{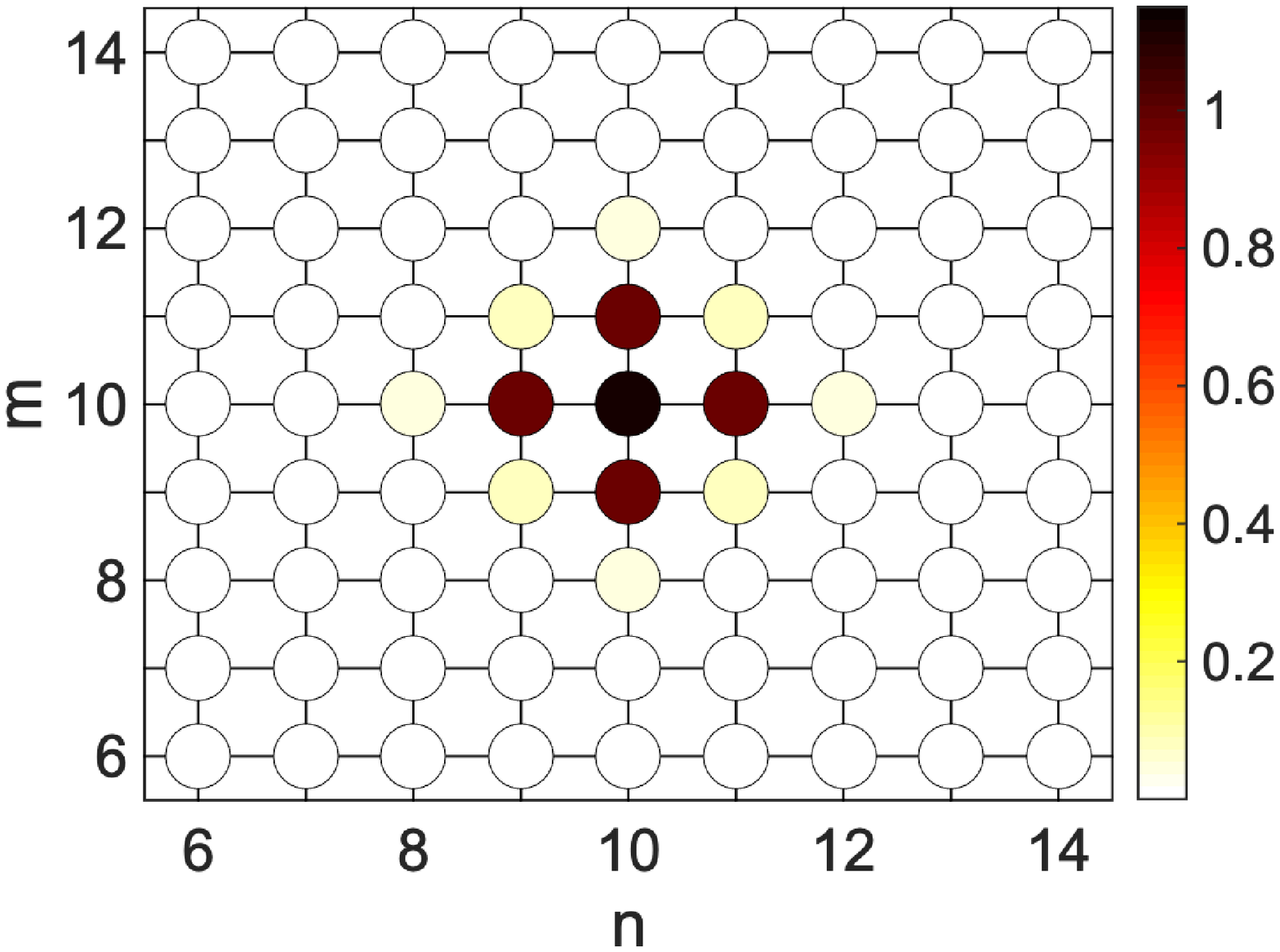}\label{subfig:prof_cp_0_05_b}}
	\subfigure[]{\includegraphics[scale=0.19]{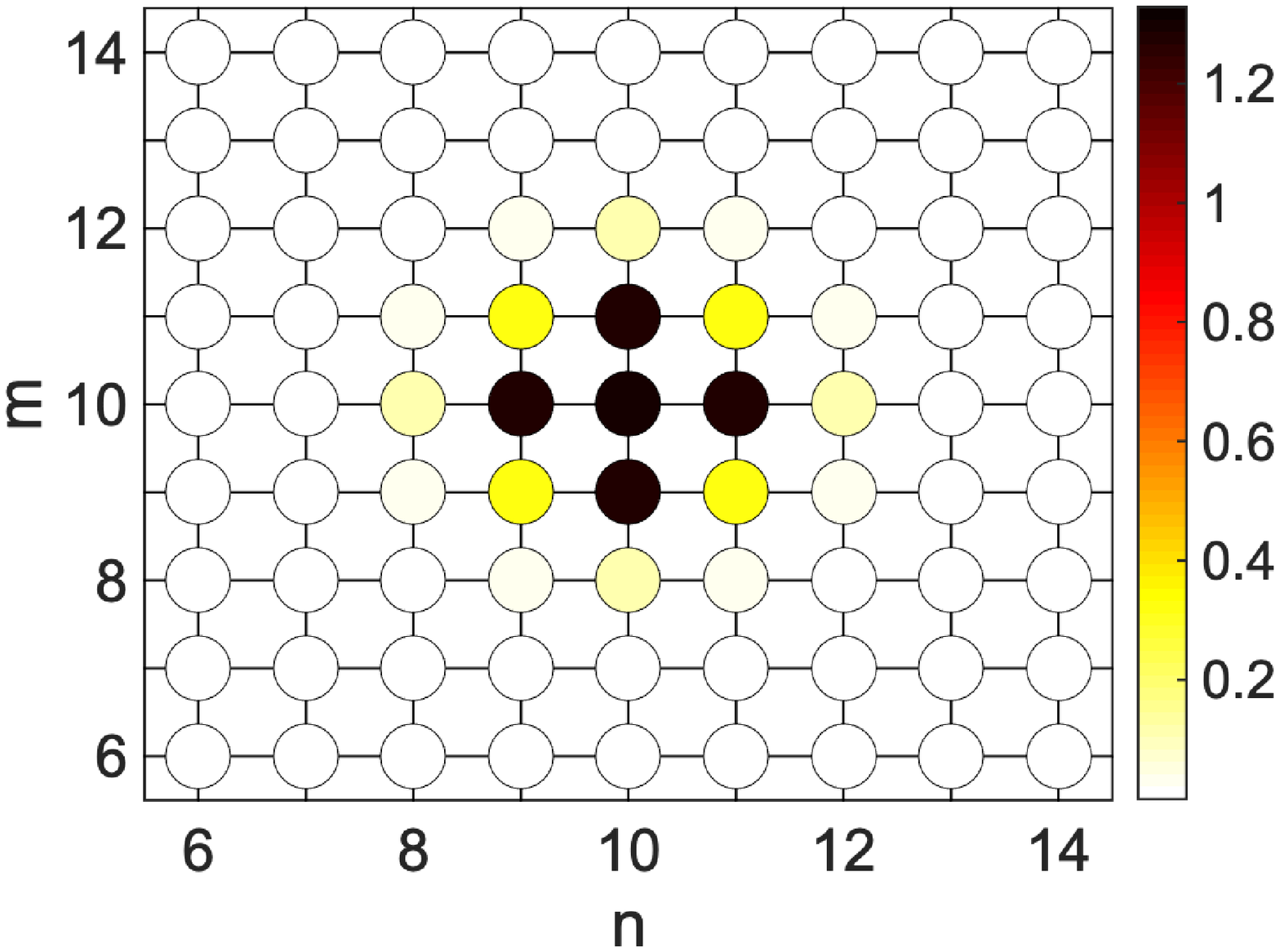}\label{subfig:prof_cp_0_05_c}}
	\subfigure[]{\includegraphics[scale=0.19]{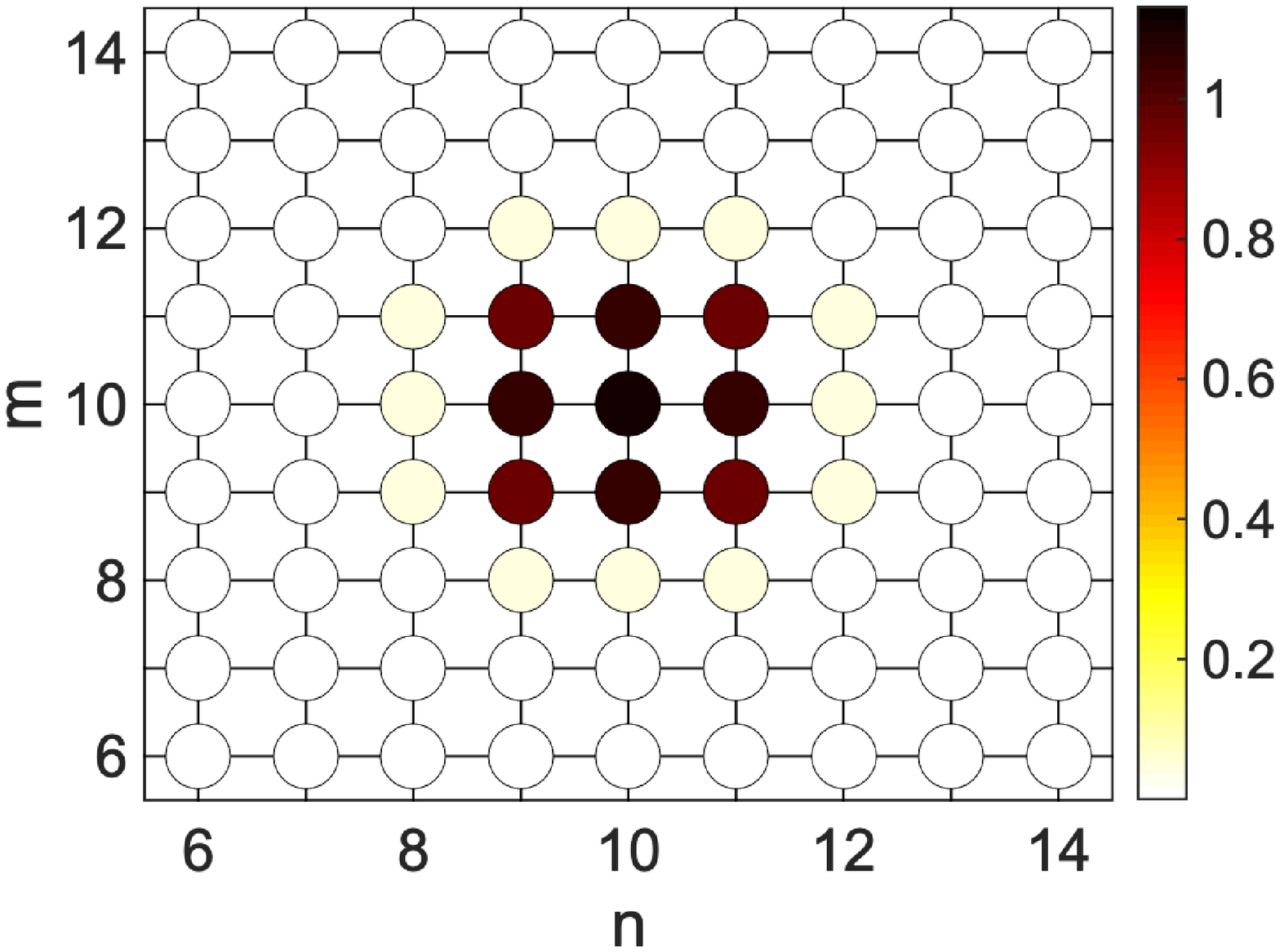}\label{subfig:prof_cp_0_05_d}}
	\subfigure[]{\includegraphics[scale=0.19]{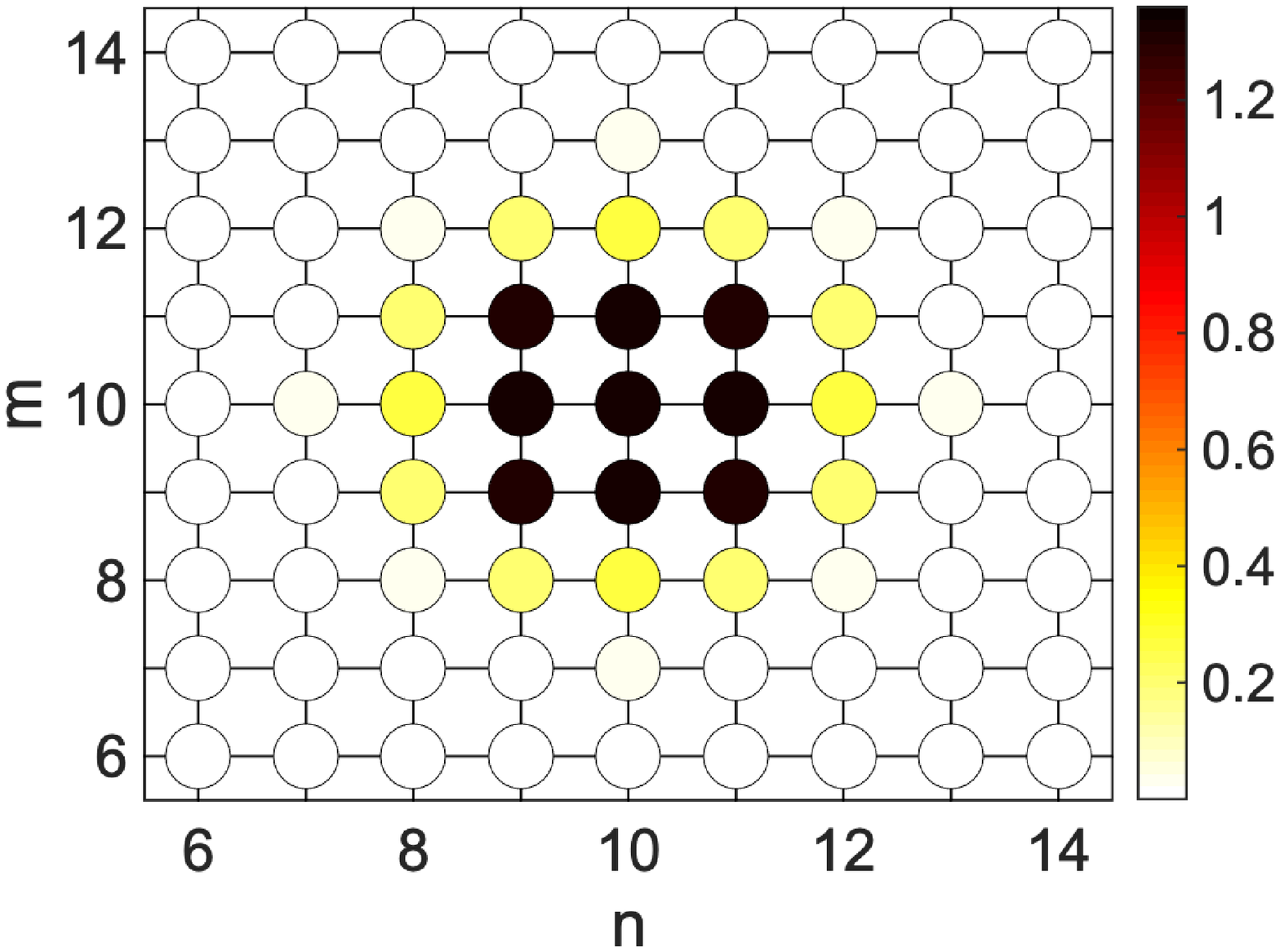}\label{subfig:prof_cp_0_05_e}}
	\subfigure[]{\includegraphics[scale=0.19]{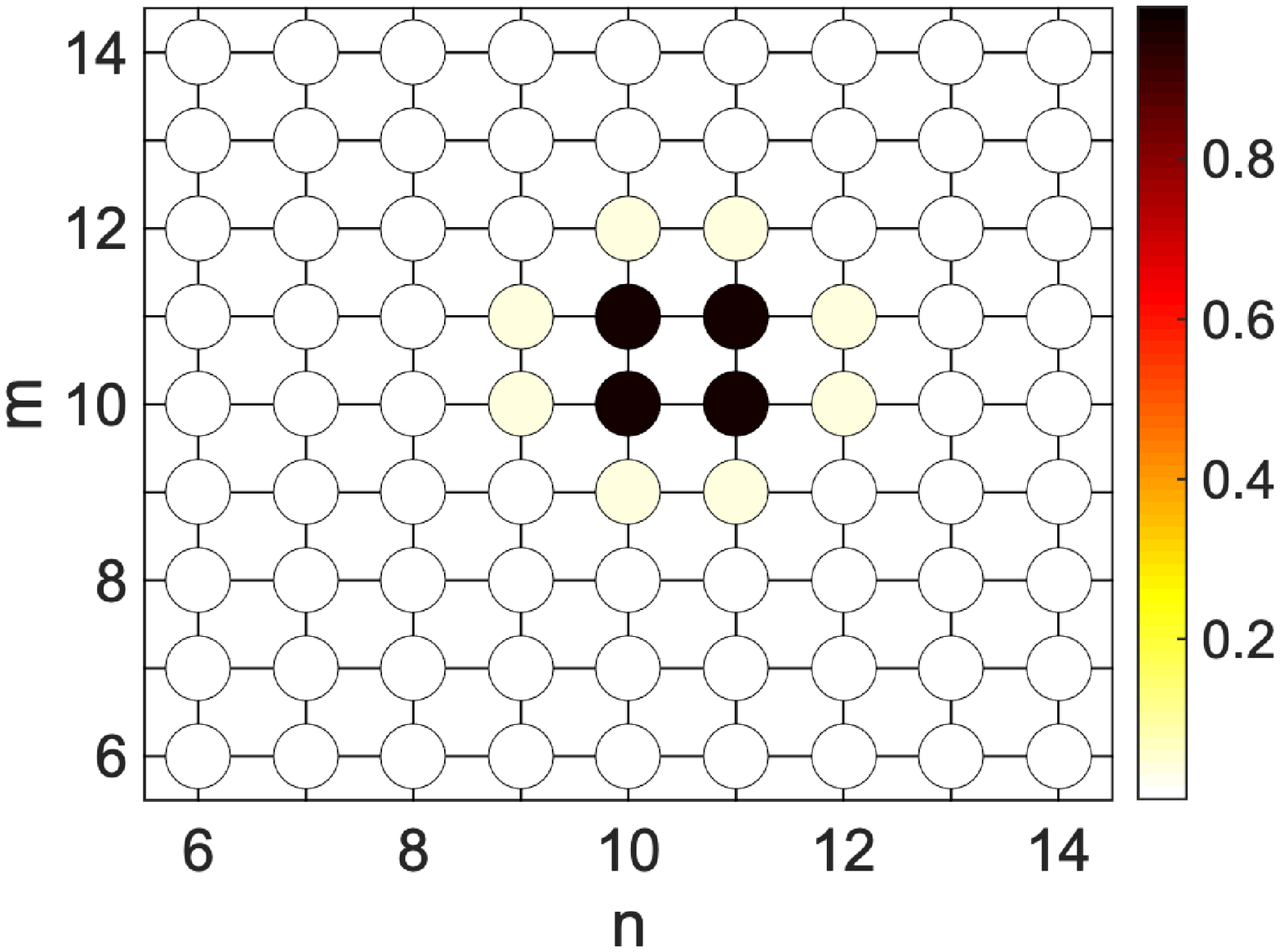}\label{subfig:prof_cp_0_05_f}}
	\subfigure[]{\includegraphics[scale=0.19]{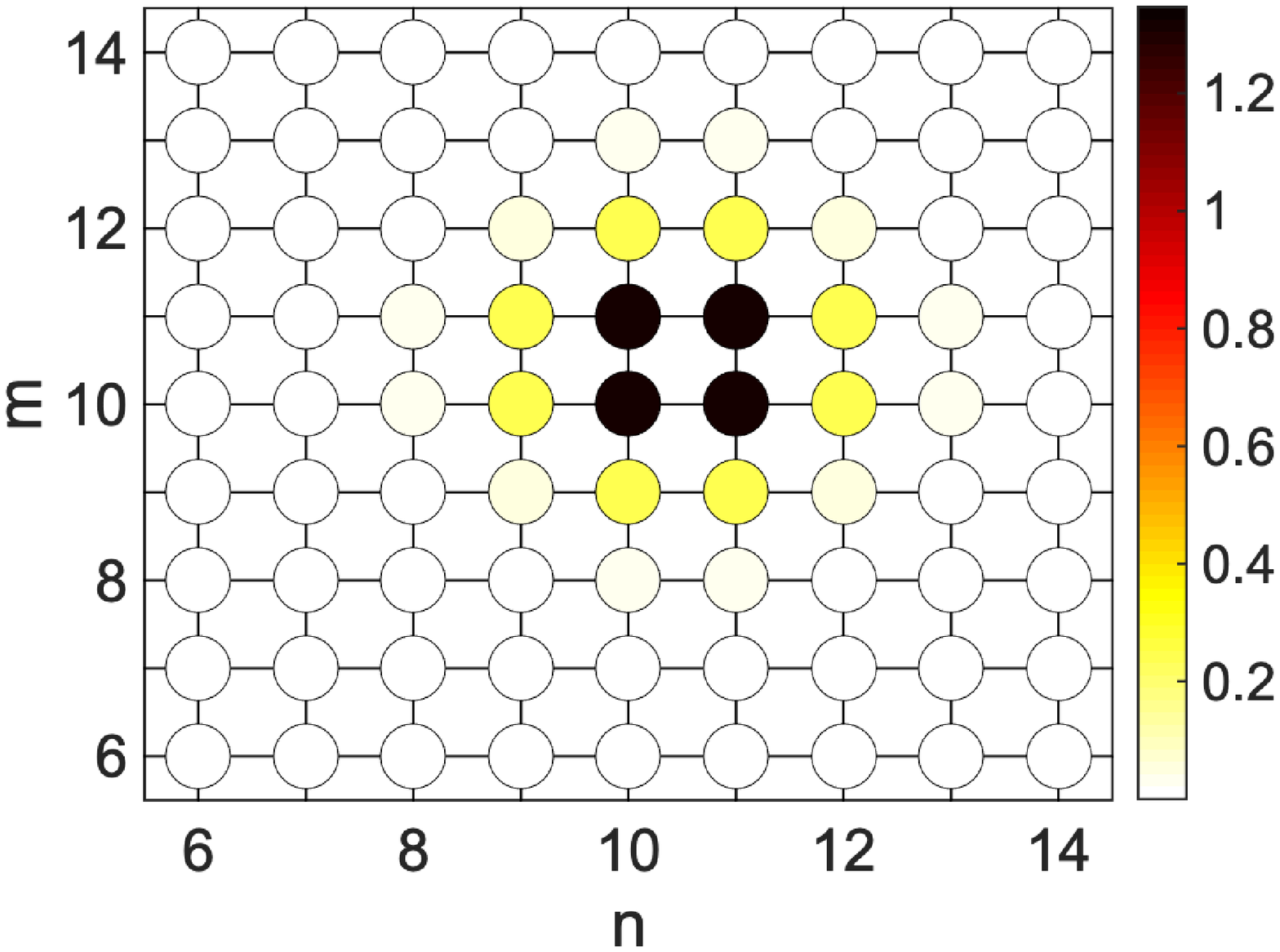}\label{subfig:prof_cp_0_05_g}}
	\caption{Top-view of localised solution profiles in square lattice that correspond to points (a)-(e) in figure \ref{subfig:bifur_cp_0_05}.
	}
	\label{fig:prof_cp}
\end{figure*}

\begin{figure*}[t!]
	\centering
	\subfigure[$c^{\Yup}=0.05$]{\includegraphics[scale=0.425]{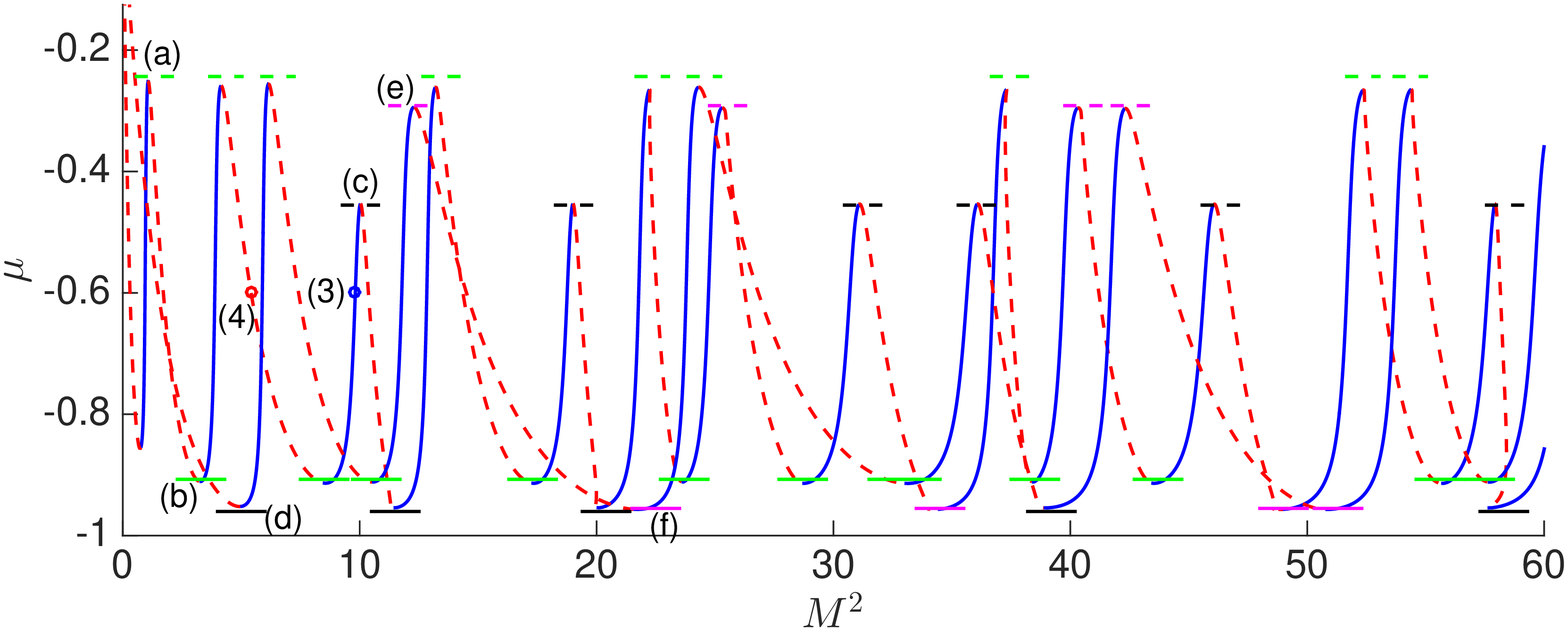}\label{subfig:bifur_ch_0_05}}
	\subfigure[$c^{\Yup}=0.15$]{\includegraphics[scale=0.425]{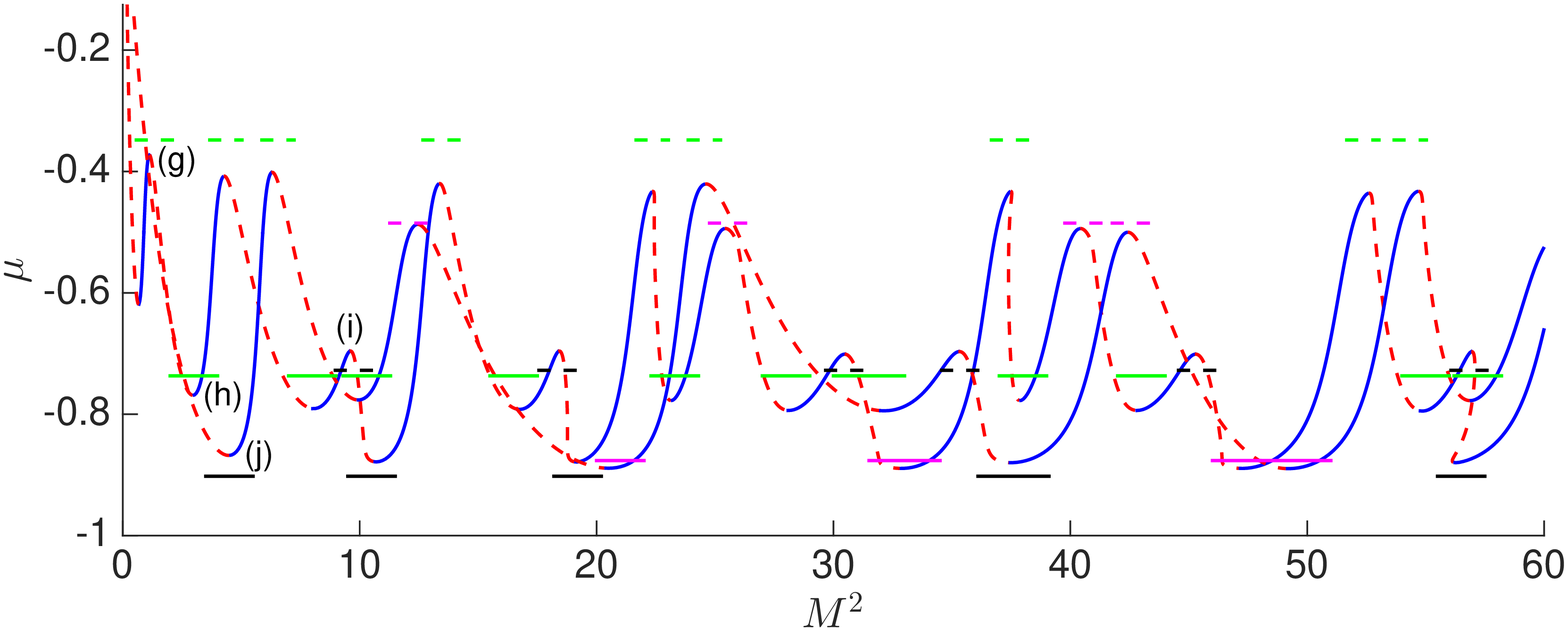}\label{subfig:bifur_ch_0_15}}
	\caption{The same as figure \ref{fig:bifur_cp}, but for honeycomb lattice. 
		The green, black, and magenta line colors correspond to our active-cell approximations of type 1, 2, and 3.}
	\label{fig:bifur_ch}
\end{figure*}

\begin{figure*}[h!]
	\centering
	\subfigure[]{\includegraphics[scale=0.23]{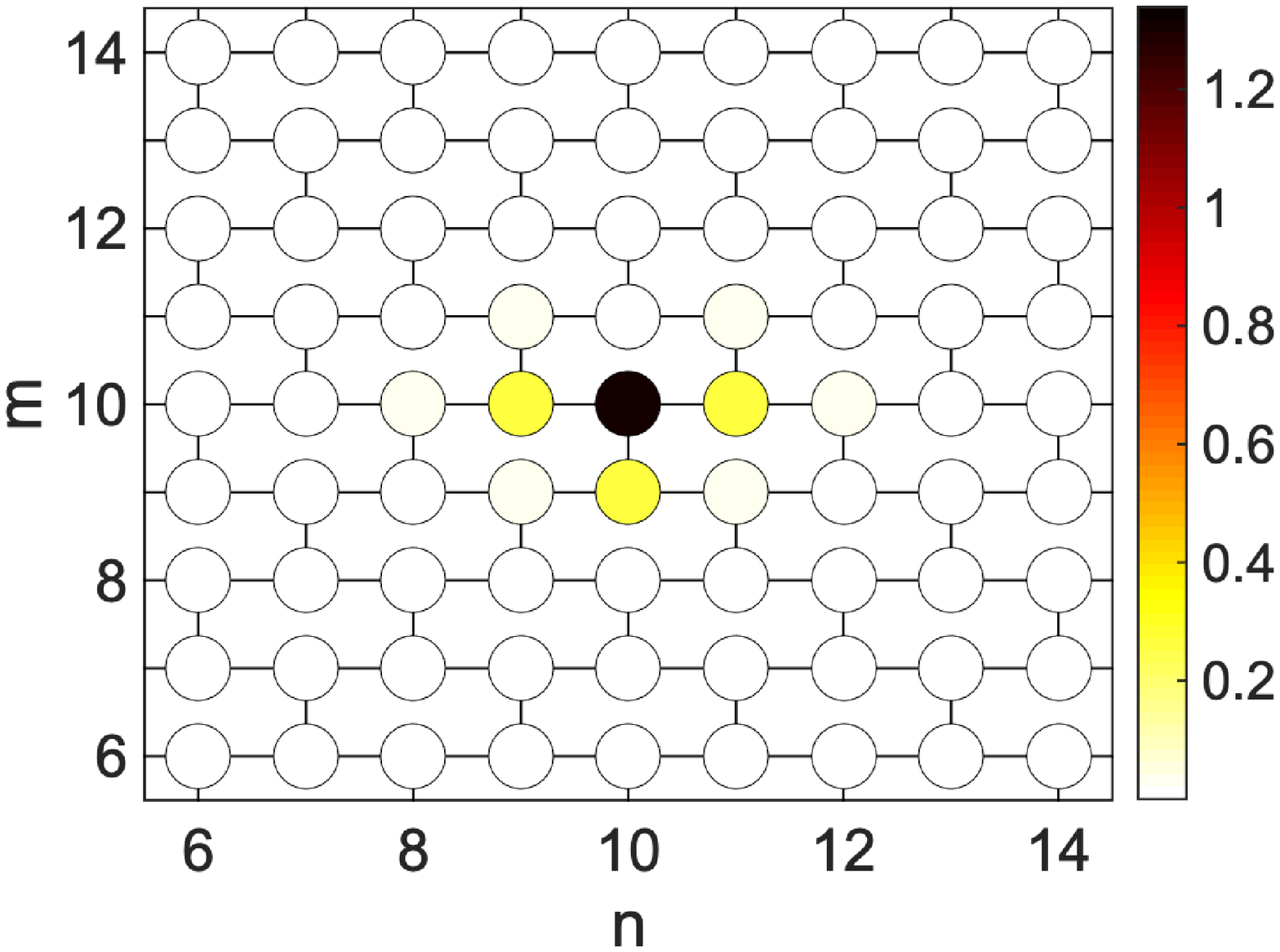}\label{subfig:prof_ch_0_05_a}}
	\subfigure[]{\includegraphics[scale=0.23]{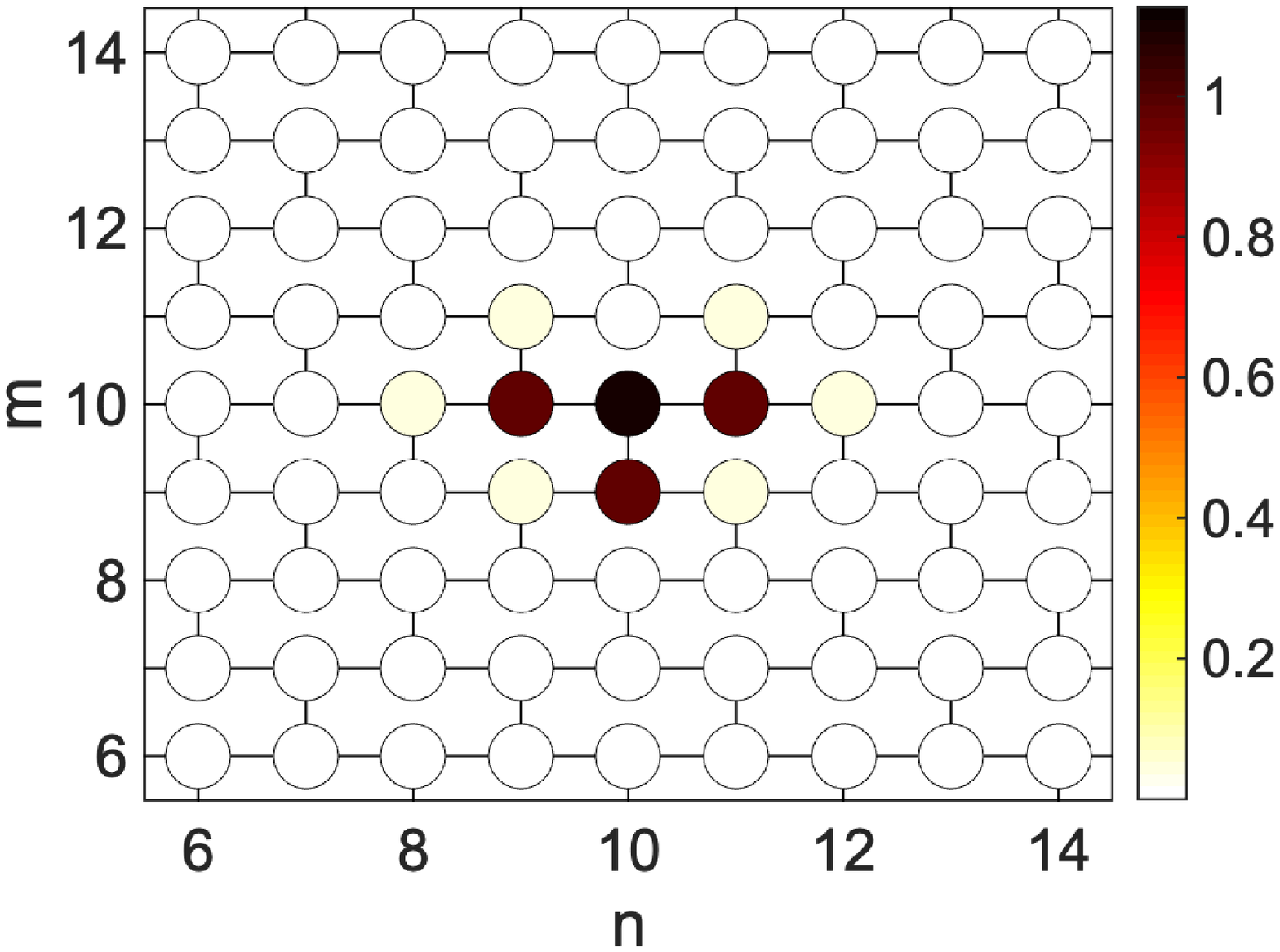}\label{subfig:prof_ch_0_05_b}}
	\subfigure[]{\includegraphics[scale=0.23]{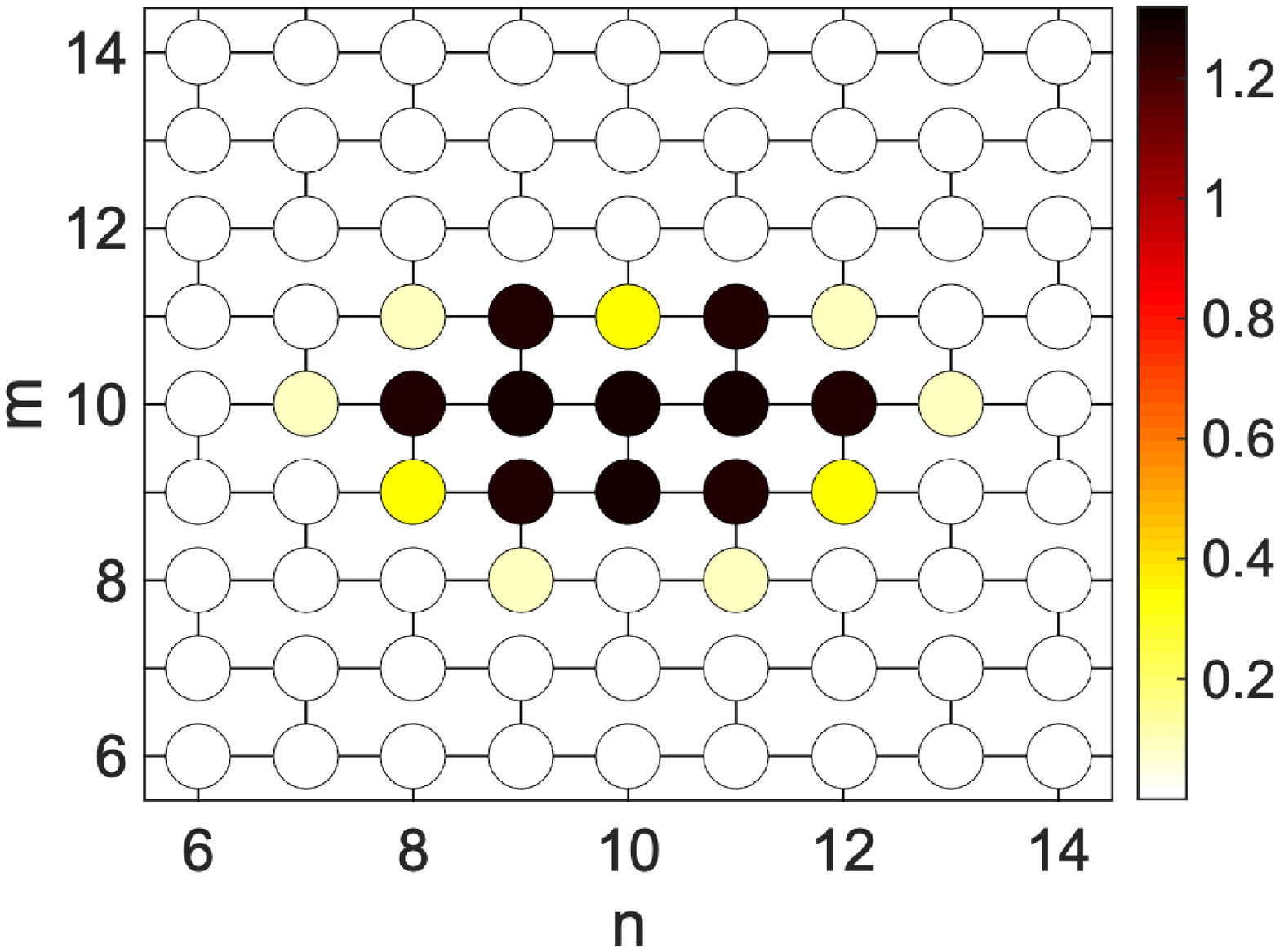}\label{subfig:prof_ch_0_05_c}}
	\subfigure[]{\includegraphics[scale=0.23]{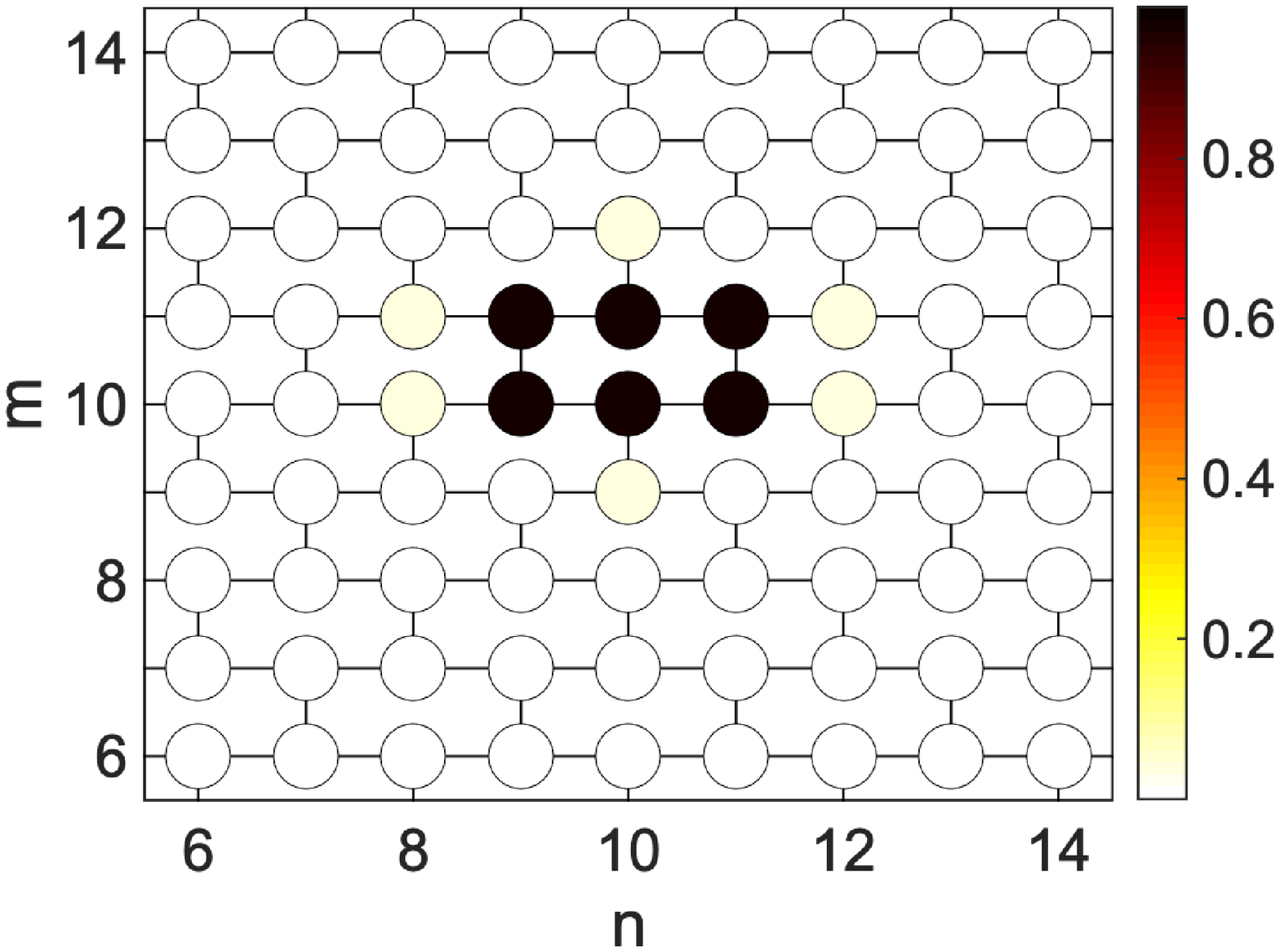}\label{subfig:prof_ch_0_05_d}}
	\subfigure[]{\includegraphics[scale=0.23]{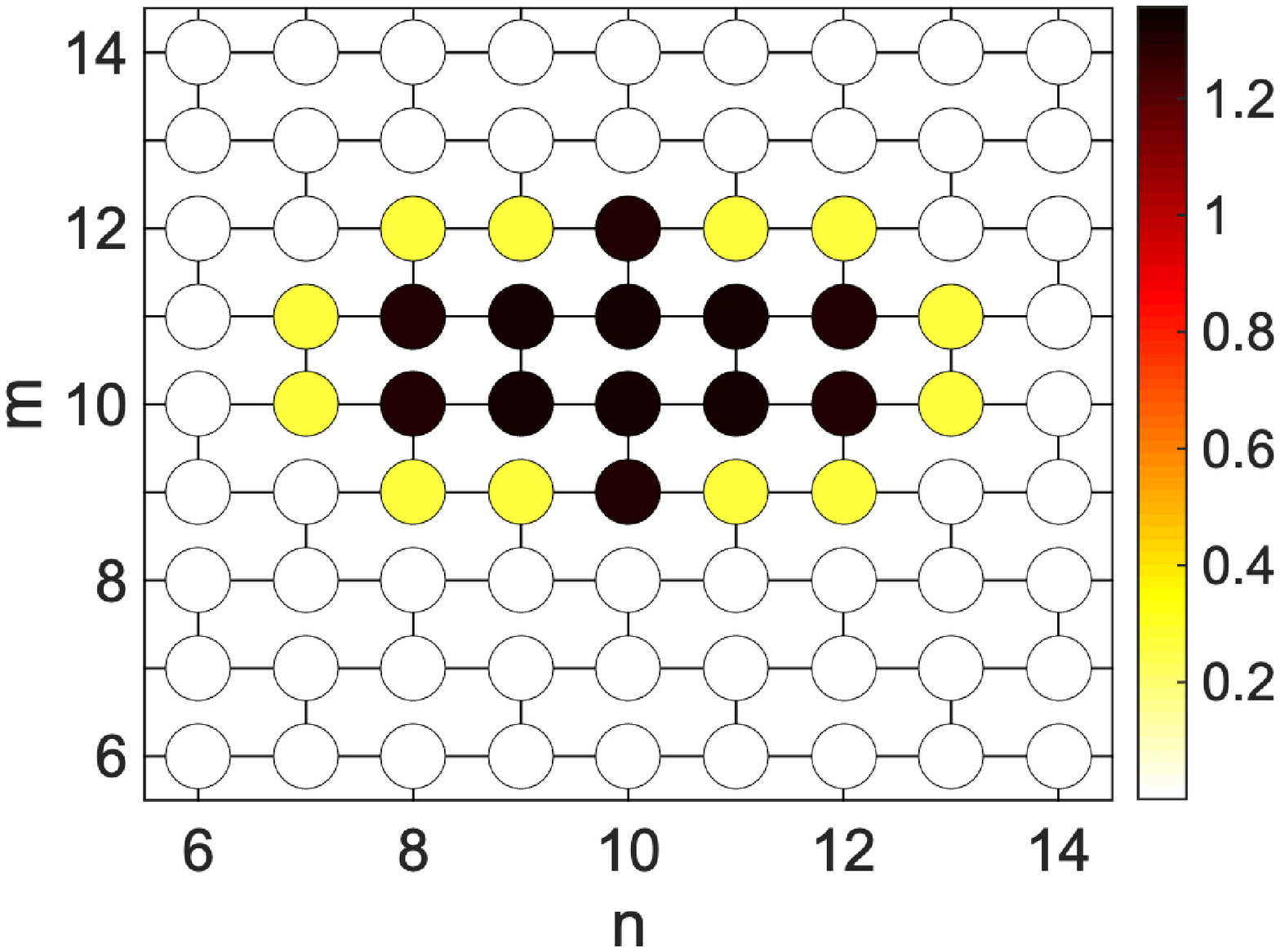}\label{subfig:prof_ch_0_05_e}}
	\subfigure[]{\includegraphics[scale=0.23]{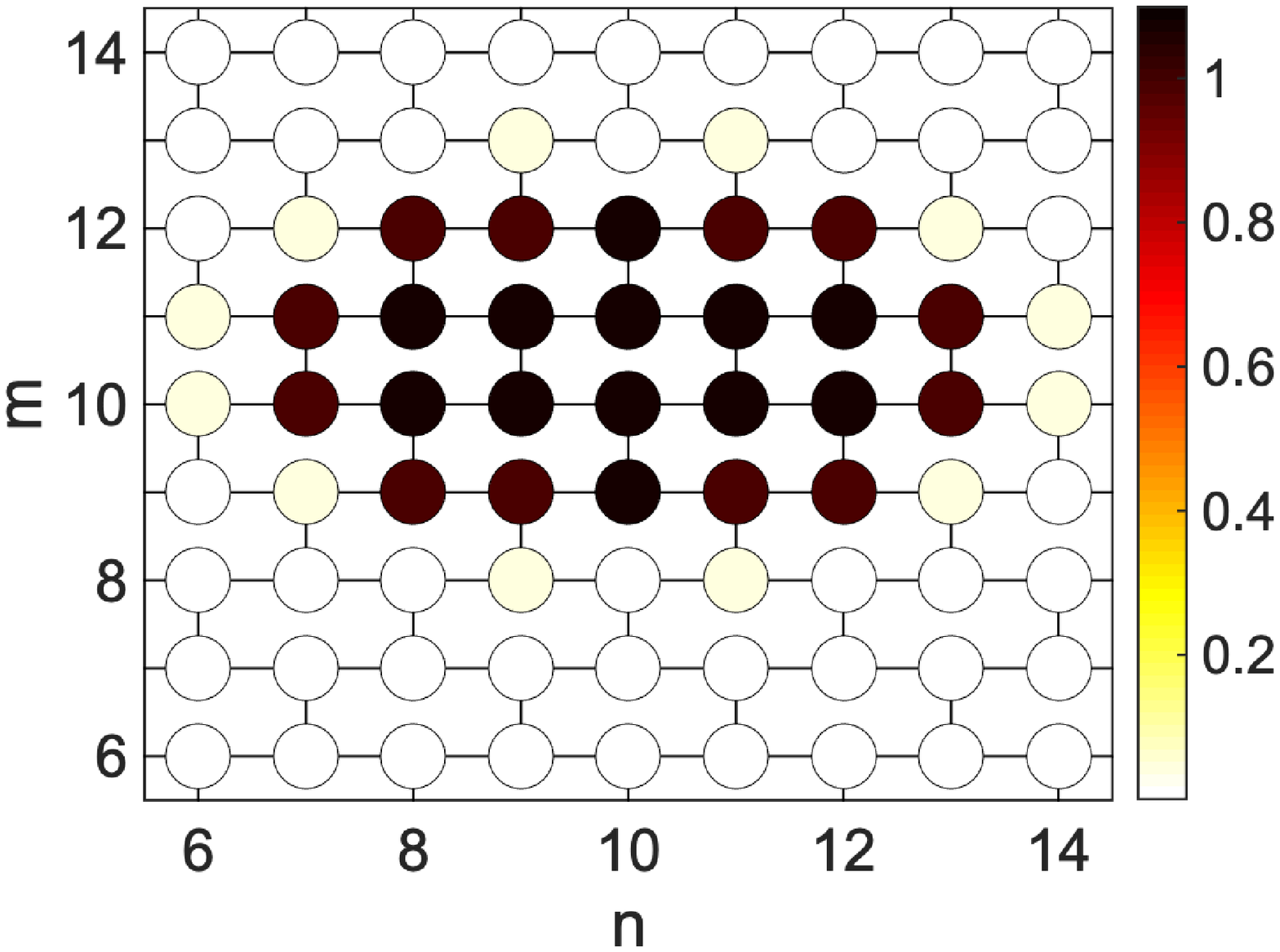}\label{subfig:prof_ch_0_05_f}}
	\caption{Top-view solution profiles in honeycomb lattice that correspond to points (a)-(f) in figure \ref{subfig:bifur_ch_0_05}.
	}
	\label{fig:prof_ch}
\end{figure*}

\begin{figure*}[t!]
	\centering
	\subfigure[{$c^{\varhexstar}=0.025$}]{\includegraphics[scale=0.425]{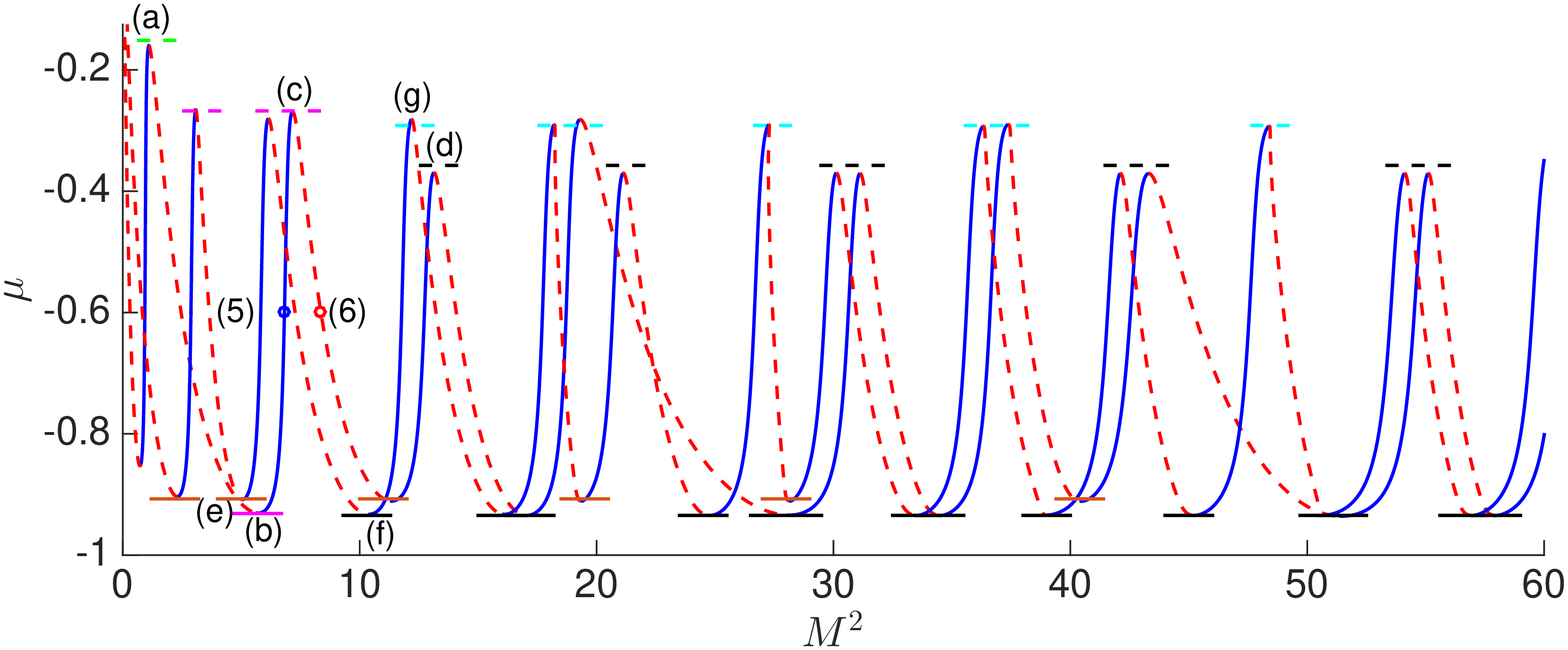}\label{subfig:bifur_ct_0_025}}
	\subfigure[{$c^{\varhexstar}=0.075$}]{\includegraphics[scale=0.425]{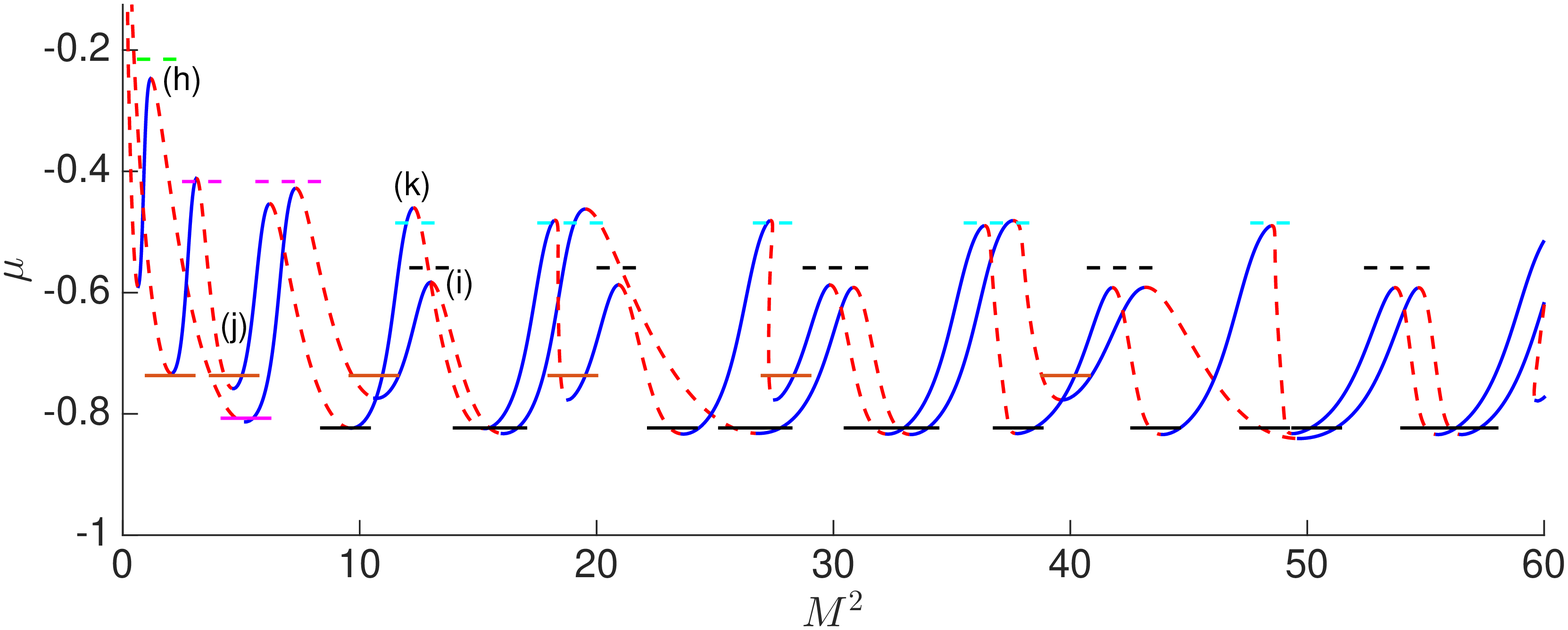}\label{subfig:bifur_ct_0_075}}
	\caption{The same as figures \ref{fig:bifur_cp} and \ref{fig:bifur_ch}, but for triangular lattice. 
		The green, black, magenta, brown, and cyan line colors correspond to our active-cell approximations of type 1, 2, 3, 4, and 5. }		
	\label{fig:bifur_ct}		
\end{figure*}

\begin{figure*}[h!]
	\centering
	\subfigure[]{\includegraphics[scale=0.19]{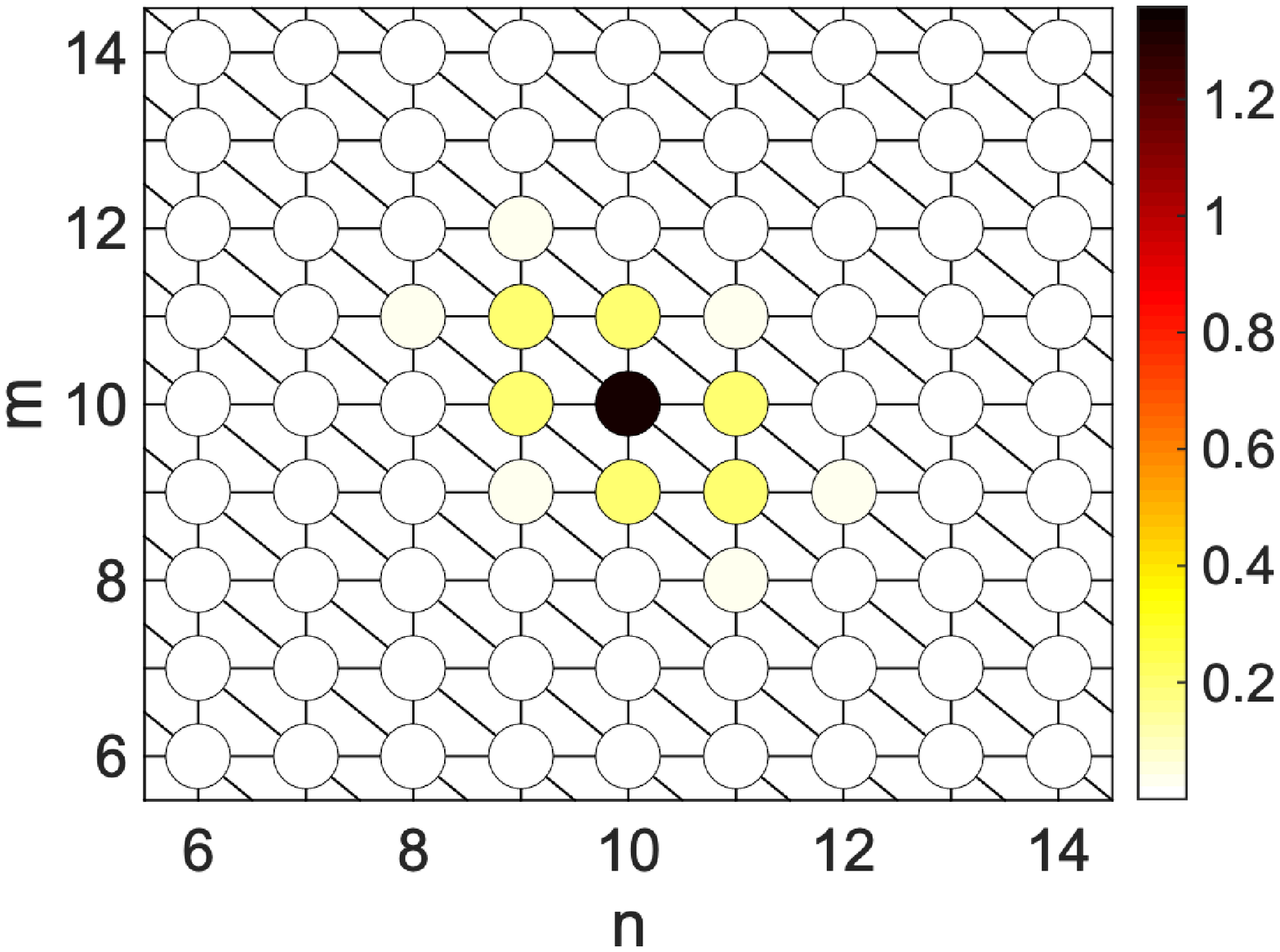}\label{subfig:prof_ct_0_025_a}}
	\subfigure[]{\includegraphics[scale=0.19]{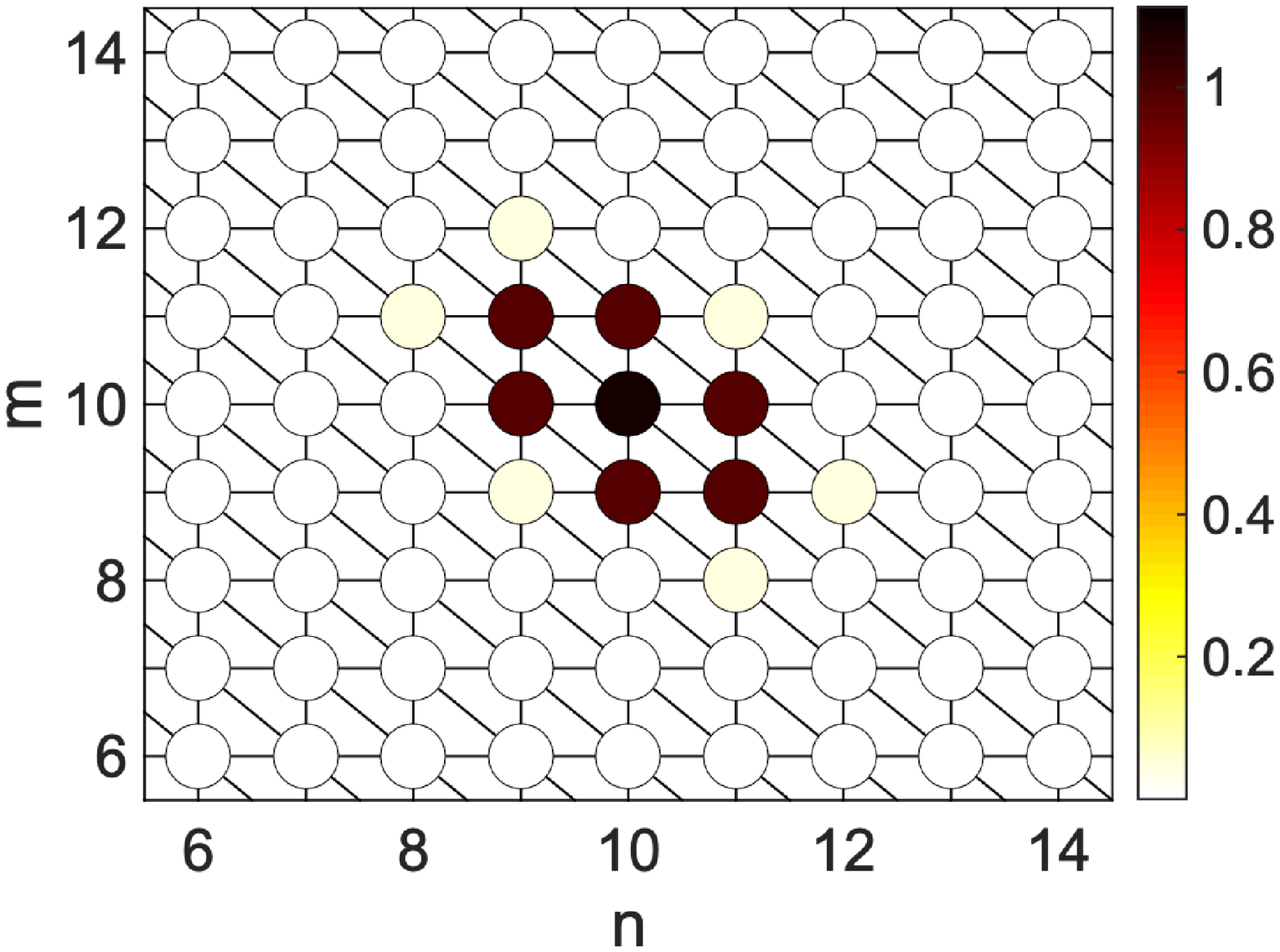}\label{subfig:prof_ct_0_025_b}}
	\subfigure[]{\includegraphics[scale=0.19]{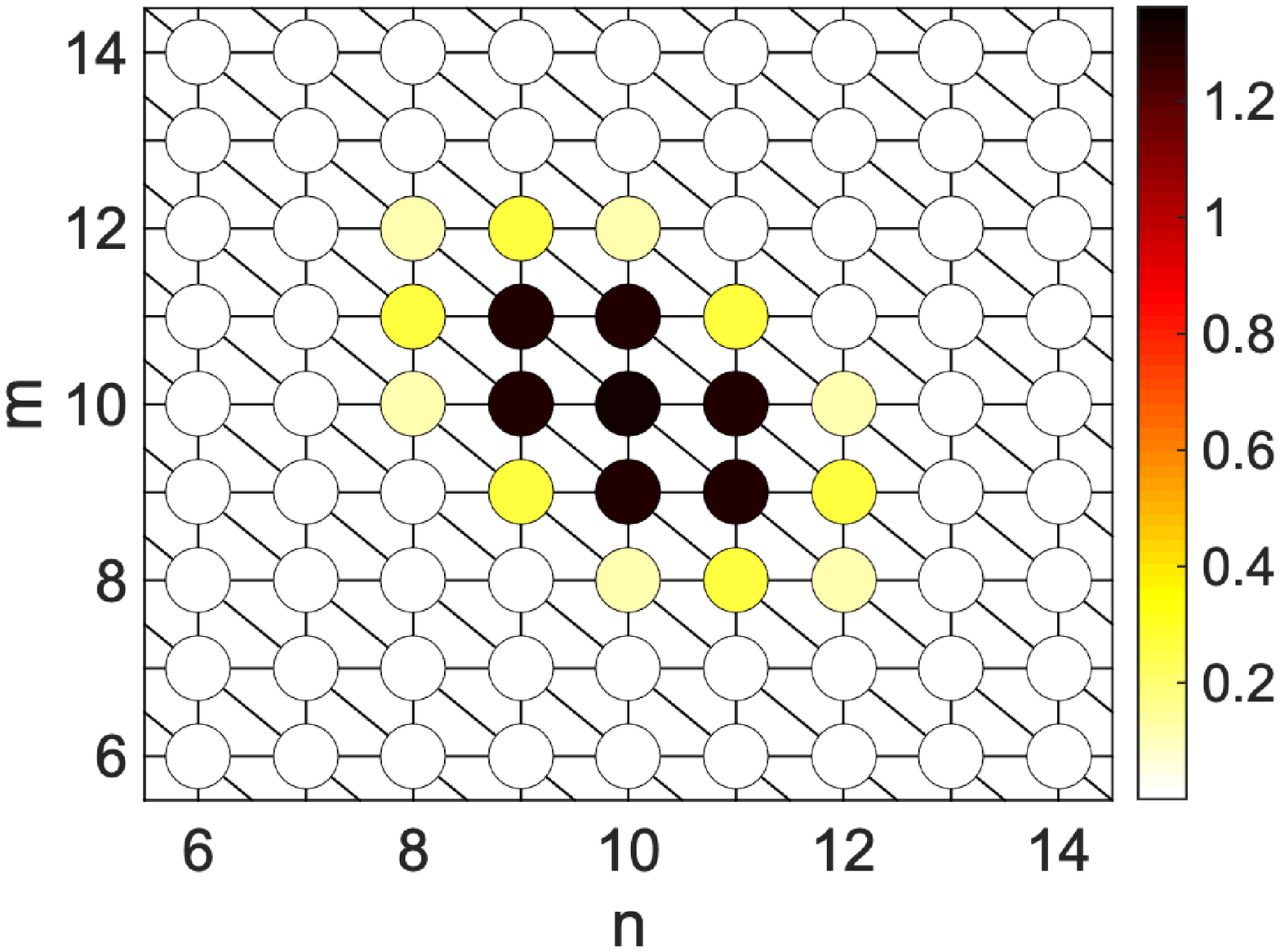}\label{subfig:prof_ct_0_025_c}}
	\subfigure[]{\includegraphics[scale=0.19]{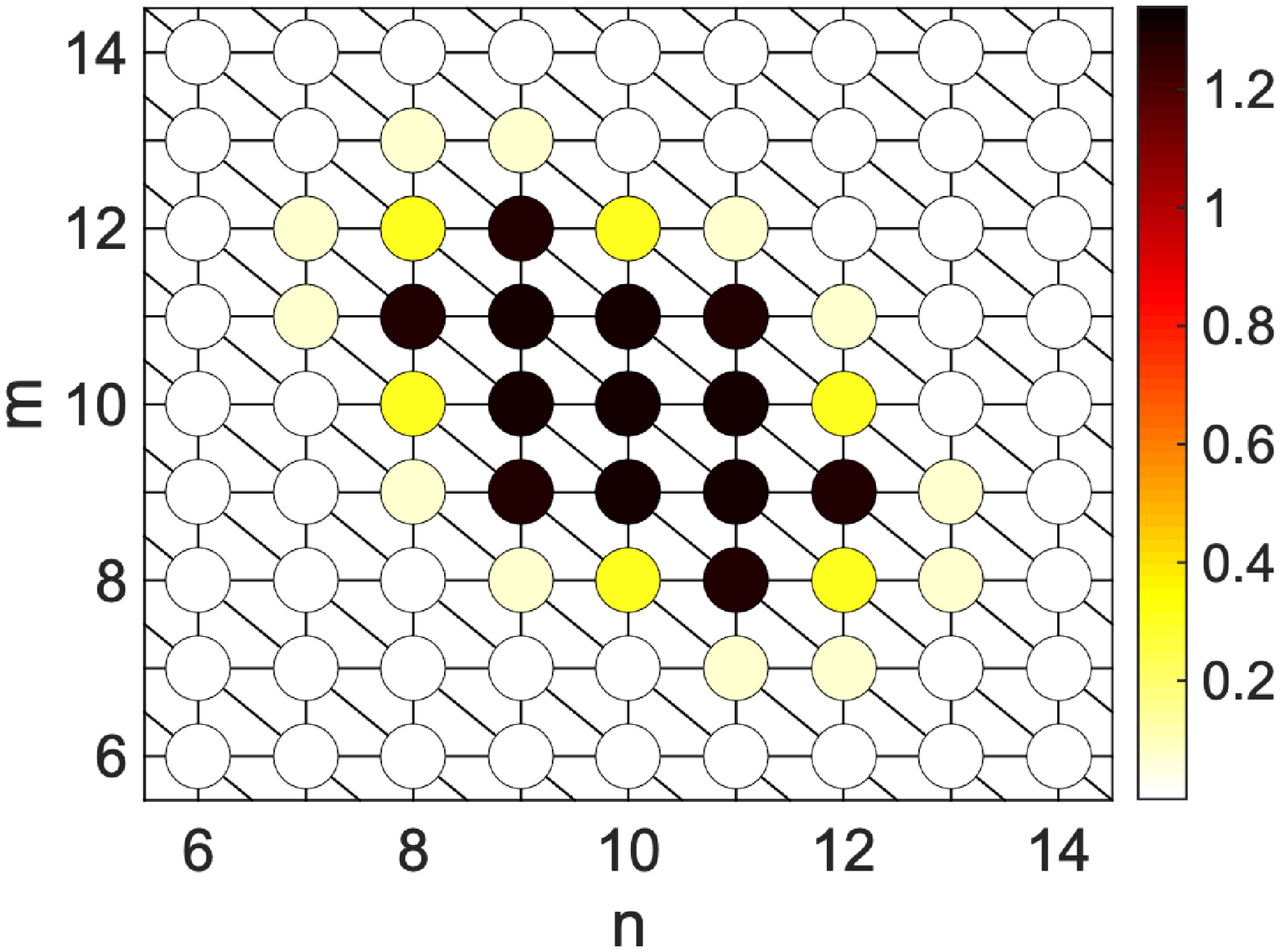}\label{subfig:prof_ct_0_025_d}}
	\subfigure[]{\includegraphics[scale=0.19]{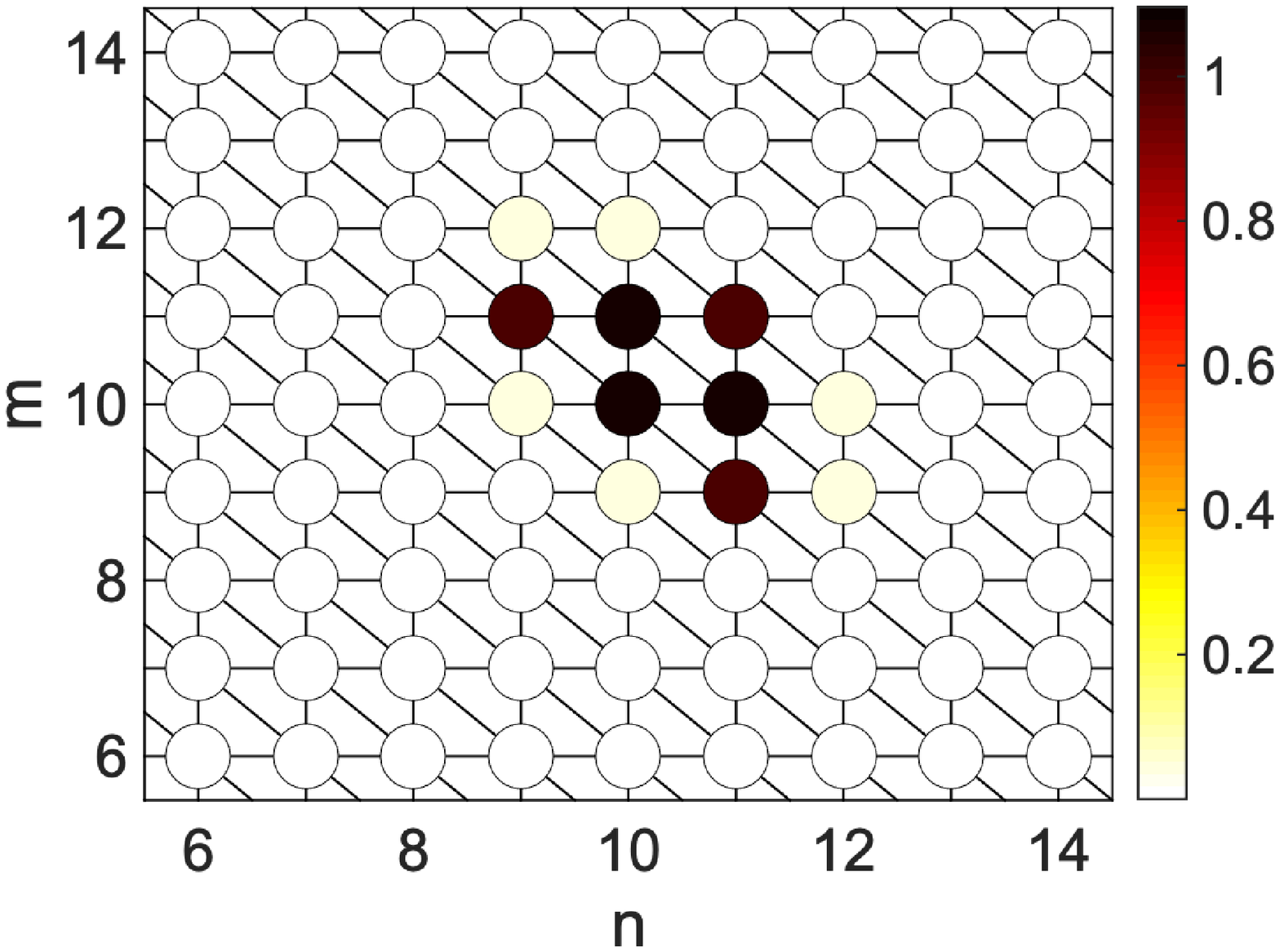}\label{subfig:prof_ct_0_025_e}}
	\subfigure[]{\includegraphics[scale=0.19]{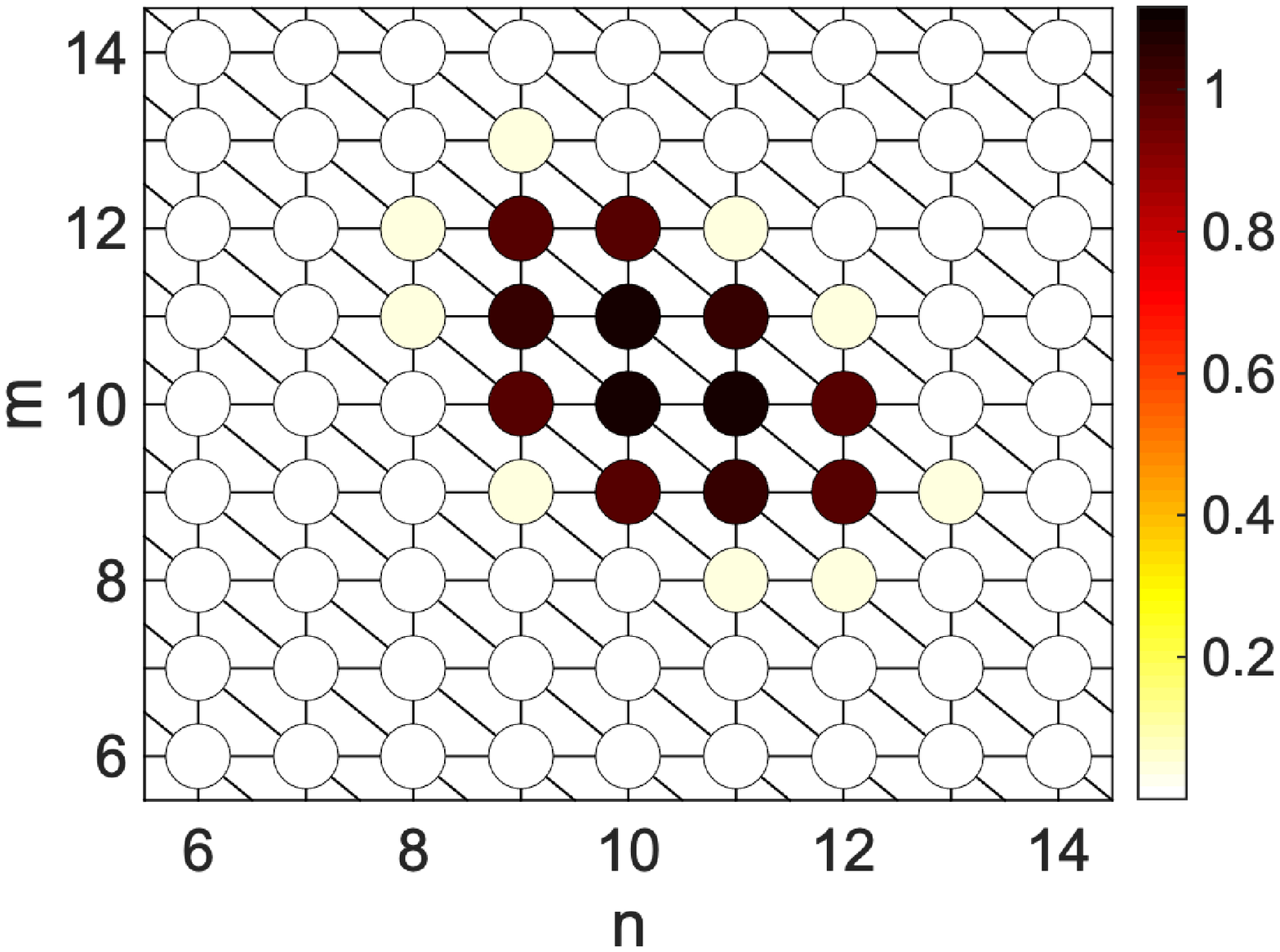}\label{subfig:prof_ct_0_025_f}}
	\subfigure[]{\includegraphics[scale=0.19]{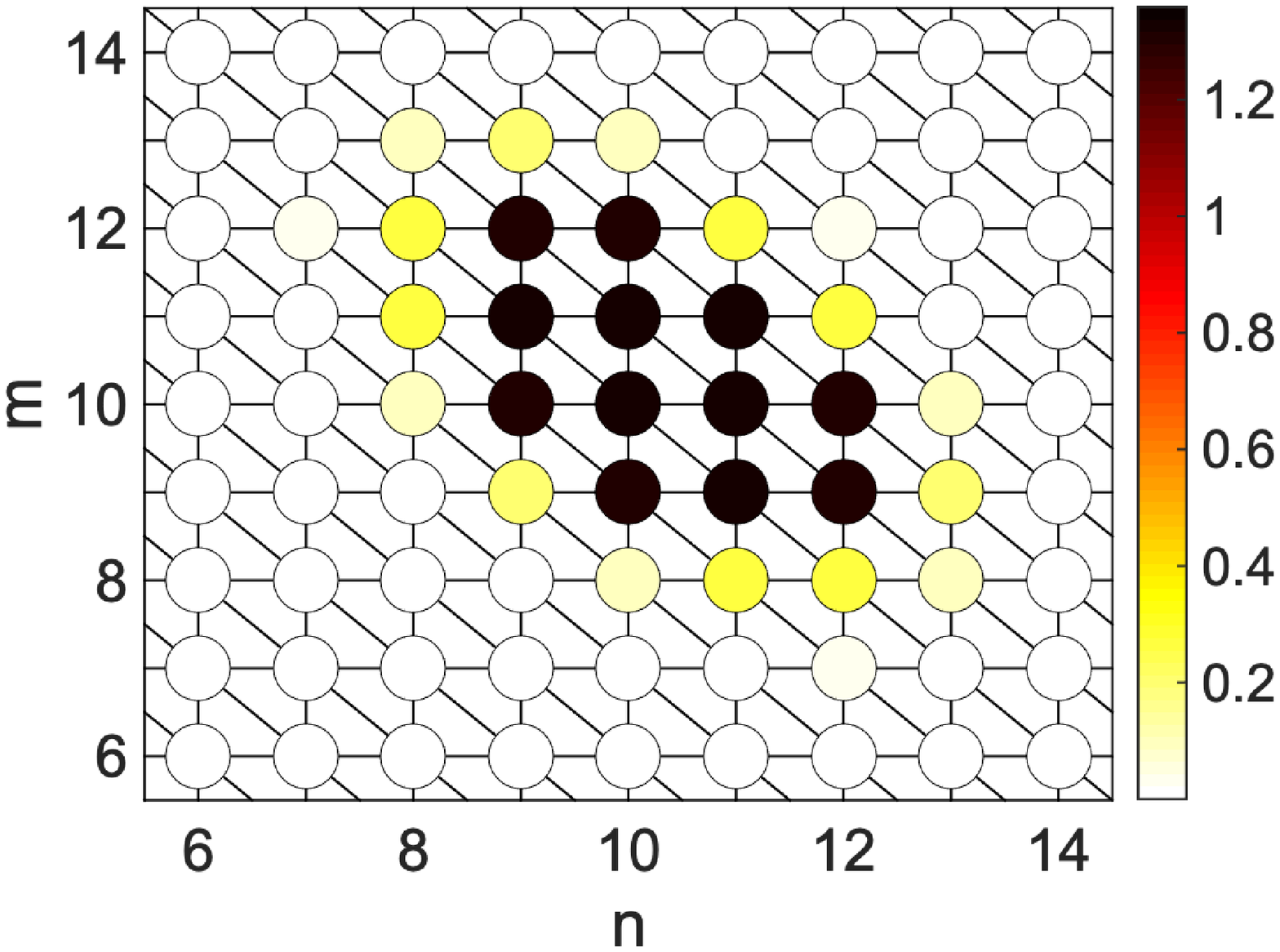}\label{subfig:prof_ct_0_025_g}}
	\caption{Top-view solution profiles in triangluar lattice that correspond to points in figure\ \ref{subfig:bifur_ct_0_025}.
	}
	\label{fig:prof_ct}
\end{figure*}

Fundamental site-centred solutions, i.e., solution profiles with odd number exited sites, of our governing equation are 
shown in figures\ \ref{subfig:site_square}, and  \ref{subfig:site_brick}, and \ref{subfig:site_triangular}.
On the other hand, bond-centred states are solutions  with the excited sites bonding with other sites and forming the simplest polygon. Our fundamental bond-centred solutions are shown in figures\ \ref{subfig:bond_square}, \ref{subfig:site_triangular}, and, \ref{subfig:bond_triangular}. Bifurcation diagrams of the fundamental localised solutions that show a snaking structure are presented in figures\ \ref{fig:bifur_cp} -  \ref{fig:prof_ct}. 

The snaking structures in the bifurcation diagrams exist at certain region called pinning region \cite{Pomeau1986}.
In 1D case, we have one pinning region, which in the limit $M\rightarrow\infty$ is bounded by two saddle-node bifurcations \cite{Taylor2010,Chong2009}.
In 2D case, saddle-node bifurcations may occur at several values of bifurcation parameter due to the presence of multiple types of saddle-node bifurcations, as we will show below. 
One can define that in 2D case, the pinning region is formed by the largest distance between the upper and lower saddle-node bifurcations.
The snaking structure in the bifurcation diagrams may also give complex snaking and isolas structures \cite{Taylor2010}.
In the next section, we will discuss the site-centred and bond-centred localised states and their snaking structures in square, honeycomb, and triangular lattices.

\subsection{Square lattice}
Figure\ \ref{fig:bifur_cp} shows bifurcation diagrams for square lattice at $c^+=0.05$ and $0.15$.
As we can see, the saddle-node bifurcations occur at several bifurcation parameters $\mu$.
Moreover, the distance between the ``upper'' and ``lower'' saddle-node bifurcations are getting smaller when the coupling strength $c^+$ increases.
In the continuum limit $c^+\rightarrow\infty$, the site-centred and bond-centred solutions merge as the snaking disappears, which also occurs in the 1D case.

Figures\ \ref{fig:prof_cp} shows several top-view (2D projection) of the solution profiles at the saddle-node bifurcations for $c^+=0.05$, which correspond to the bifurcation diagrams in figure\ \ref{fig:bifur_cp}.
One can see that, as the norm $M$ increases, the ``upper'' state invades ``lower'' state of the localised solution. 
In the 1D case, the mechanism of ``upper'' state invading ``lower'' state occurs around the fronts and it has two directions. 
In the square lattice, the patches clearly have four directions as one can deduce from the Laplacian operator $\Delta^+$. 

\subsection{Honeycomb lattice}
Figure\ \ref{fig:bifur_ch} shows bifurcation diagrams for honeycomb lattice at $c^\Yup=0.05$ and $0.15$.
In general, the properties of the snaking in the bifurcation diagrams are the same as square lattice.
However, at the same value of coupling strength, it has larger pinning regions compared to the square lattice.
It happens because the Laplacian operator $\Delta^{\Yup}$ for the honeycomb lattice has fewer interfaces, i.e., three ones, than the square lattice that connect the ``upper'' and ``lower'' states.

Figure\ \ref{fig:prof_ch} shows several top-view of the solution profiles at the saddle-node bifurcations for $c^\Yup=0.05$, which correspond to the snaking bifurcations in figure\ \ref{subfig:bifur_ch_0_05}.
The localised solution behaviour also has the same mechanism as that in the square lattice, where the ``upper'' state invades the ``lower'' state as the norm $M$ increases.
\subsection{Triangular lattice}
Figure\ \ref{fig:bifur_ct} shows bifurcation diagrams for triangular lattice at {$c^{\varhexstar}=0.025$ and $0.075$}.
The main difference between the square, honeycomb and triangular lattices is that the triangular lattice has relatively the smallest pinning region at the same value of coupling strength.
It happens because the triangular lattice has six interfaces in the Laplacian operator $\Delta^{\varhexstar}$.
Several top-view solution profiles at the saddle-node bifurcations are shown in figure\ \ref{fig:prof_ct} for {$c^{\varhexstar}=0.025$}, which correspond to bifurcation diagrams in figure\ \ref{subfig:bifur_ct_0_025}.

In summary, the number of cells that involve in the Laplacian operator determines the 2D lattice interface direction.
One can say that the width of the pinning region is inversely proportional to the number of interfaces, which is not the same as in the 1D case.


\section{Saddle-node bifurcation analysis}\label{sec2d:saddle_anal}
In general, when the coupling strength is quite small (weakly coupled), the solution consists of only three states, i.e., ``upper''  state $U_1$, ``lower'' state $U_0$, and interface (active-cell).
By using the assumption, we can assume that there are only three states that involve in the dynamics.
Hence, we can re-write equation \eqref{eq:dnls_all} into a simple ordinary differential equation \cite{Kusdiantara2017}
\begin{equation}
\dot{x}=F(x)=\mu x + 2x^3-x^5+Z(x),
\label{eq:active_cell_fun}
\end{equation}
where
\begin{equation}
Z(x)=c^\square\left(aU_{1}-bx\right),
\end{equation}
and $x$ is the interface.
The coefficients $a$ and $b$ are determined by the type of lattice and the number of ``upper" state $U_1$, ``lower" state $U_0$, and active-cell at the interface.
The list of coefficients $a$ and $b$ are shown in Table \ref{tab:coeff_active_cell} for the square, honeycomb, and triangular lattices. 
\begin{table}[tbhp!]
	\centering
	\footnotesize
	\begin{tabular}{|l|l|l|}
		\multicolumn{3}{c}{Square lattice} \\ 
		\hline
		\multicolumn{1}{|c|}{Type} & \multicolumn{1}{C{0.75cm}|}{$a$} & \multicolumn{1}{C{0.75cm}|}{$b$} \\ 
		\hline
		\multicolumn{1}{|c|}{1} & \multicolumn{1}{c|}{1} & \multicolumn{1}{c|}{4} \\ 
		\multicolumn{1}{|c|}{2} & \multicolumn{1}{c|}{2} & \multicolumn{1}{c|}{4} \\ 
		\multicolumn{1}{|c|}{3} & \multicolumn{1}{c|}{1} & \multicolumn{1}{c|}{3} \\ 
		\multicolumn{1}{|c|}{4} & \multicolumn{1}{c|}{1} & \multicolumn{1}{c|}{2} \\ 
		\hline
	\end{tabular}
\quad\quad
	\begin{tabular}{|l|l|l|}
		\multicolumn{3}{c}{Honeycomb lattice} \\ 
		\hline
		\multicolumn{1}{|c|}{Type} & \multicolumn{1}{C{0.75cm}|}{$a$} & \multicolumn{1}{C{0.75cm}|}{$b$} \\ 
		\hline
		\multicolumn{1}{|c|}{1} & \multicolumn{1}{c|}{1} & \multicolumn{1}{c|}{3} \\ 
		\multicolumn{1}{|c|}{2} & \multicolumn{1}{c|}{2} & \multicolumn{1}{c|}{3} \\ 
		\multicolumn{1}{|c|}{3} & \multicolumn{1}{c|}{1} & \multicolumn{1}{c|}{2} \\ 
		\hline
	\end{tabular}
\quad\quad
	\begin{tabular}{|l|l|l|}
		\multicolumn{3}{c}{Triangular lattice} \\ 
		\hline
		\multicolumn{1}{|c|}{Type} & \multicolumn{1}{C{0.75cm}|}{$a$} & \multicolumn{1}{C{0.75cm}|}{$b$} \\ 
		\hline
		\multicolumn{1}{|c|}{1} & \multicolumn{1}{c|}{1} & \multicolumn{1}{c|}{4} \\ 
		\multicolumn{1}{|c|}{2} & \multicolumn{1}{c|}{2} & \multicolumn{1}{c|}{5} \\ 
		\multicolumn{1}{|c|}{3} & \multicolumn{1}{c|}{2} & \multicolumn{1}{c|}{5} \\ 
		\multicolumn{1}{|c|}{4} & \multicolumn{1}{c|}{2} & \multicolumn{1}{c|}{6} \\ 
		\multicolumn{1}{|c|}{5} & \multicolumn{1}{c|}{2} & \multicolumn{1}{c|}{4} \\ 
		\hline
	\end{tabular}
	\caption{List of coefficients in the active-cell approximations for all lattices.}
	\label{tab:coeff_active_cell}
\end{table}
\begin{figure}[t!]
	\centering
	\includegraphics[scale=0.6]{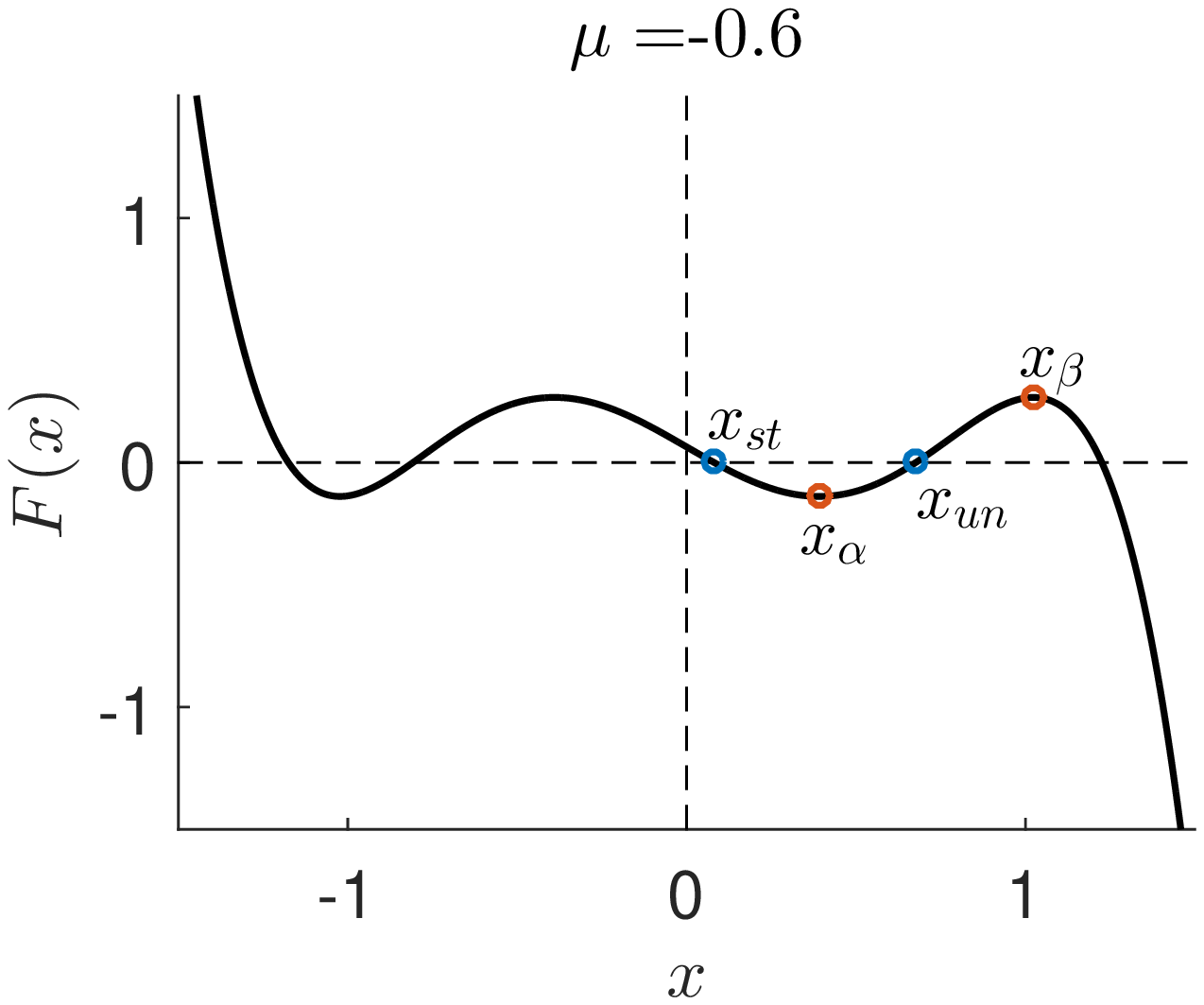}
	\caption{Active-cell function type 1 for square lattice at $c^+=0.05$.
		$x_\alpha$ and $x_\beta$ indicate as lower and upper saddle-node bifurcations.
		$x_{st}$ and $x_{un}$ represent the stable and unstable cell solution.
	}
	\label{fig:active_cell}
\end{figure}
In general, $F(x)$ can have five real roots.
Note that only two of them are related to the snaking as they correspond to the ``upper'' and ``lower'' saddle-node bifurcations. 
One can recognise that a saddle-node bifurcation is a condition when $F(x)$ at the local minimum $x=x_\alpha$ and local maximum $x=x_\beta$ vanishes, which correspond to the ``lower'' and ``upper'' saddle-node bifurcations, respectively.
It is quite straightforward to obtain that
\begin{equation}
x_{\alpha,\beta}=\left(\frac{3}{5}\pm\frac{1}{5}\sqrt{9+5\left(\mu-c^\square b\right)}\right)^\frac{1}{2}.
\end{equation}

We found that there are several types of saddle-node bifurcations in the snaking diagrams. 
By identifying the types of saddle-node bifurcations, we can apply the active-cell approximation to the solution profiles.
In particular, we have four, three, and five types of saddle-node bifurcations for the square, honeycomb, and triangular lattices, which are classified by the numbers and positions of the ``upper'' state $U_1$, ``lower'' state $U_0$, and active-cell in their solution profiles, see figures\ \ref{fig:active_cell_sq}, \ref{fig:active_cell_hc}, and \ref{fig:active_cell_tr}.
One also can say that the active-cell approximation is a rotation invariant at their center or axes.
The approximations for all of the saddle-node bifurcation results are shown in figures\ \ref{fig:bifur_cp}, \ref{fig:bifur_ch}, and \ref{fig:bifur_ct}.


\subsection{Square lattice}
\begin{figure*}[h!]
	\centering
	\subfigure[$Z(x)=c^+\left(U_1-4x\right)$]{\includegraphics[scale=0.34]{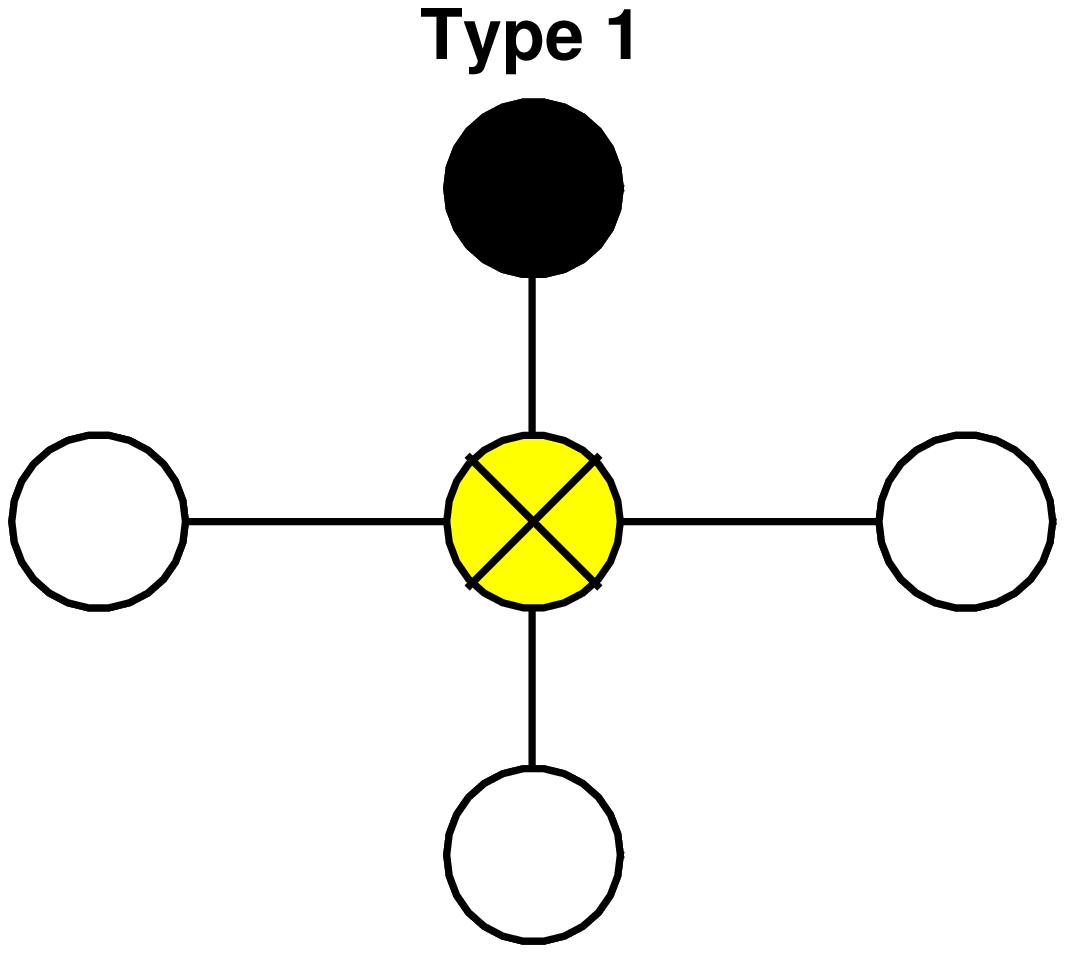}\label{subfig:sq_tipe1}}
	\subfigure[$Z(x)=c^+\left(2U_1-4x\right)$]{\includegraphics[scale=0.34]{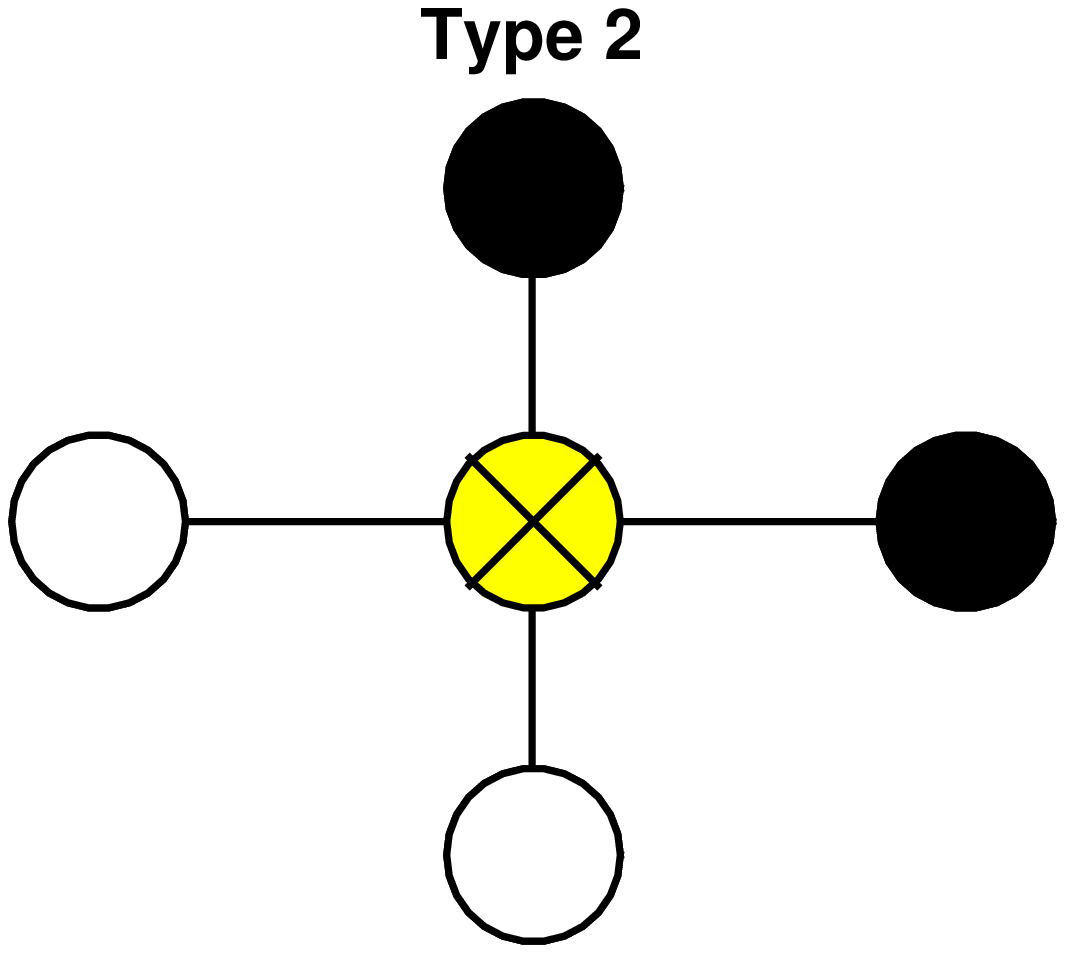}\label{subfig:sq_tipe2}}
	\subfigure[$Z(x)=c^+\left(U_1-3x\right)$]{\includegraphics[scale=0.34]{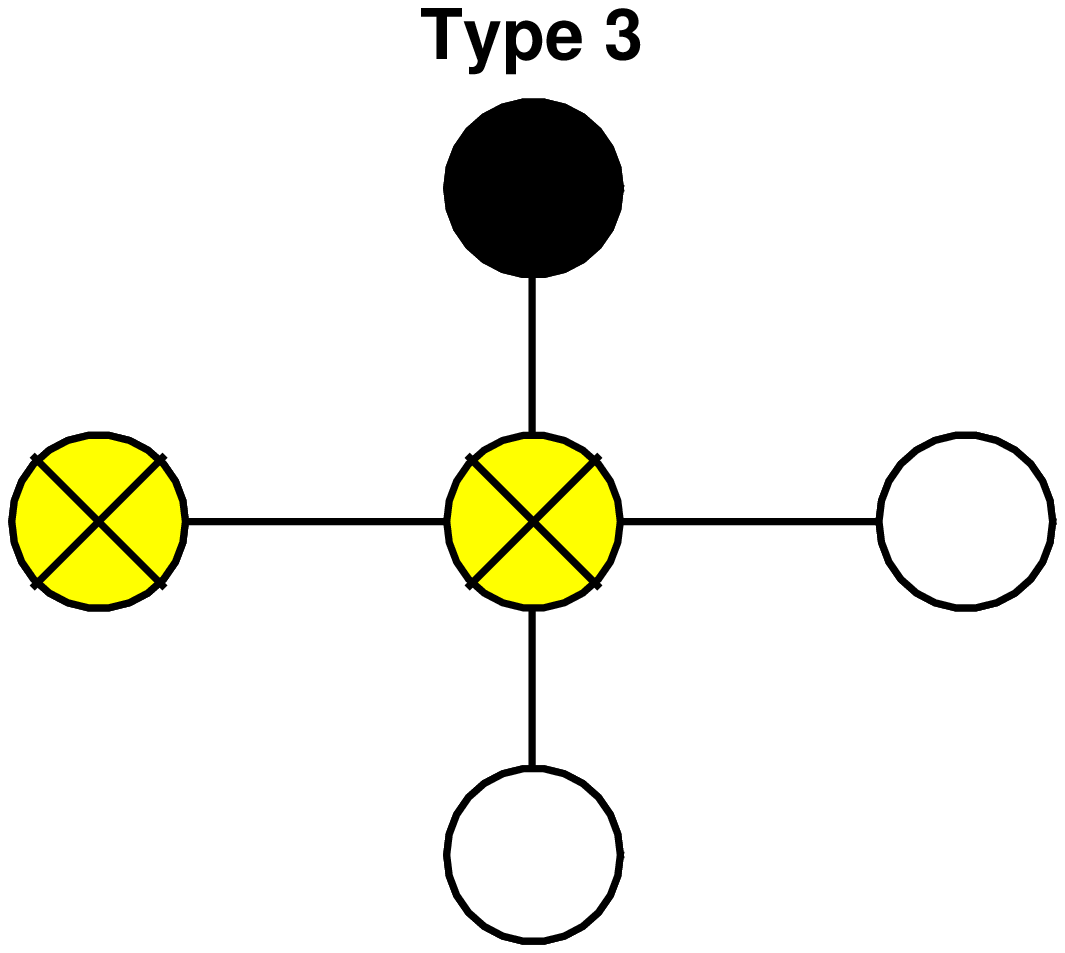}\label{subfig:sq_tipe3}}
	\subfigure[$Z(x)=c^+\left(U_1-2x\right)$]{\includegraphics[scale=0.34]{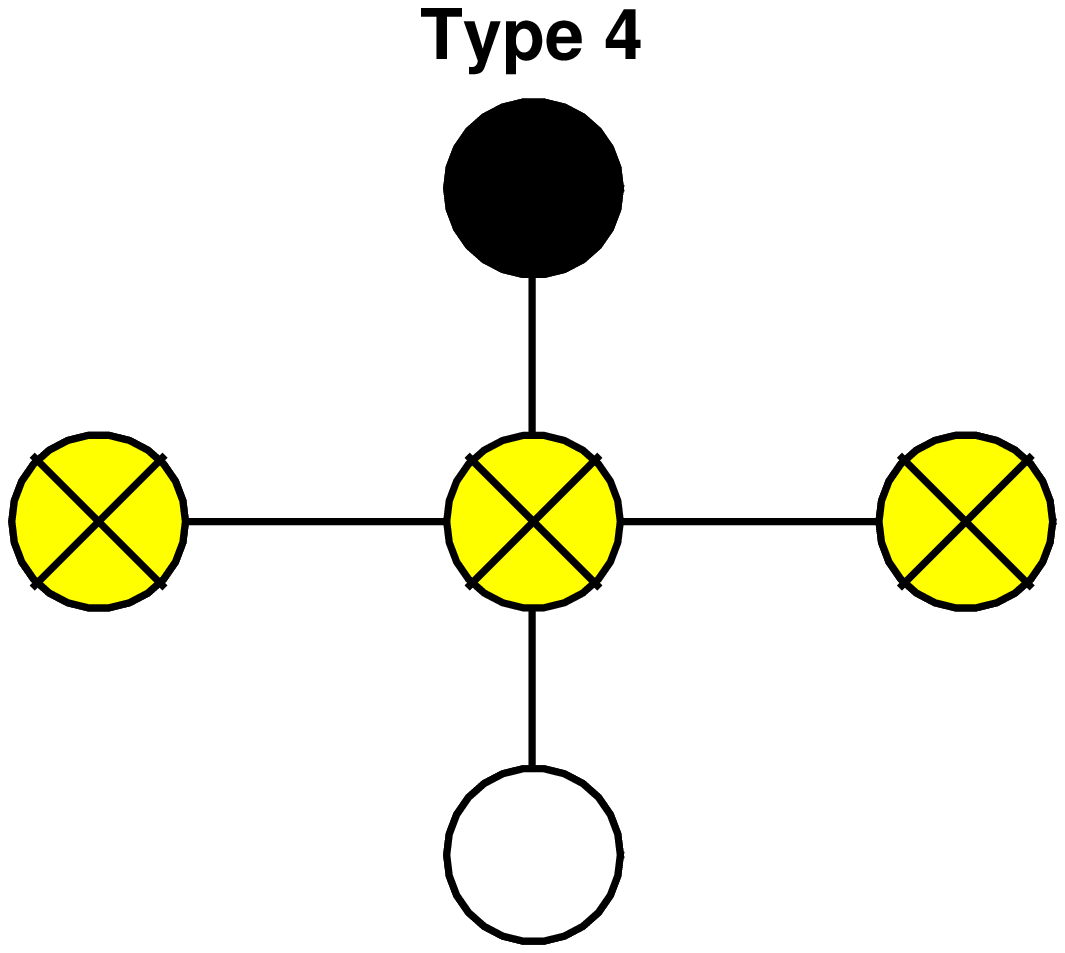}\label{subfig:sq_tipe4}}
	\includegraphics[scale=0.34]{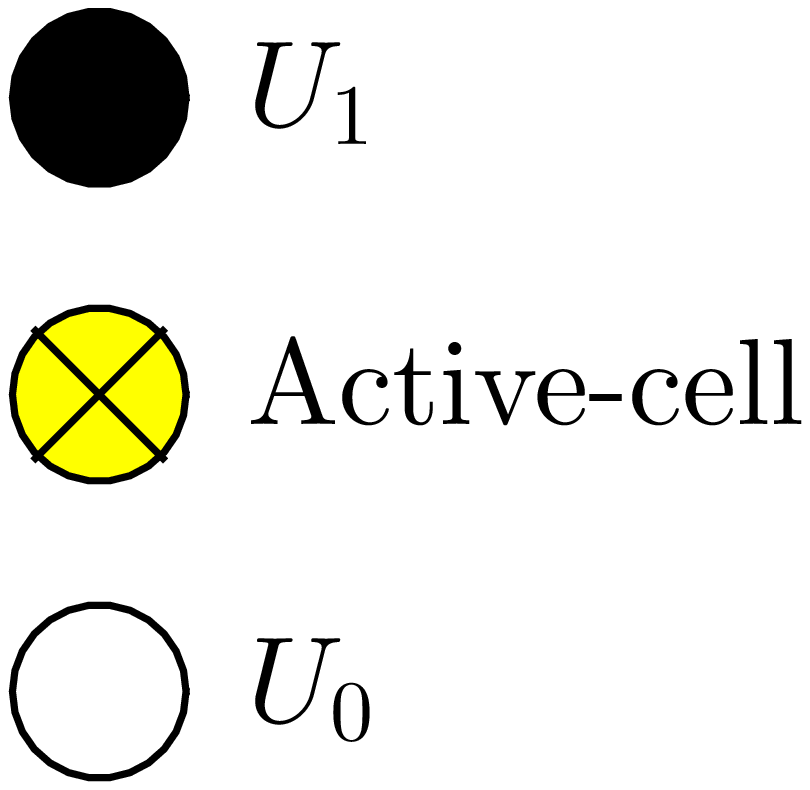}
	\caption{Types of active-cell approximations for square lattice.}
	\label{fig:active_cell_sq}
\end{figure*}
Figure \ref{subfig:bifur_cp_0_05} shows several types of saddle-node bifurcations and their approximations for the square lattice at $c^+=0.05$.
In general, there are four types of saddle-node bifurcations for the square lattice, 
see figure \ref{fig:active_cell_sq}.

The bifurcations at points (a) and (b), (c) and (d), (f) and (g), and (e) belong to type 1, 2, 3, and 4, respectively.
The saddle-node bifurcations of type 1 and 3 only appear in site-centred solutions. Meanwhile, type 2 and 4 may appear in both site-centred and bond-centred solutions. The approximations for all of the types give good agreement for the ``lower'' and ``upper'' saddle-node bifurcations. %
Note that type 2 and 4 mostly appear in the large value of norm $M$.

Figure\ \ref{subfig:bifur_cp_0_15} shows the approximation results of the saddle-node bifurcations for square lattice at $c^+=0.15$.
By comparing between $c^+=0.05$ and $0.15$, one can see that the active-cell approximations give better results at smaller coupling strength. 
As we can see, the active-cell approximations fail to approximate points (h), (i), (j), and (k).
These happen because the patch interfaces do not satisfy the active-cell approximation assumption. 
As the coupling is getting larger, one will have more cells with different amplitudes around the interfaces that are also excited. 
\subsection{Honeycomb lattice}

\begin{figure*}[htbp]
	\centering
	\subfigure[$Z(x)=c^\Yup\left(U_1-3x\right)$]{\includegraphics[scale=0.34]{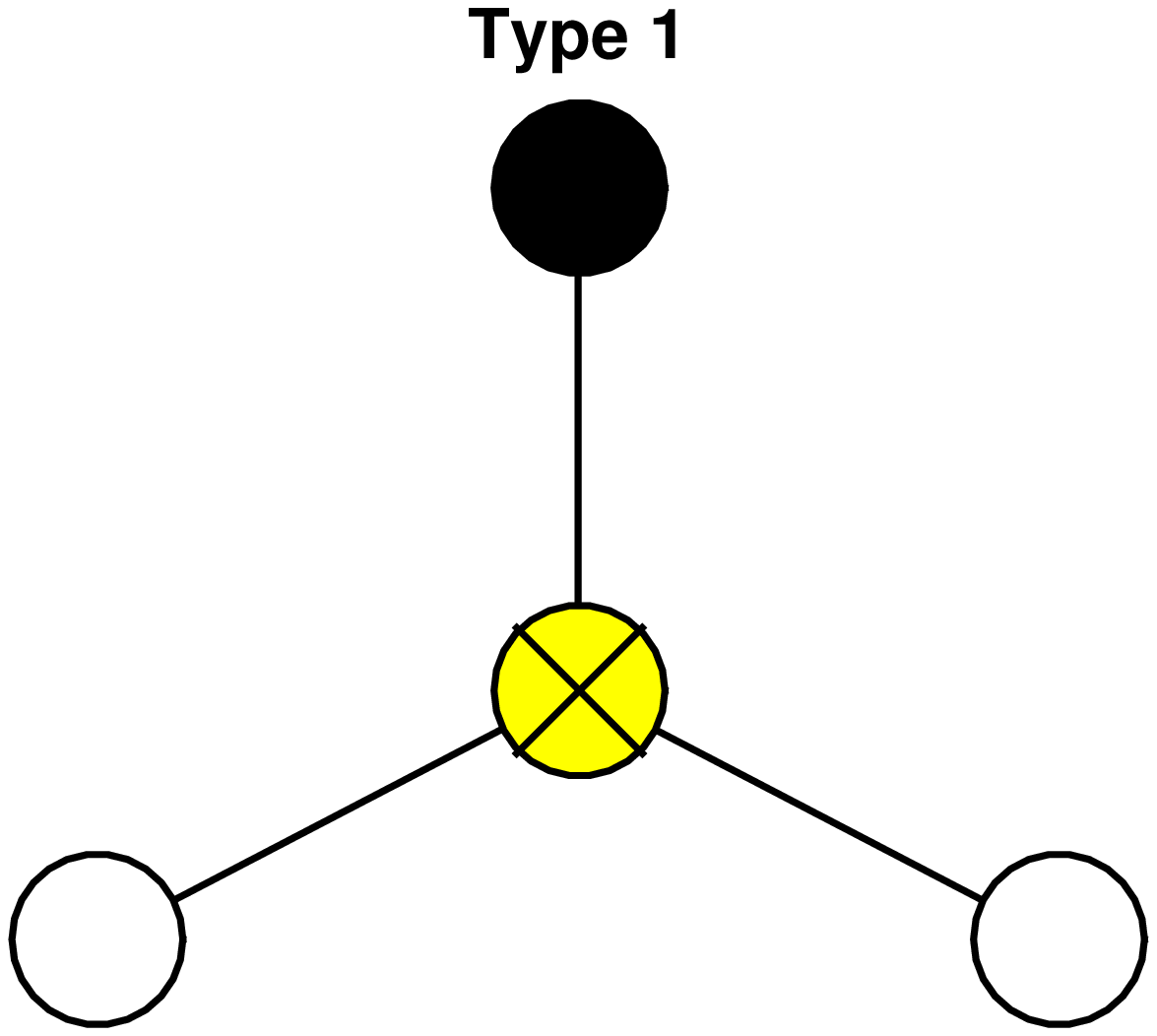}\label{subfig:hc_tipe1}}
	\subfigure[$Z(x)=c^\Yup\left(2U_1-3x\right)$]{\includegraphics[scale=0.34]{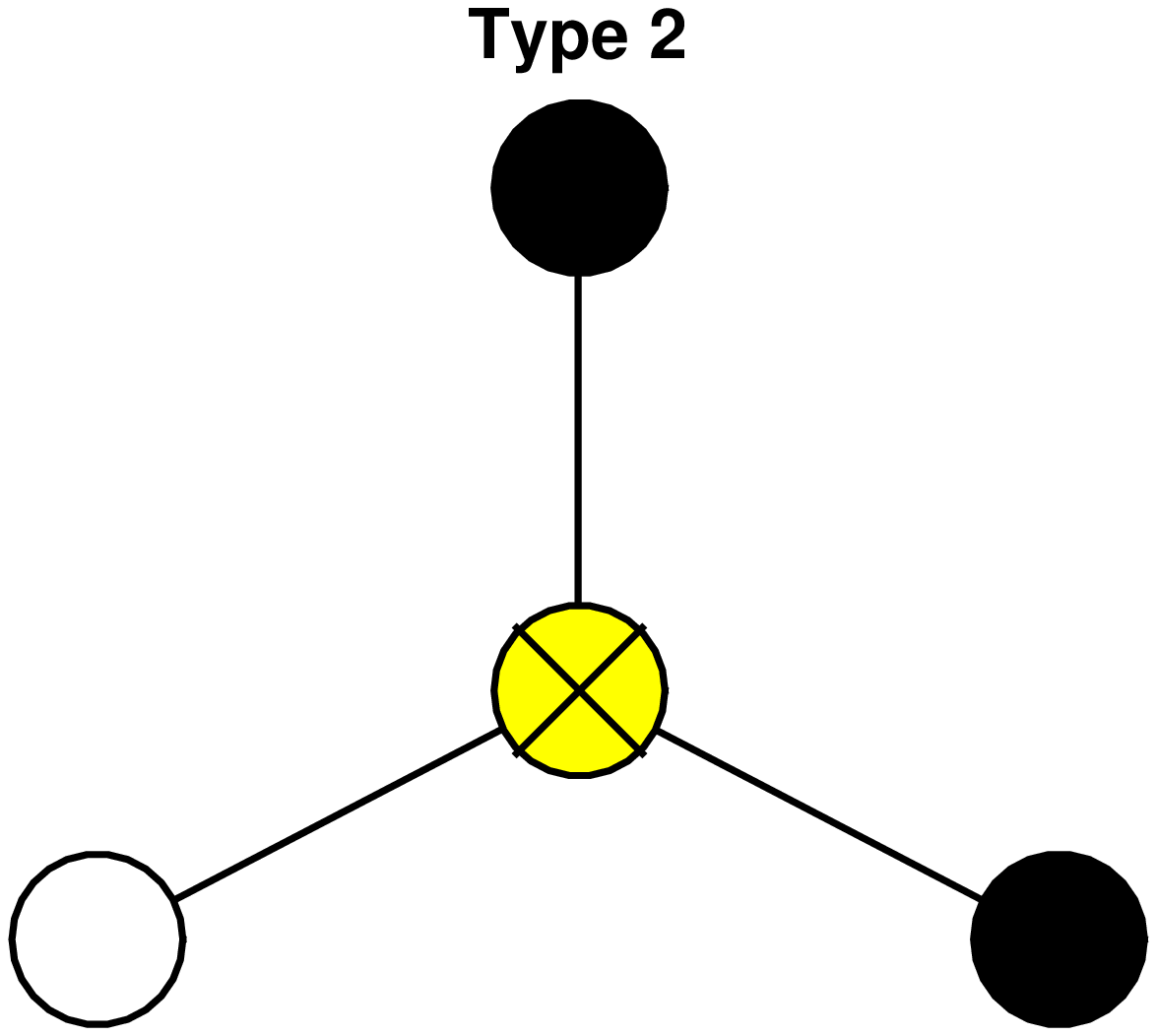}\label{subfig:hc_tipe2}}
	\subfigure[$Z(x)=c^\Yup\left(U_1-2x\right)$]{\includegraphics[scale=0.34]{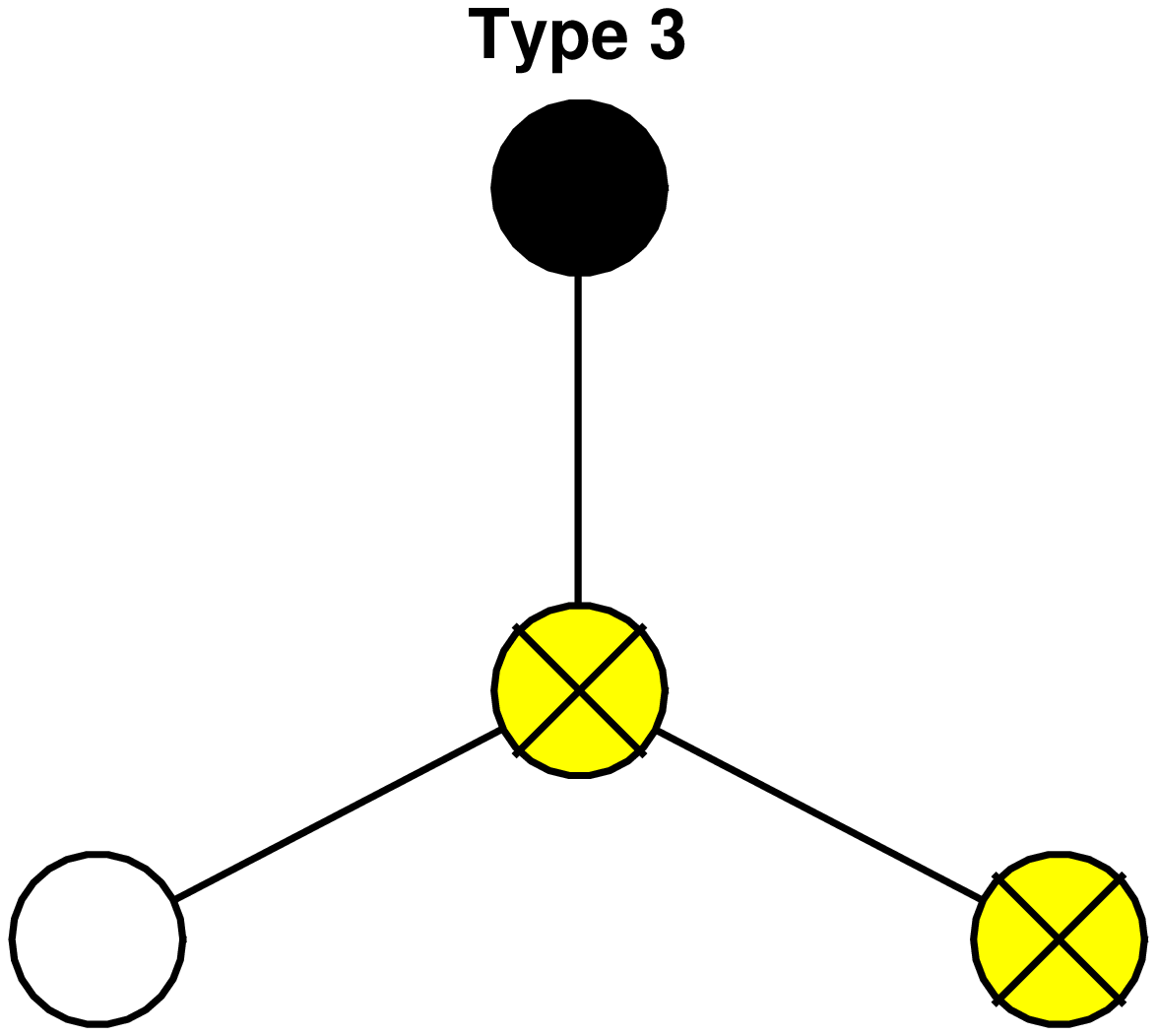}\label{subfig:hc_tipe3}}
	\caption{Types of active-cell approximations for honeycomb lattice.}
	\label{fig:active_cell_hc}
\end{figure*}

Figure \ref{subfig:bifur_ch_0_05} shows several types of saddle-node bifurcations and their approximations for the honeycomb lattice at $c^\Yup=0.05$.
In general, there are three types of saddle-node bifurcations for the honeycomb lattice, 
see figure \ref{fig:active_cell_hc}.

%
%
%

The bifurcations at points (a) and (b), (c) and (d), and (f) and (g) are belong to type 1, 2, and 3, respectively.
All of the types of saddle-node bifurcations appear in site-centred and bond-centred solutions.
Generally, the approximation for all of the types give good agreement for the ``lower'' and ``upper'' saddle-node bifurcations. %

Figure\ \ref{subfig:bifur_ch_0_15} shows the approximation results of the saddle-node bifurcations for honeycomb lattice at $c^\Yup=0.15$.
By comparing between $c^\Yup=0.05$ and $0.15$, one can see that the active-cell approximations give better results at smaller coupling strength. 
As we can see, the active-cell approximations fail to approximate points (g), (h), (i), and (j).
These also happen due to the solution interfaces that no longer satisfy the active-cell approximation assumption. As the coupling is getting larger, we have more cells with different amplitudes around the interfaces that are also excited, which also happen in square lattice. 


%

\subsection{Triangular lattice}
\begin{figure*}[htbp]
	\centering
	\subfigure[$Z(x)=c^{\varhexstar}\left(U_1-4x\right)$]{\includegraphics[scale=0.34]{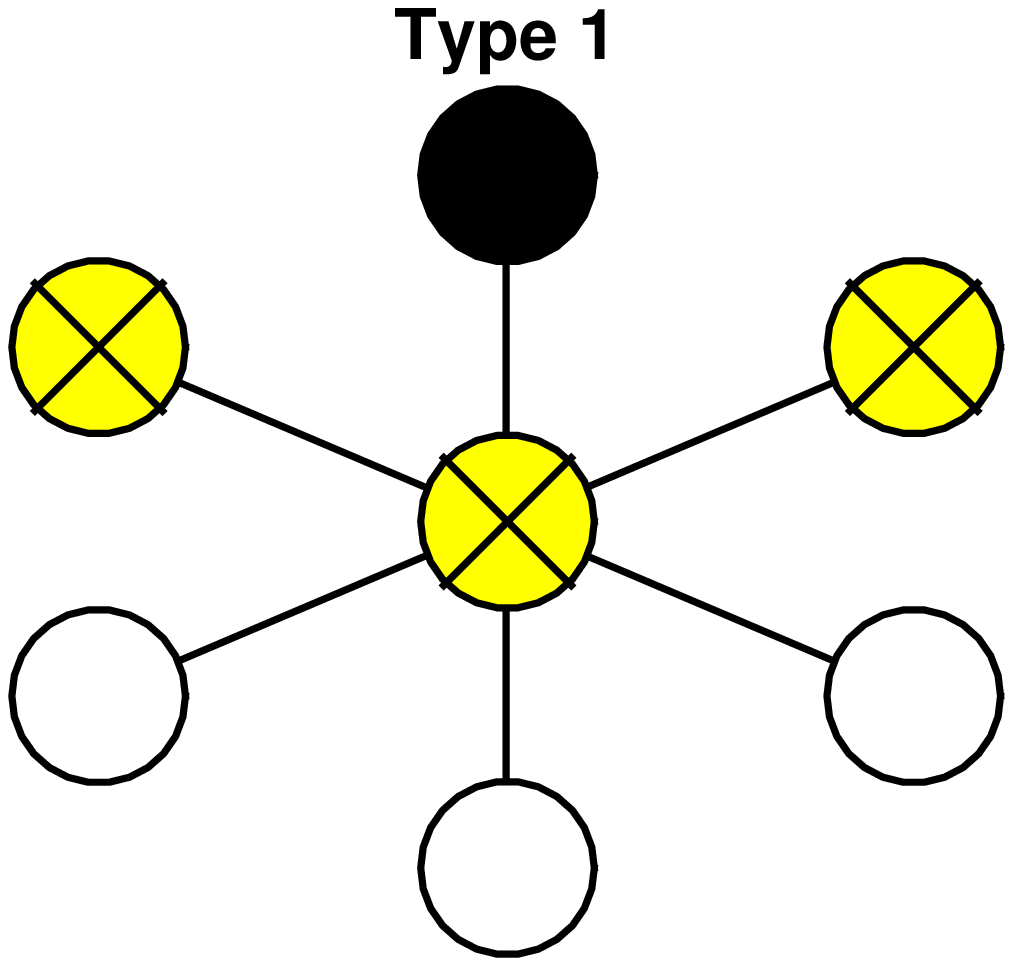}\label{subfig:tr_tipe1}}
	\subfigure[$Z(x)=c^{\varhexstar}\left(3U_1-6x\right)$]{\includegraphics[scale=0.34]{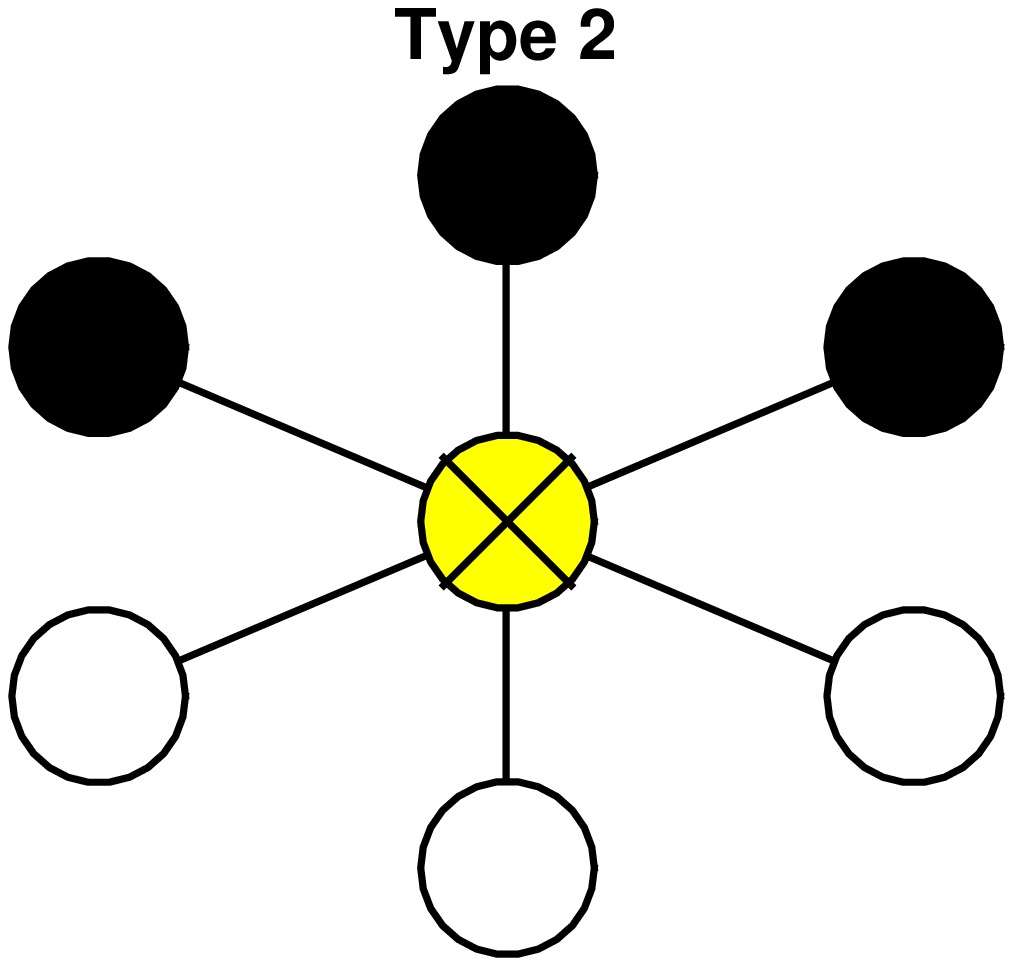}\label{subfig:tr_tipe2}}
	\subfigure[$Z(x)=c^{\varhexstar}\left(2U_1-5x\right)$]{\includegraphics[scale=0.34]{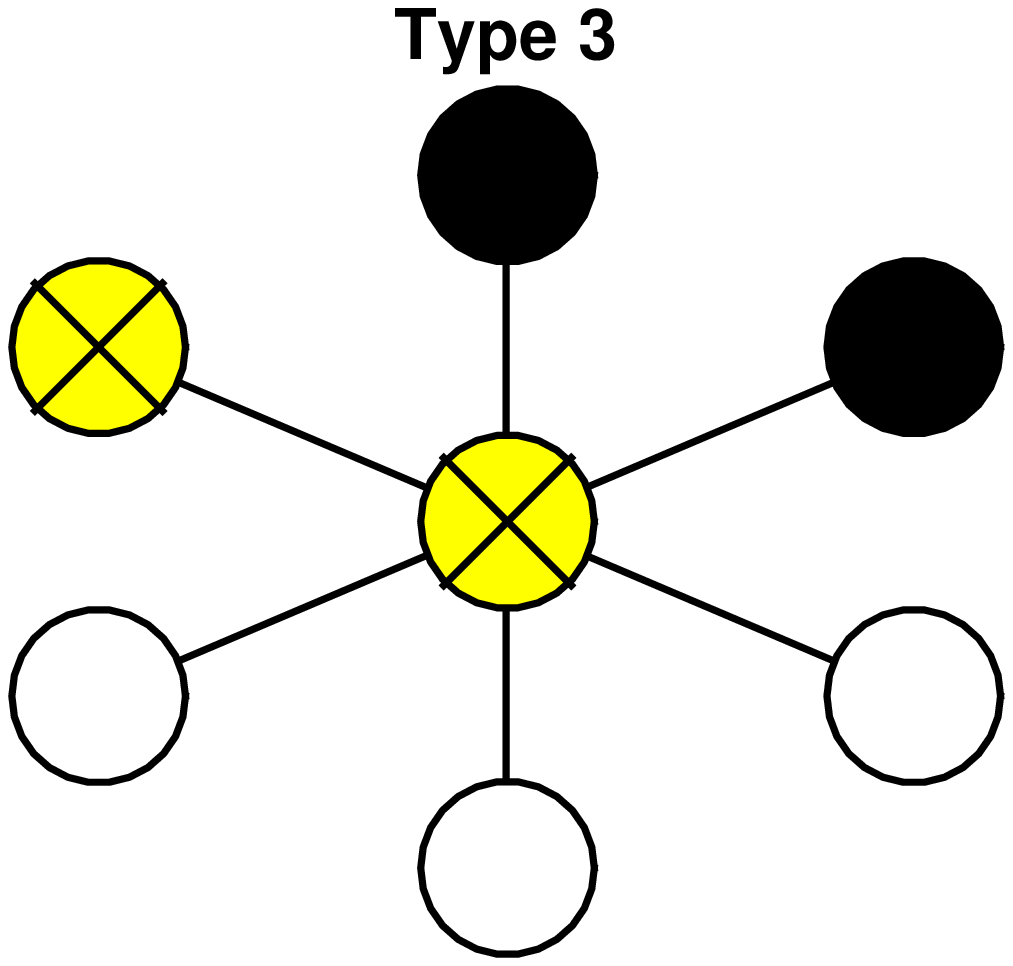}\label{subfig:tr_tipe3}}
	\subfigure[$Z(x)=c^{\varhexstar}\left(2U_1-6x\right)$]{\includegraphics[scale=0.34]{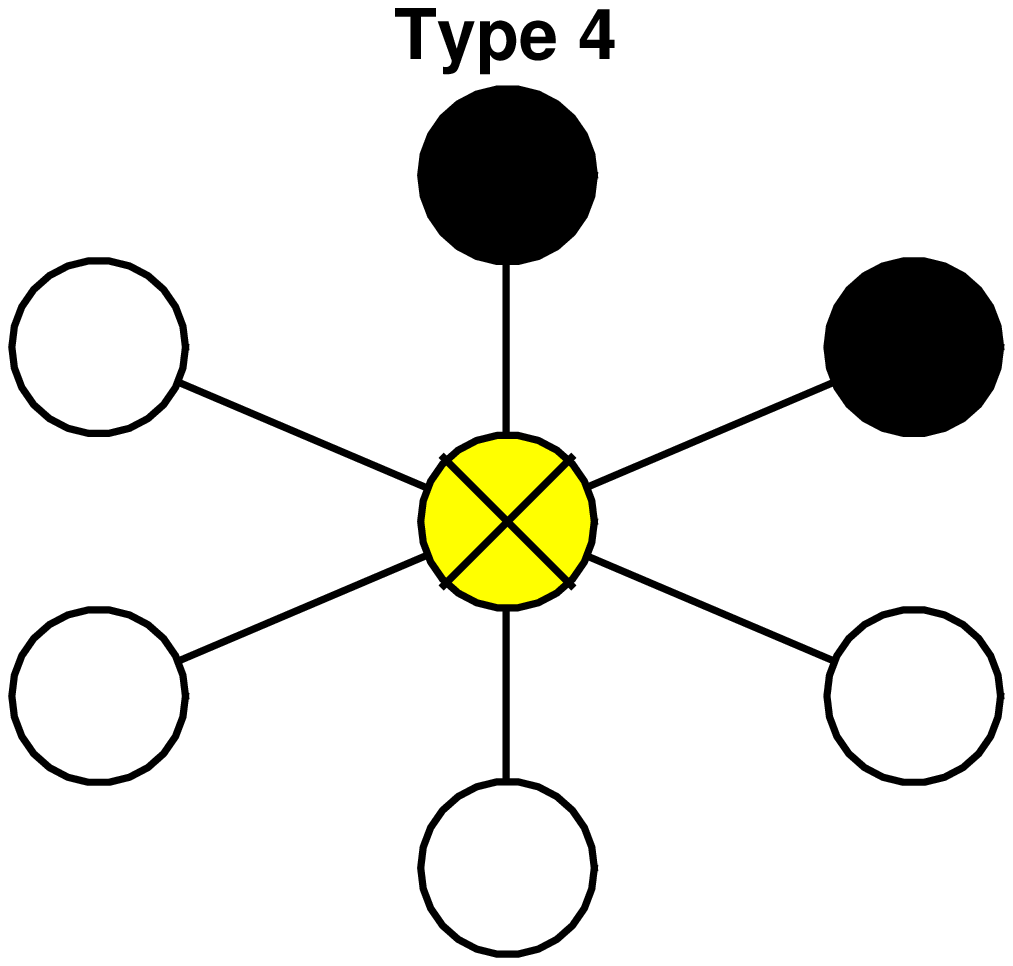}\label{subfig:tr_tipe4}}
	\subfigure[$Z(x)=c^{\varhexstar}\left(2U_1-4x\right)$]{\includegraphics[scale=0.34]{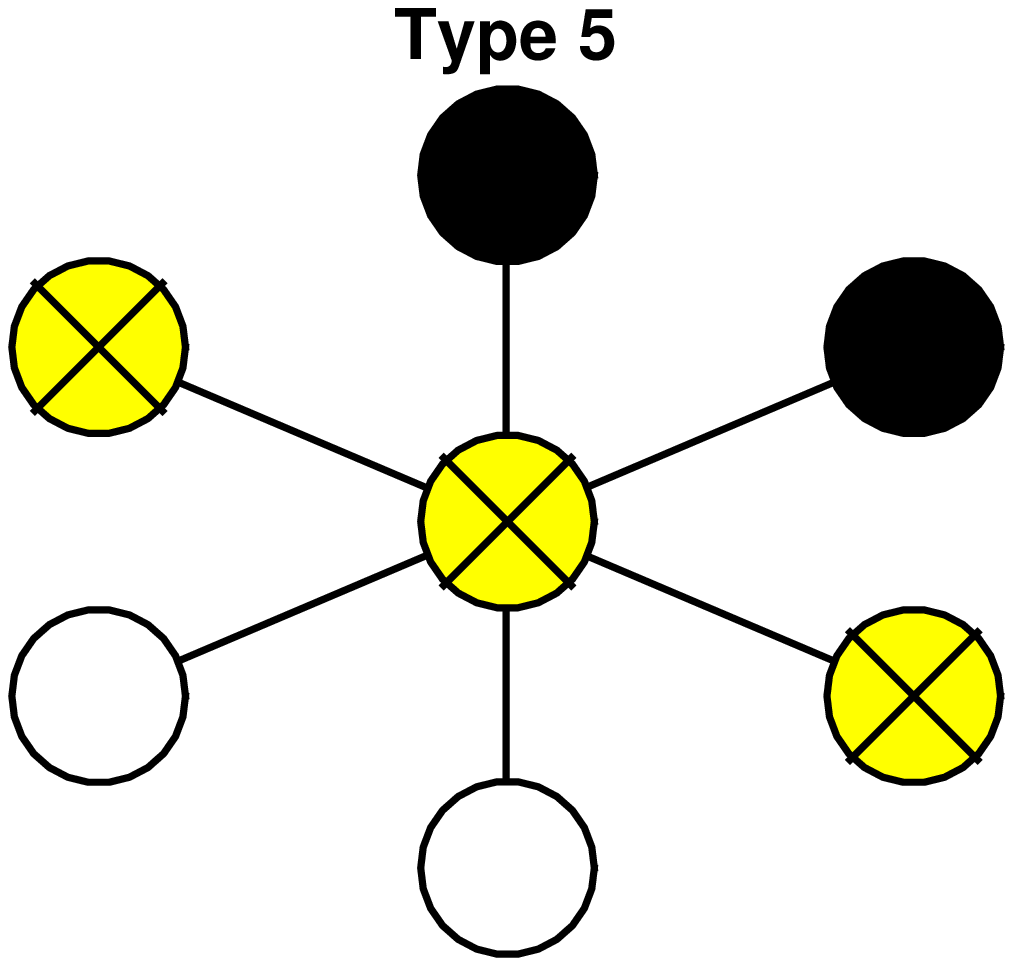}\label{subfig:tr_tipe5}}
	\caption{Types of active-cell approximations for triangular lattice.}
	\label{fig:active_cell_tr}
\end{figure*}

Figure \ref{subfig:bifur_ct_0_025} shows several types of saddle-node bifurcations and their approximations for the triangular lattice at {$c^{\varhexstar}=0.025$}.
In general, there are five types of saddle-node bifurcations for the triangular lattice, 
see figure \ref{fig:active_cell_tr}.

The bifurcations at points (a), (d) and (f), (b) and (c), (e), and (g) belong to type 1, 2, 3, 4, and 5, respectively.
All of the types of saddle-node bifurcations appear in site-centred and bond-centred solutions.
In general, the approximation for all of the types give good agreement for the ``lower'' and ``upper'' saddle-node bifurcations. %
Note that type 4 and 5 only appear in the ``lower'' and ``upper'' saddle-node bifurcations, respectively.

Figure\ \ref{subfig:bifur_ct_0_075} shows the approximation results of the saddle-node bifurcations for triangular lattice at {$c^{\varhexstar}=0.075$}.
By comparing between {$c^{\varhexstar}=0.025$ and $0.075$}, one can see that the active-cell approximations also give better results at smaller coupling strength. 
As we can see, the active-cell approximations fail to approximate points (h), (j), (k), (k), (l), and (n).
These also happen due to the solution interfaces that no longer satisfy the active-cell approximation assumption. 
As the coupling is getting larger, one will have more cells with different amplitudes around the interfaces that are also excited, which also happens in square and honeycomb lattices. 

\begin{figure*}[t!]
	\begin{minipage}[b]{.34\linewidth}
		\centering
		\includegraphics[scale=0.255]{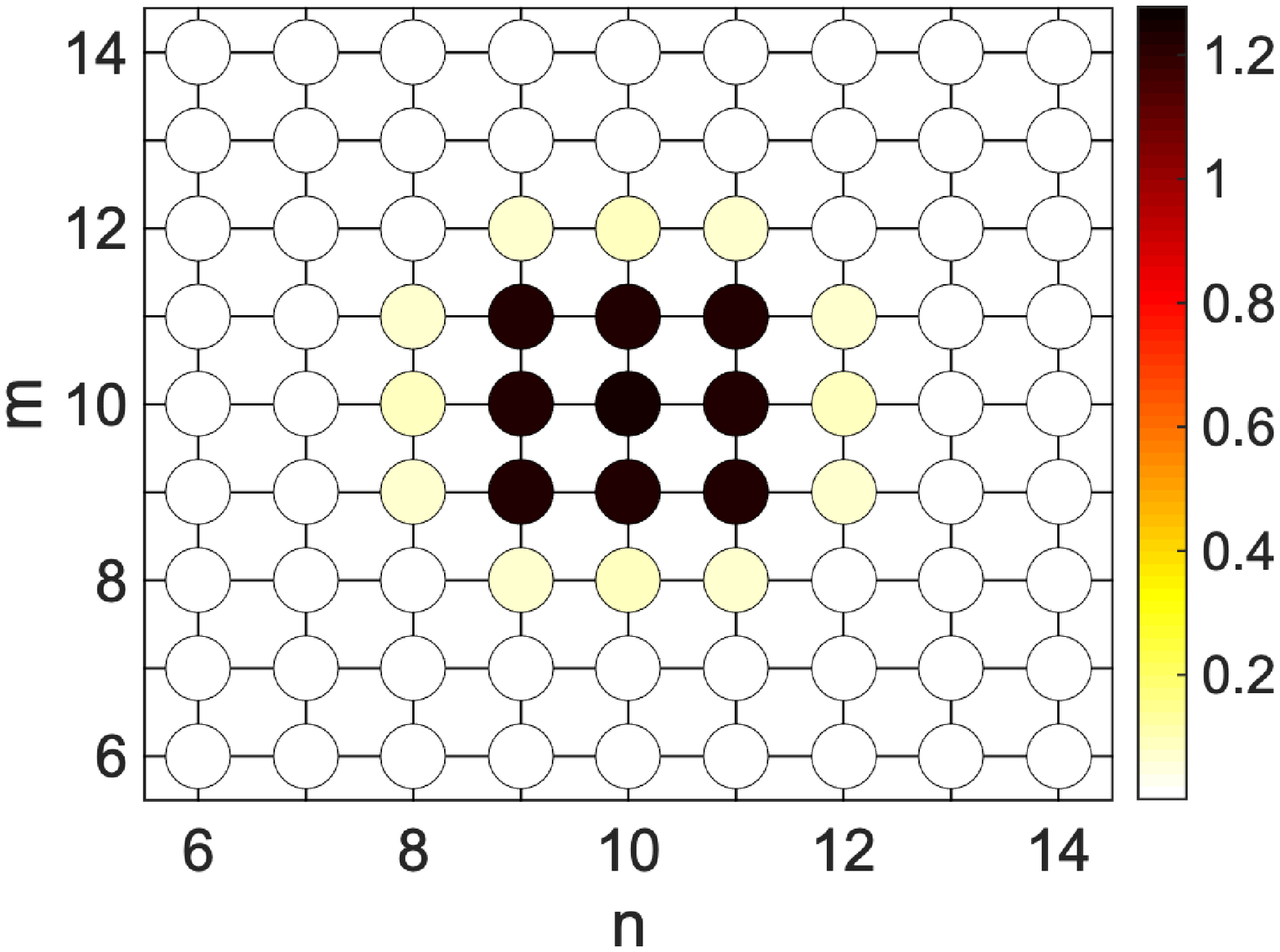}\label{subfig:prof_cp_0_05_1}\\ 
		\scriptsize{(1)}
	\end{minipage}%
	\begin{minipage}[b]{.34\linewidth}
		\centering
		\includegraphics[scale=0.255]{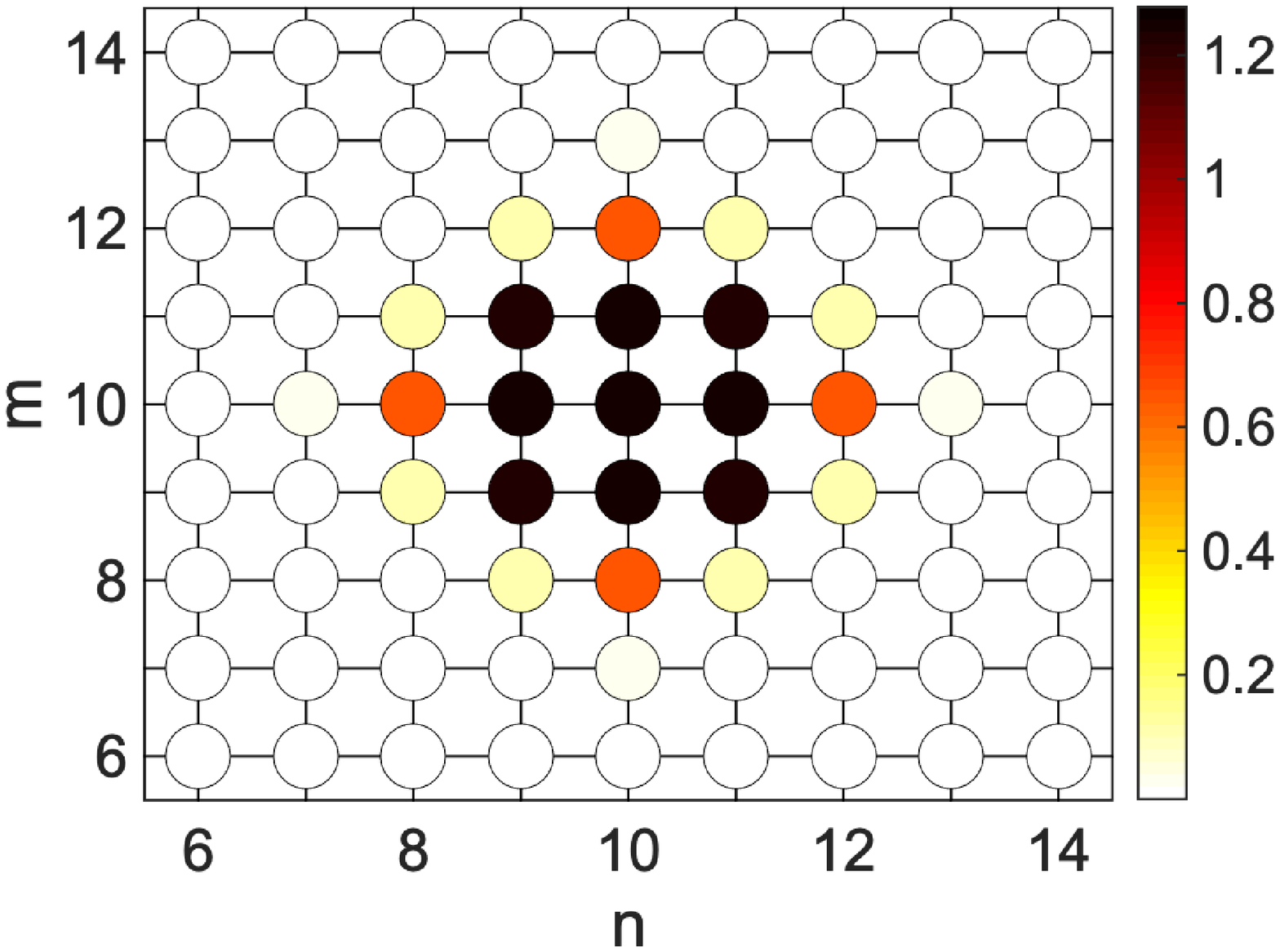}\label{subfig:prof_cp_0_05_2}\\
		\scriptsize{(2)}
	\end{minipage}%
	\begin{minipage}[b]{.34\linewidth}
		\centering
		\includegraphics[scale=0.255]{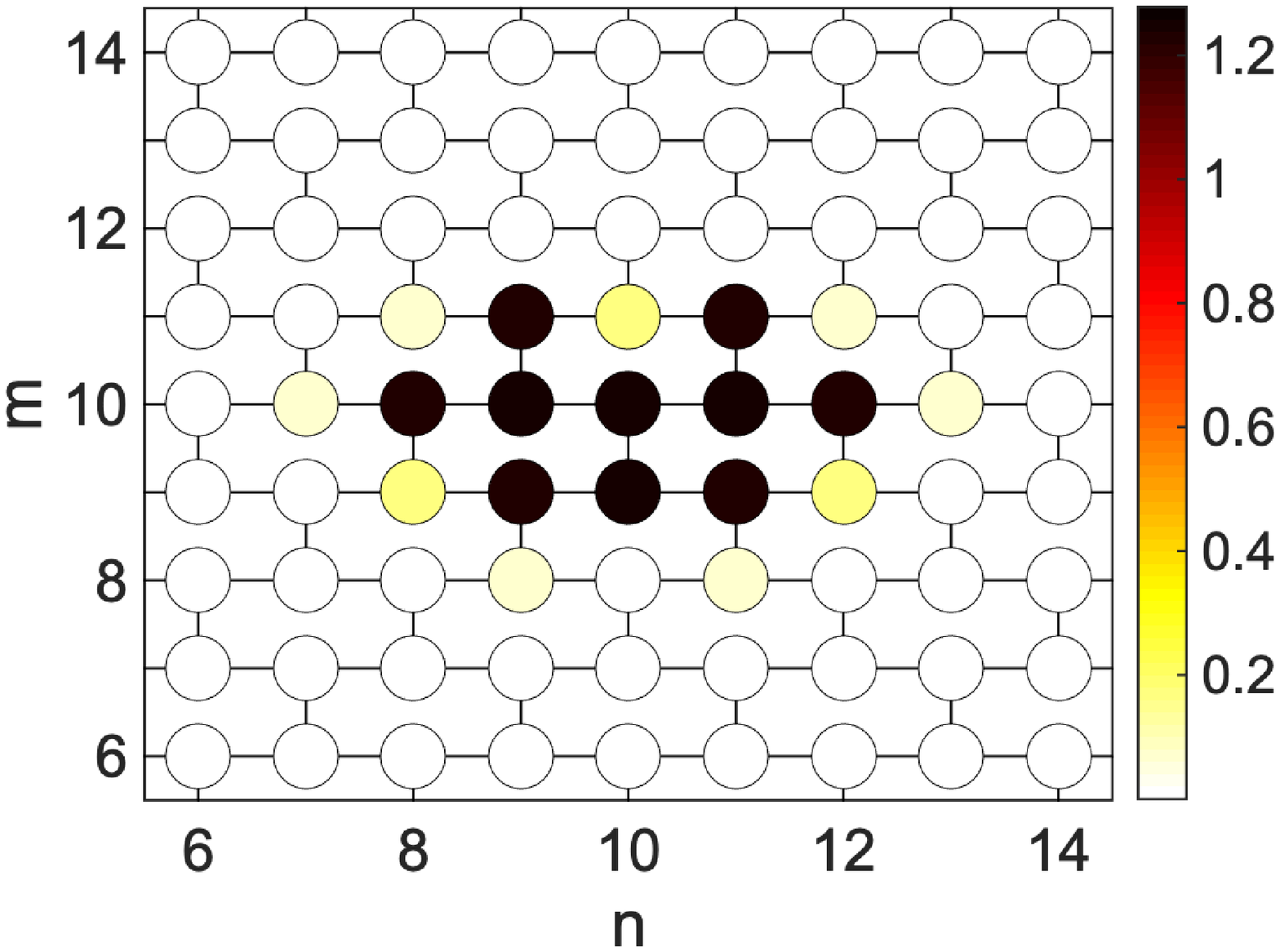}\label{subfig:prof_ch_0_05_3}\\ 
		\scriptsize{(3)}
	\end{minipage}%
\\
	\begin{minipage}[b]{.34\linewidth}
		\centering
		\includegraphics[scale=0.255]{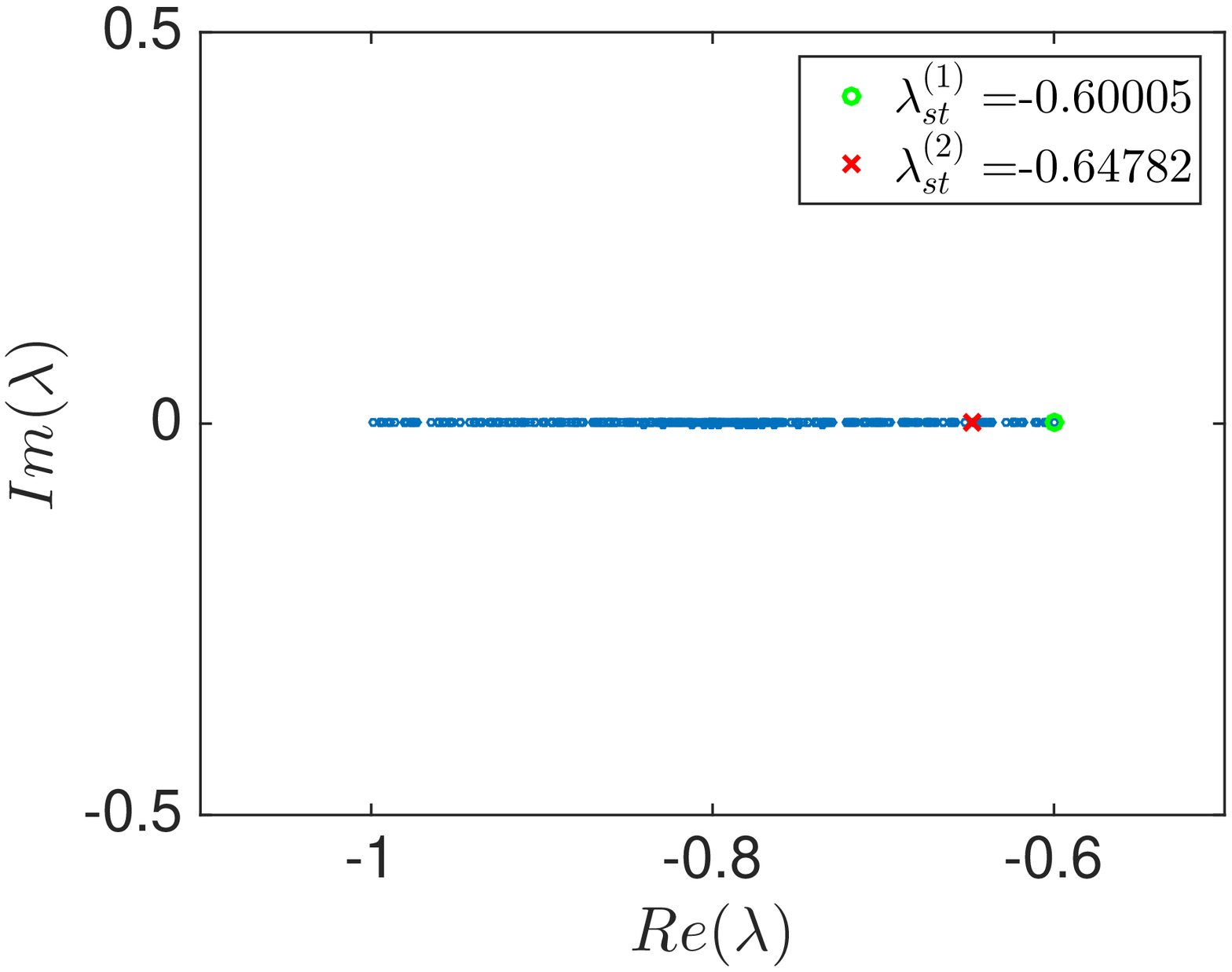}\label{subfig:eig_cp_0_05_1}\\ 
		\scriptsize{(1)}
	\end{minipage}%
	\begin{minipage}[b]{.34\linewidth}
		\centering
		\includegraphics[scale=0.255]{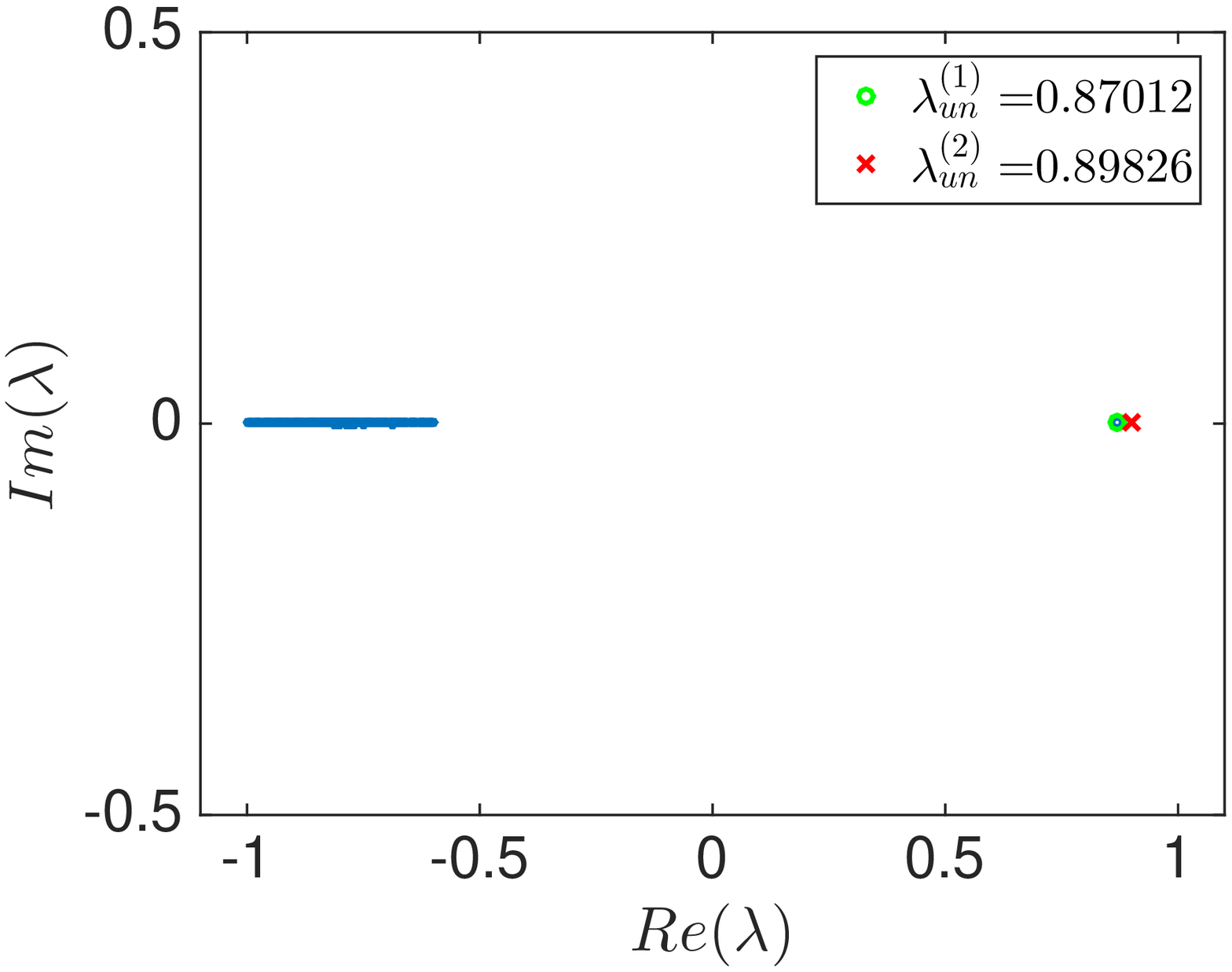}\label{subfig:eig_cp_0_05_2}\\ 
		\scriptsize{(2)}
	\end{minipage}%
	\begin{minipage}[b]{.34\linewidth}
		\centering
		\includegraphics[scale=0.255]{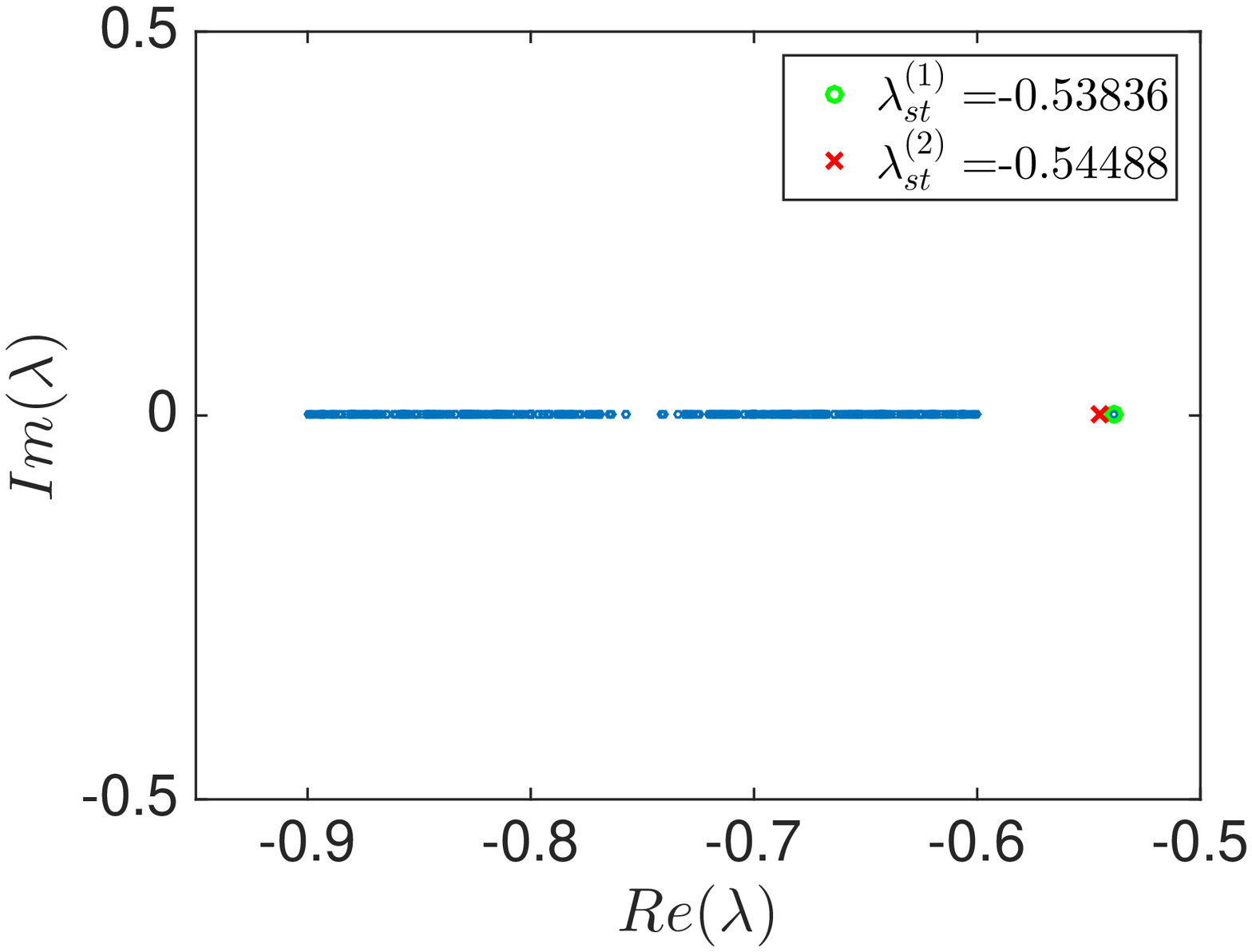}\label{subfig:eig_ch_0_05_3}\\ 
		\scriptsize{(3)}
	\end{minipage}%
	\\
	\begin{minipage}[b]{.34\linewidth}
		\centering
		\includegraphics[scale=0.255]{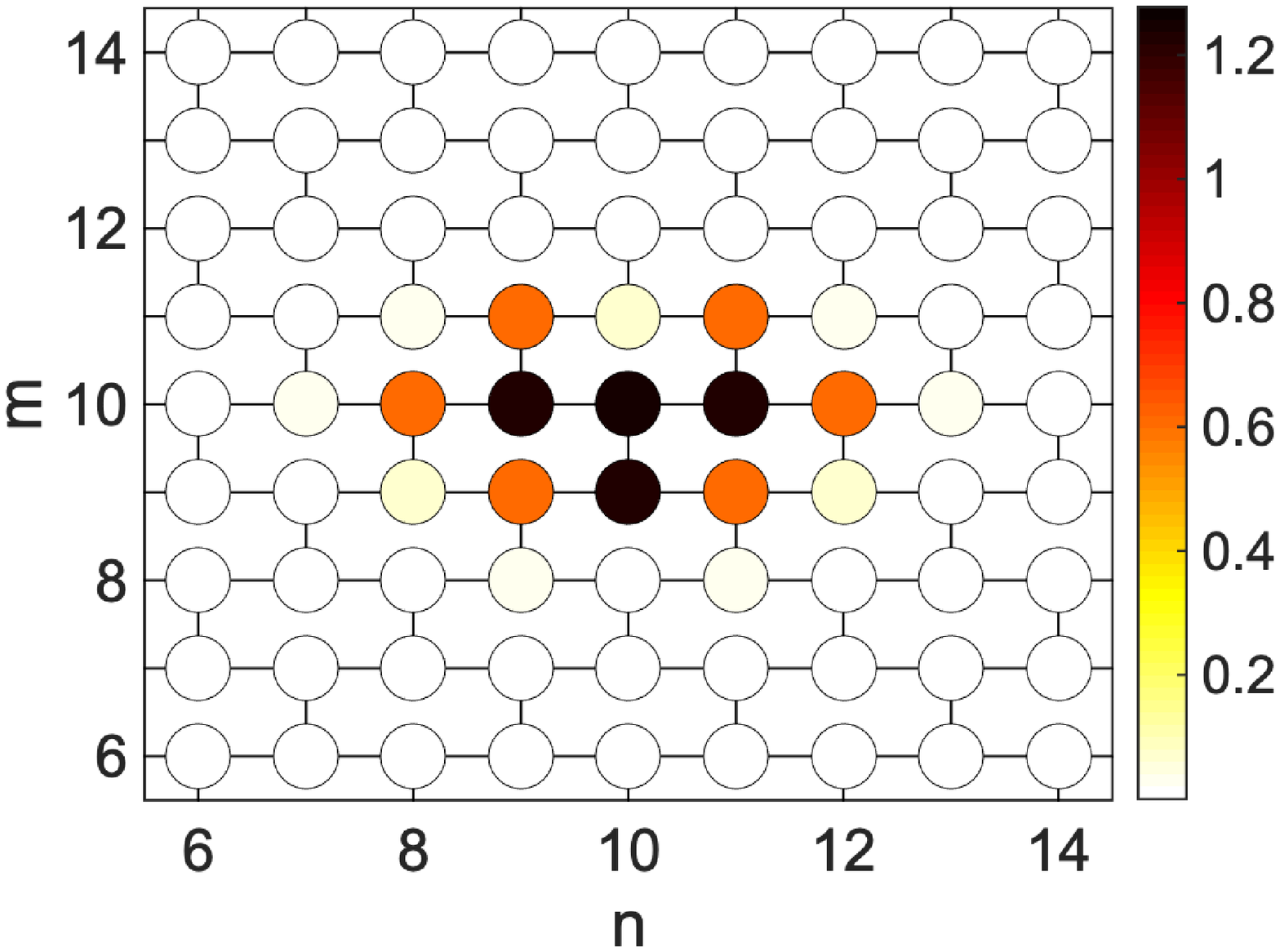}\label{subfig:prof_ch_0_05_4}\\ 
		\scriptsize{(4)}
	\end{minipage}%
	\begin{minipage}[b]{.34\linewidth}
		\centering
		\includegraphics[scale=0.255]{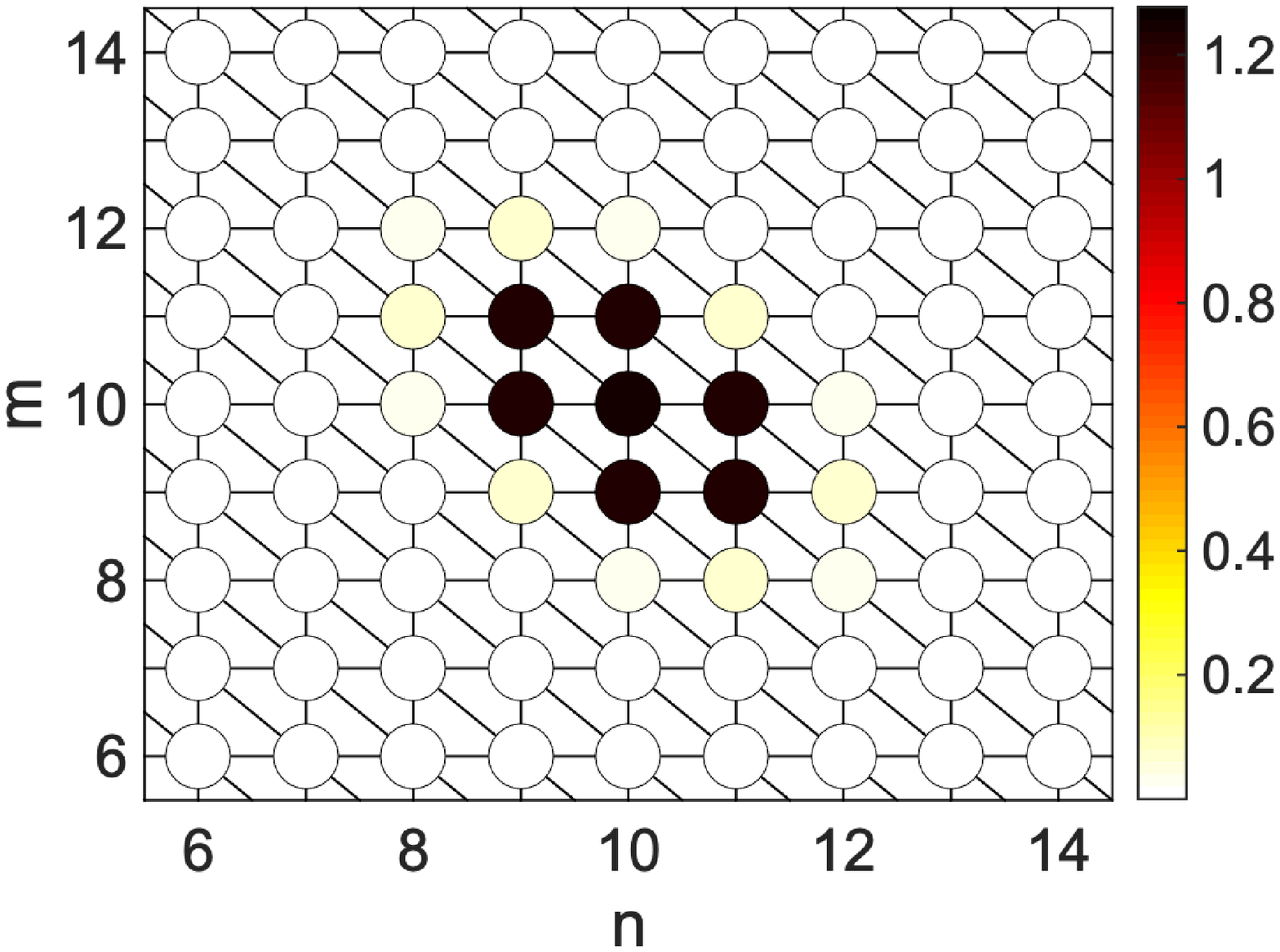}\label{subfig:prof_ct_0_06_5}\\ 
		\scriptsize{(5)}
	\end{minipage}%
	\begin{minipage}[b]{.34\linewidth}
		\centering
		\includegraphics[scale=0.255]{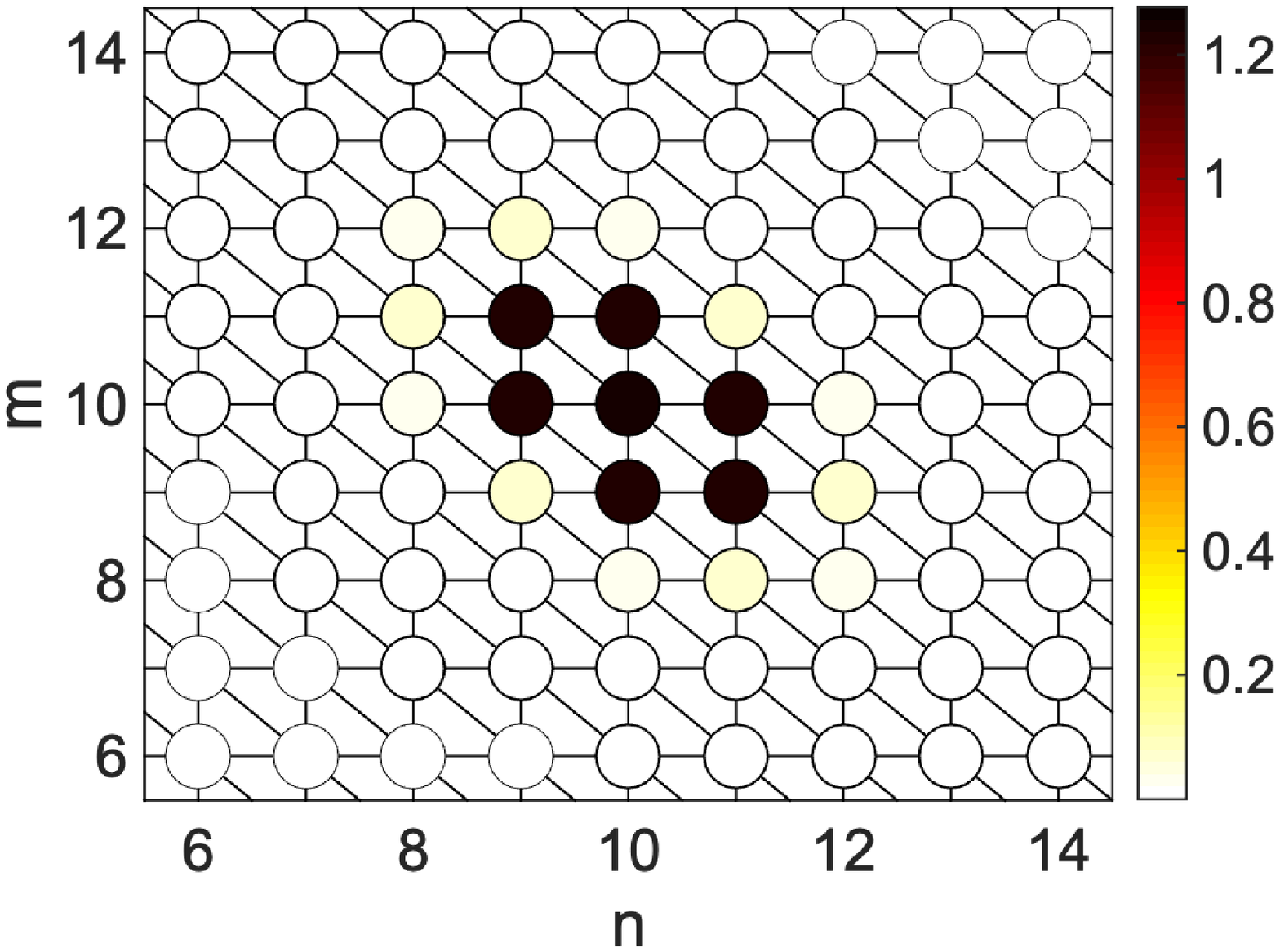}\label{subfig:prof_ct_0_06_6}\\
		 \scriptsize{(6)}
	\end{minipage}%
	\\
	\begin{minipage}[b]{.34\linewidth}
		\centering
		\includegraphics[scale=0.255]{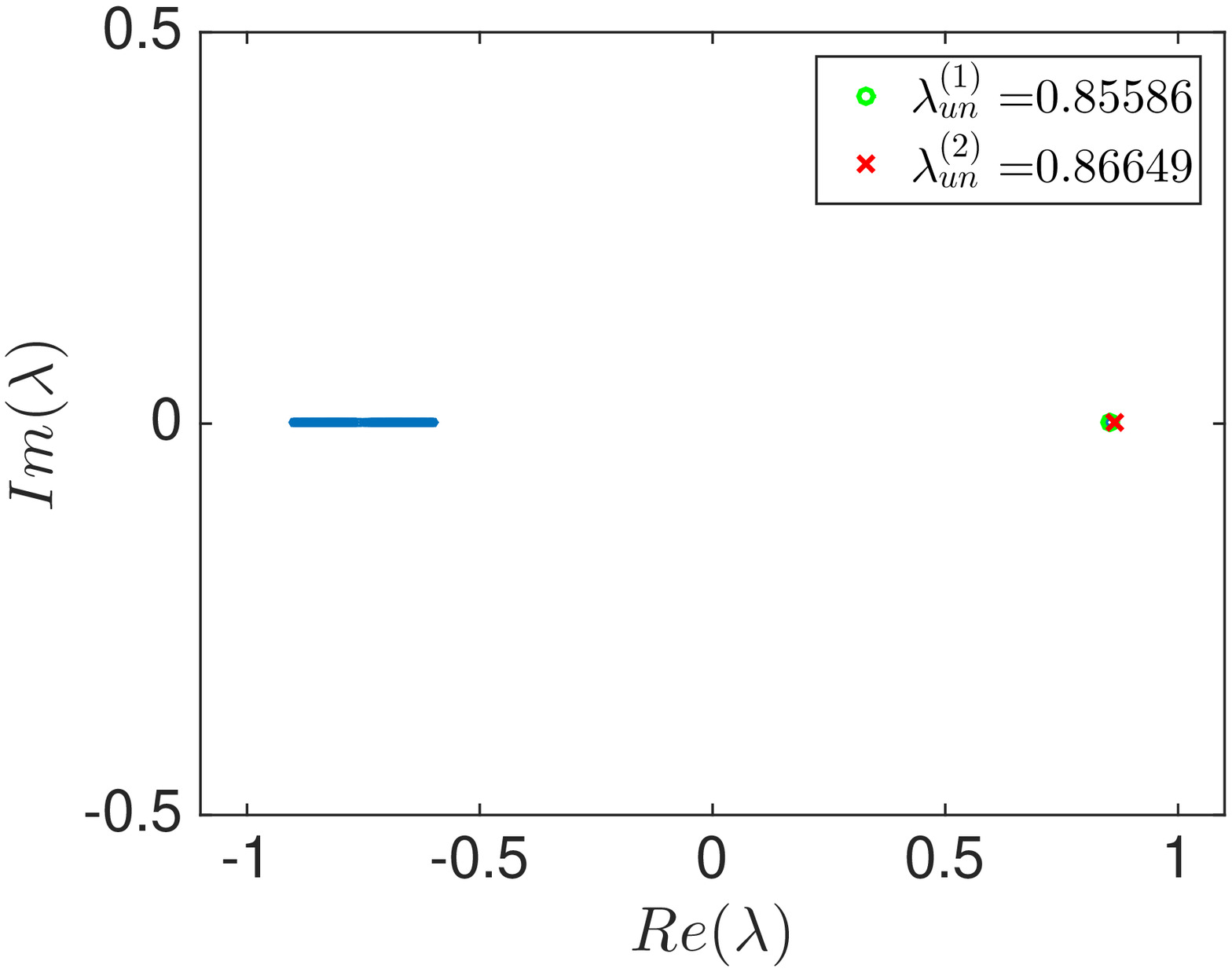}\label{subfig:eig_ch_0_05_4}\\ 
		\scriptsize{(4)}
	\end{minipage}%
	\begin{minipage}[b]{.34\linewidth}
		\centering
		\includegraphics[scale=0.255]{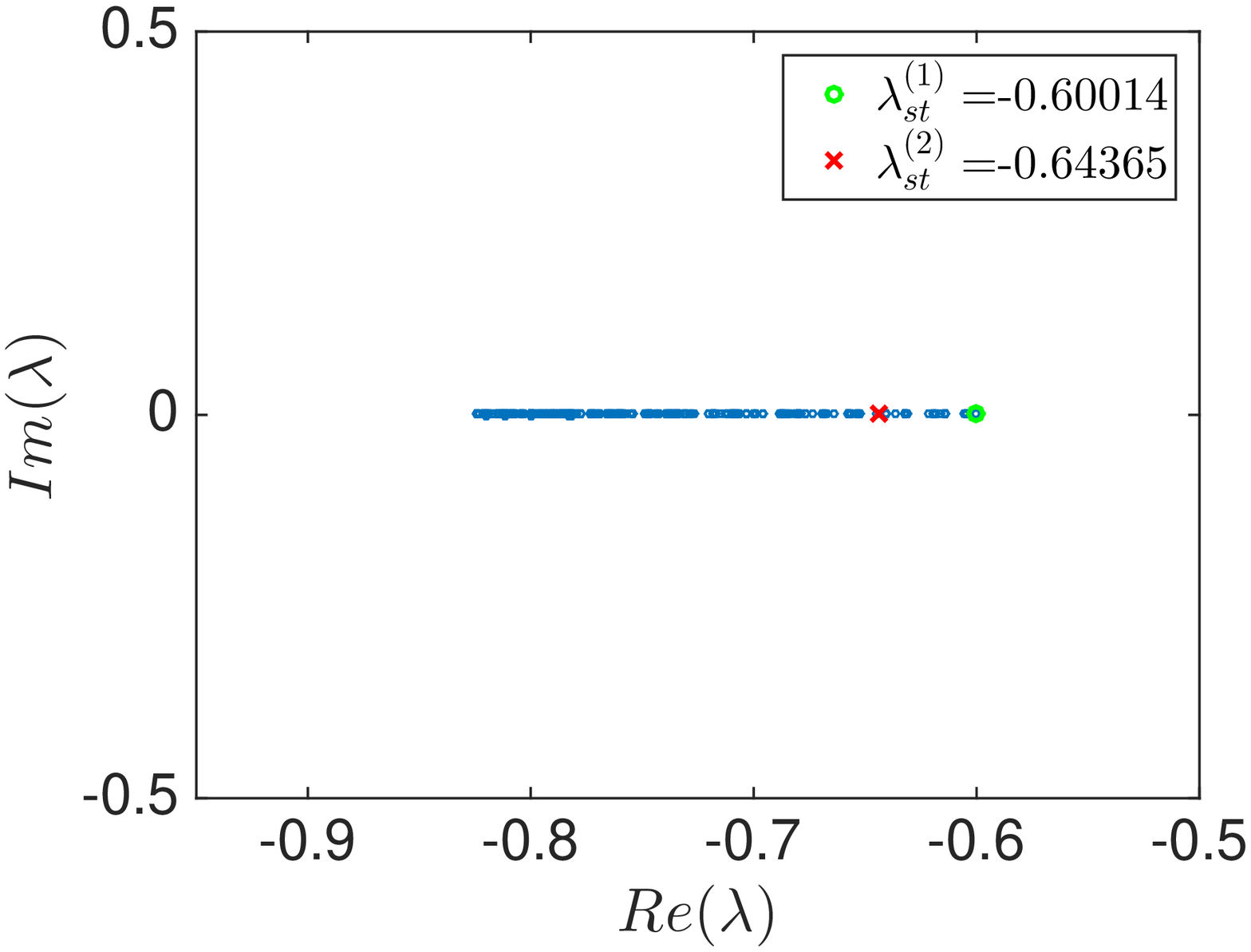}\label{subfig:eig_ct_0_06_5}\\ 
		\scriptsize{(5)}
	\end{minipage}%
	\begin{minipage}[b]{.34\linewidth}
		\centering
		\includegraphics[scale=0.255]{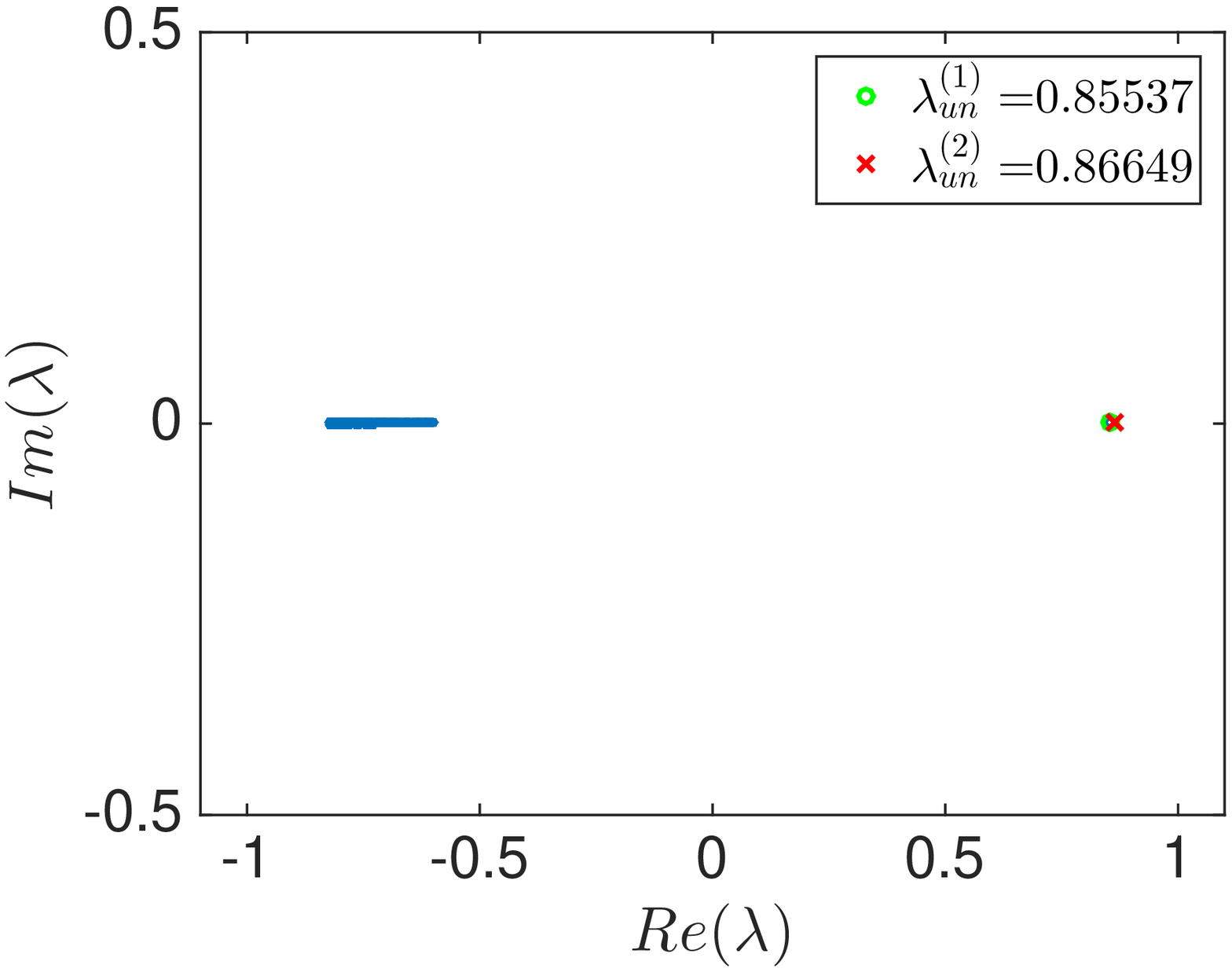}\label{subfig:eig_ct_0_06_6}\\ \scriptsize{(6)}
	\end{minipage}%
	\caption{
		Plot of the localised states, their corresponding numerical eigenvalues, and critical one obtained the active-cell approximation indicated as points (1)-(2) in figure\ \ref{subfig:bifur_cp_0_05}, (3)-(4) in figure \ref{subfig:bifur_ch_0_05}, and (5)-(6) in figure \ref{subfig:bifur_ct_0_025} for stable and unstable site-centered solutions for all of the lattice types. 
	}
	\label{fig:active_cell_compare}
\end{figure*}


All in all, at a relatively small coupling strength, the active-cell approximation is a good approximator, especially when we have a fewer number of interfaces in the Laplacian operator.
Thus, at the same coupling strength, our approximation gives the best result in the honeycomb lattice as it has three interfaces only, compared to the square and triangular lattices that have four and six interfaces, respectively.
Note that the first saddle-node bifurcation that appear right after the branching points of localised solutions for all of the lattice types cannot be approximated by our method because the solutions are rather close to the continuum limit. 

\subsection{Eigenvalue approximation}
The active-cell approximation also can to be used to approximate the critical eigenvalue of the equation \eqref{eq:dnls_all} for all of the lattice domains.
By considering our assumption in equation \eqref{eq:active_cell_fun}, it is straightforward that from the linearisation, one can obtain the eigenvalue problem
\begin{equation}
\lambda x = \left.\frac{{d}}{{d}x}F(x)\right|_{x=x_{st,un}}x,
\end{equation}
i.e., $\lambda$ is given by
\begin{equation}
\lambda(\mu)=\mu  + 6x_{st,un}^2-5x_{st,un}^4+\frac{\partial Z(x_{st,un})}{\partial x_{st,un}}.
\end{equation}
Our approximation of the critical eigenvalue at points (1)-(6) indicated in figures\ \ref{subfig:bifur_cp_0_05}, \ref{subfig:bifur_ch_0_05}, and \ref{subfig:bifur_ct_0_025} is shown in figure\ \ref{fig:active_cell_compare}, where good results are obtained when the coupling is weak.

\section{Planar fronts}\label{secadd}

\begin{figure*}[tbhp!]
	\centering
	\subfigure[]{\includegraphics[scale=0.3]{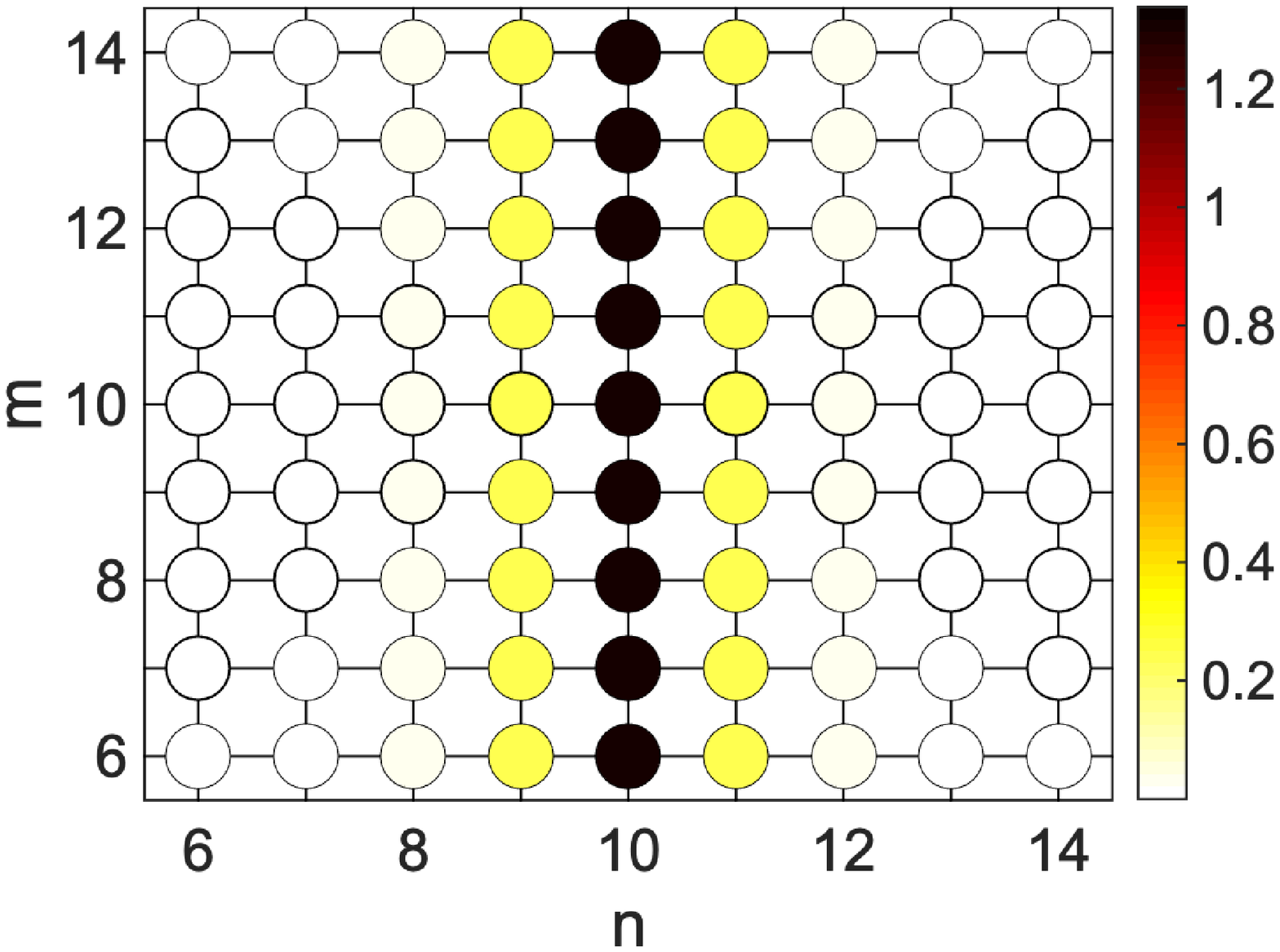}\label{subfig:prof_1D_all_cp_ch_0_05_ct_0_025_a}}
	\subfigure[]{\includegraphics[scale=0.3]{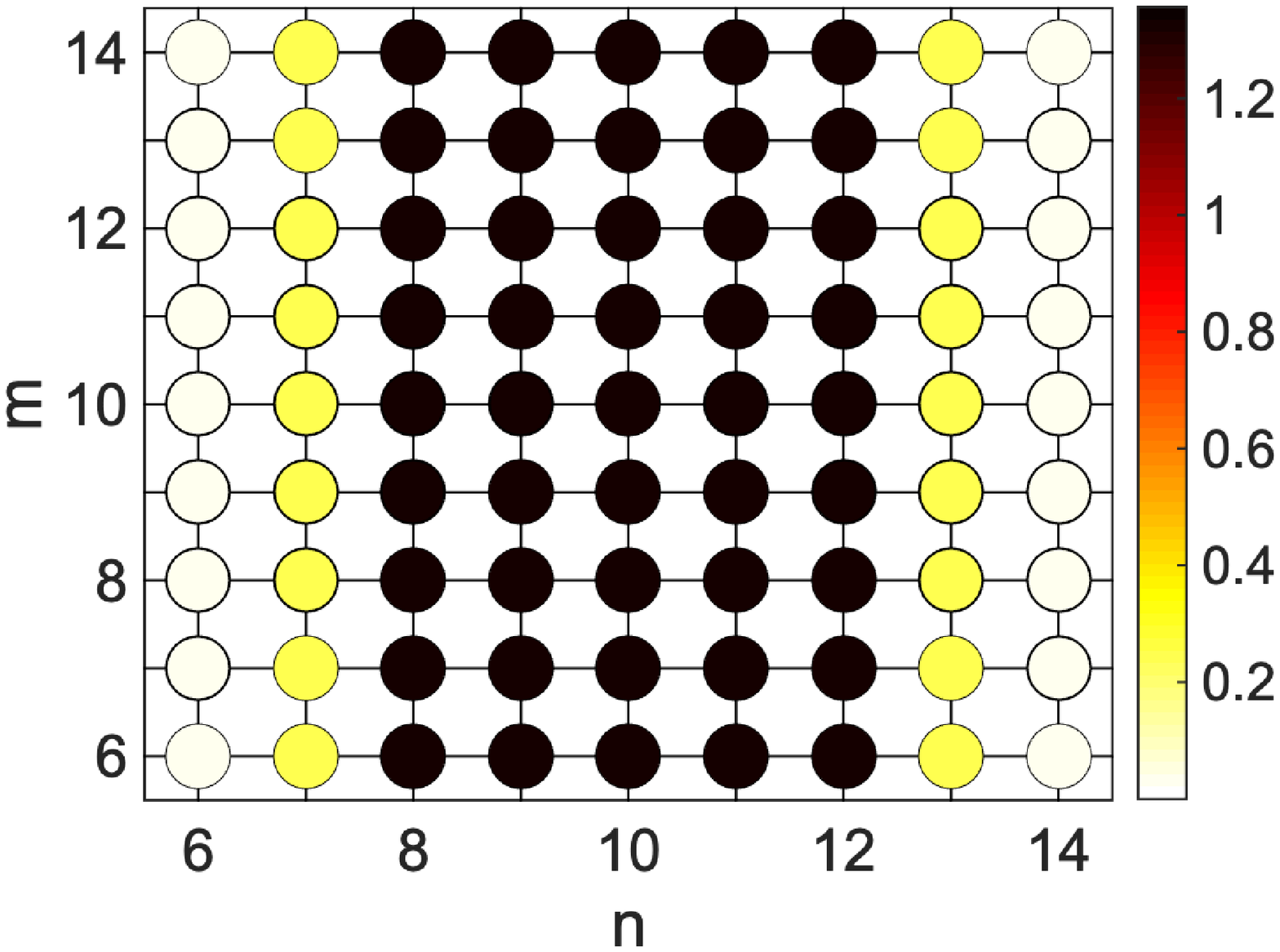}\label{subfig:prof_1D_all_cp_ch_0_05_ct_0_025_b}}
	\subfigure[]{\includegraphics[scale=0.3]{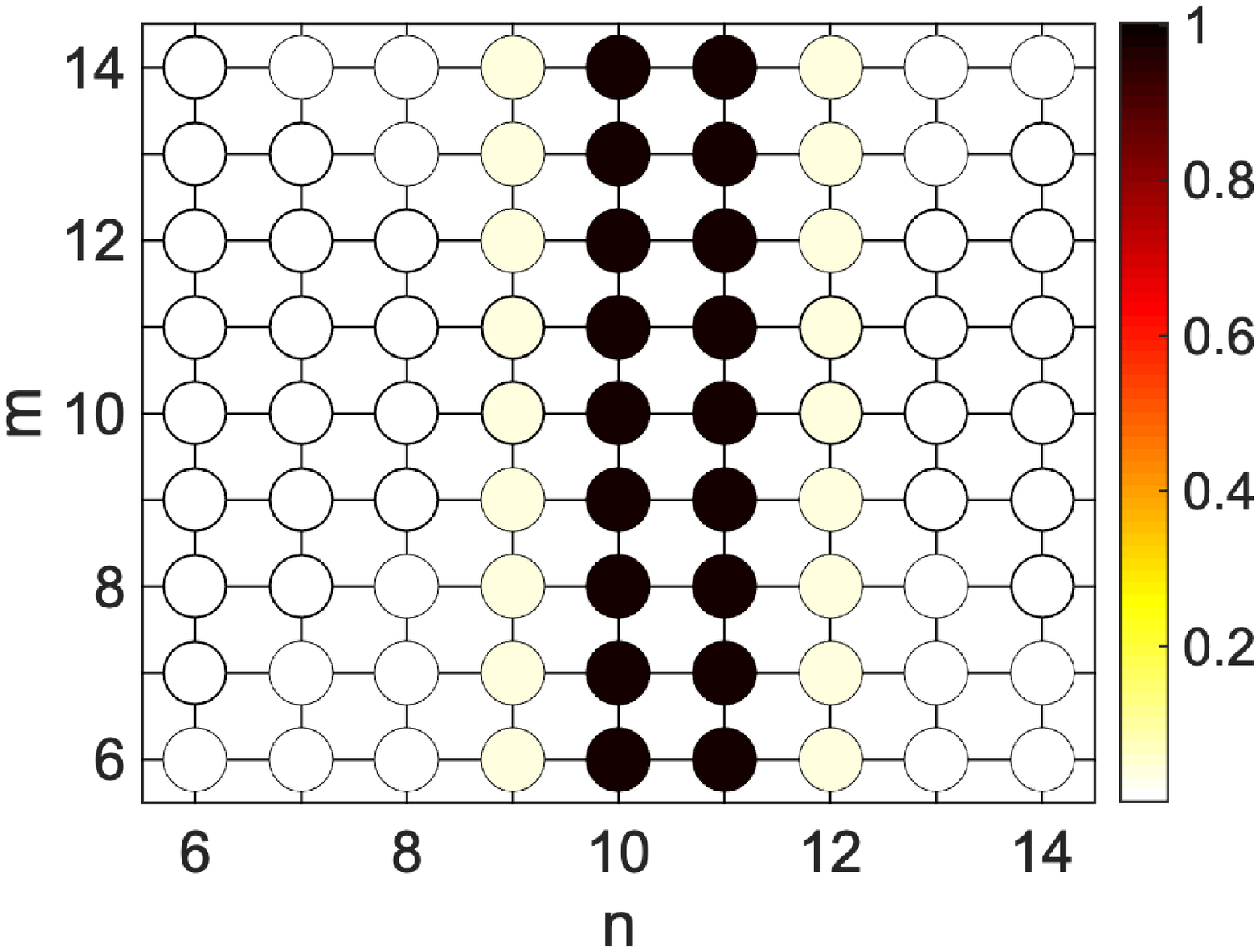}\label{subfig:prof_1D_all_cp_ch_0_05_ct_0_025_c}}
	\subfigure[]{\includegraphics[scale=0.3]{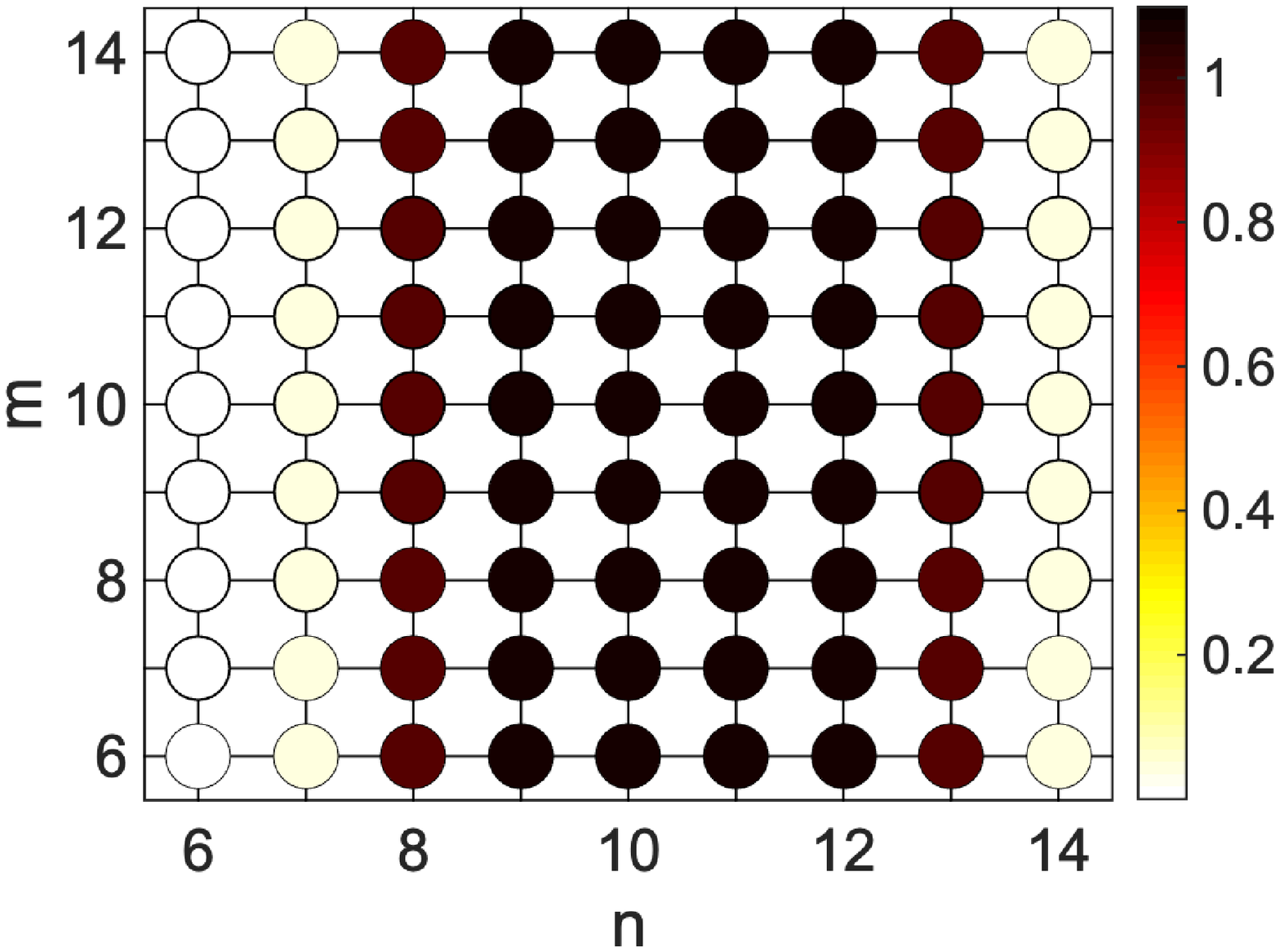}\label{subfig:prof_1D_all_cp_ch_0_05_ct_0_025_d}}
\caption{{Top-view of planar fronts, i.e., solutions that are localised in the $n$-direction, but constant in the $m$-direction, in the discrete system with the square lattice. Panels (a), (b) are on-site fronts, while (c) and (d) are intersite ones. Here, $c^{+}=0.05$.  
	}}
	\label{fig:prof_1D_loc}
\end{figure*}

\begin{figure*}[tbhp!]
	\centering
	{\includegraphics[scale=0.425]{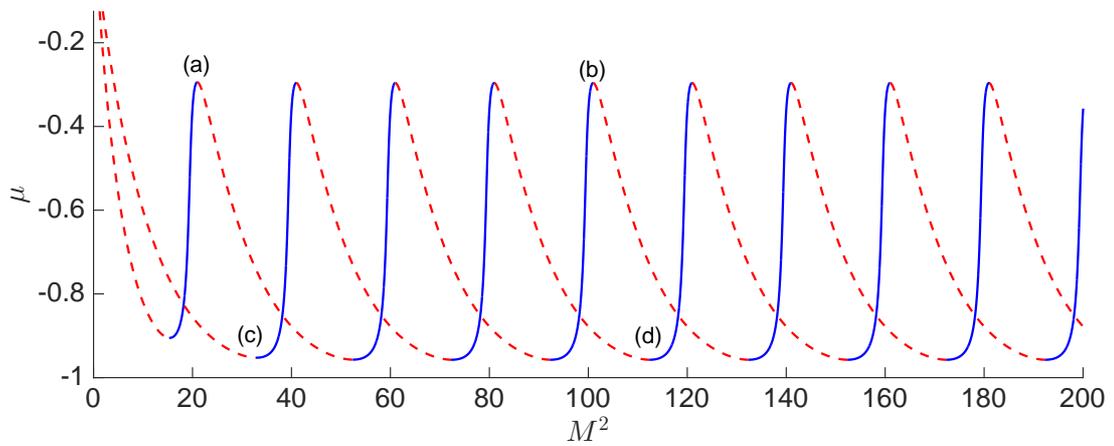}}
	\caption{{Bifurcation diagram of the localised solutions in figure \ref{fig:prof_1D_loc}. Points labelled as (a)-(d) show the location of the profiles in figure \ref{fig:prof_1D_loc} on the diagram.}
	}
	\label{fig:bifur_1D_loc}
\end{figure*}

\begin{figure*}[h!]
	\centering
	\subfigure[]{\includegraphics[scale=0.3]{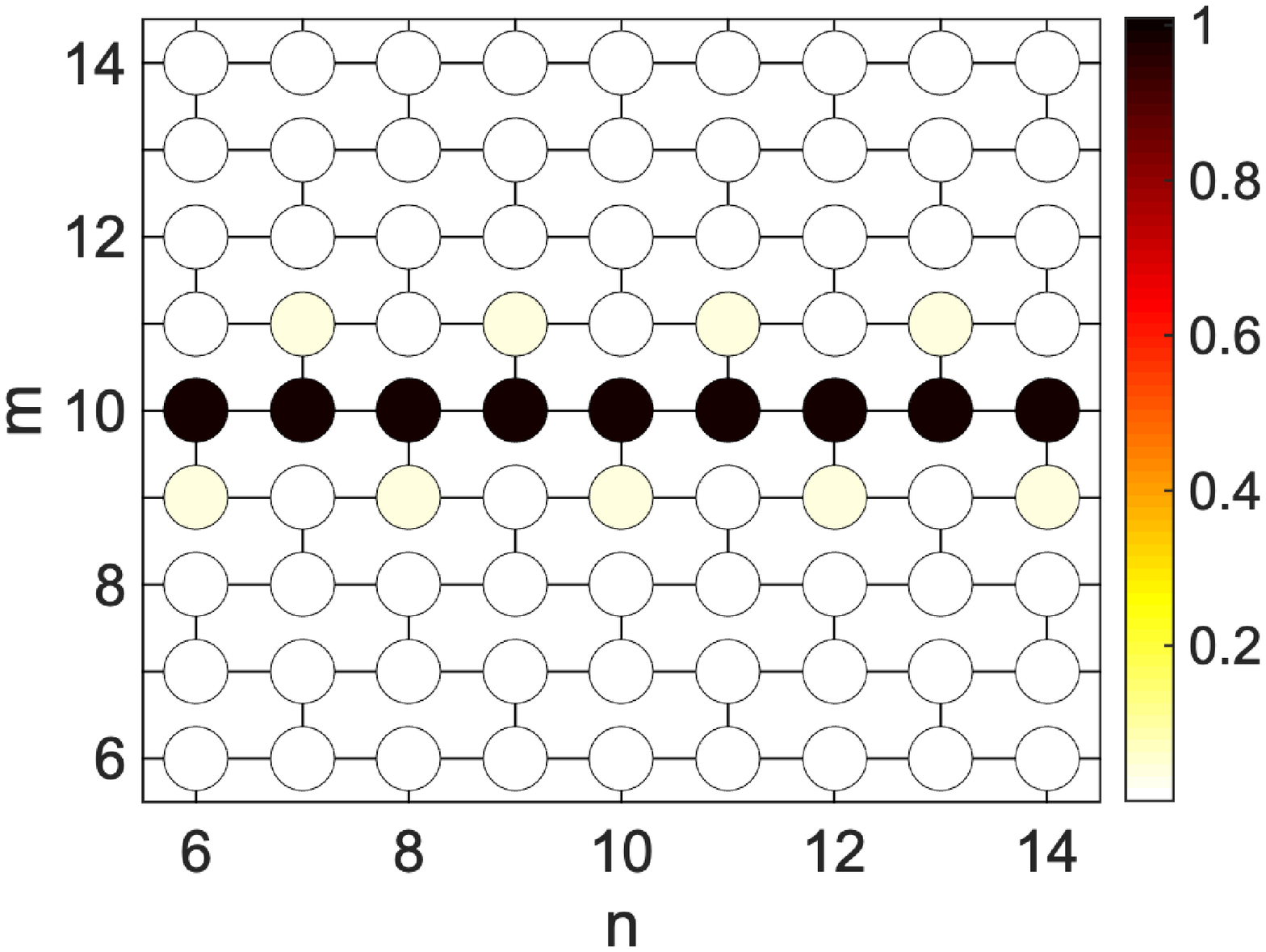}\label{subfig:prof_1D_ch_t2_0_05_a}}
	\subfigure[]{\includegraphics[scale=0.3]{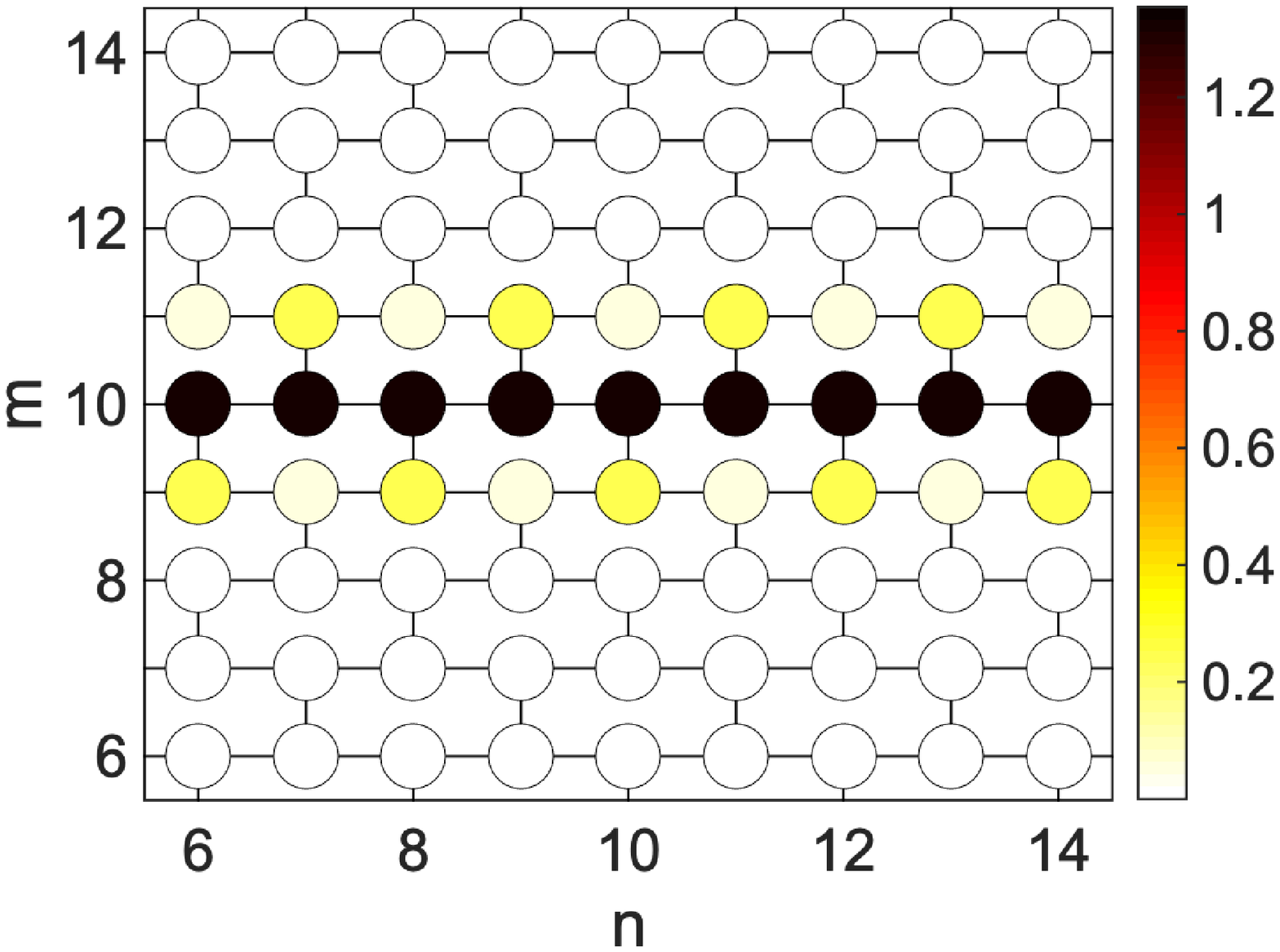}\label{subfig:prof_1D_ch_t2_0_05_b}}
	\subfigure[]{\includegraphics[scale=0.3]{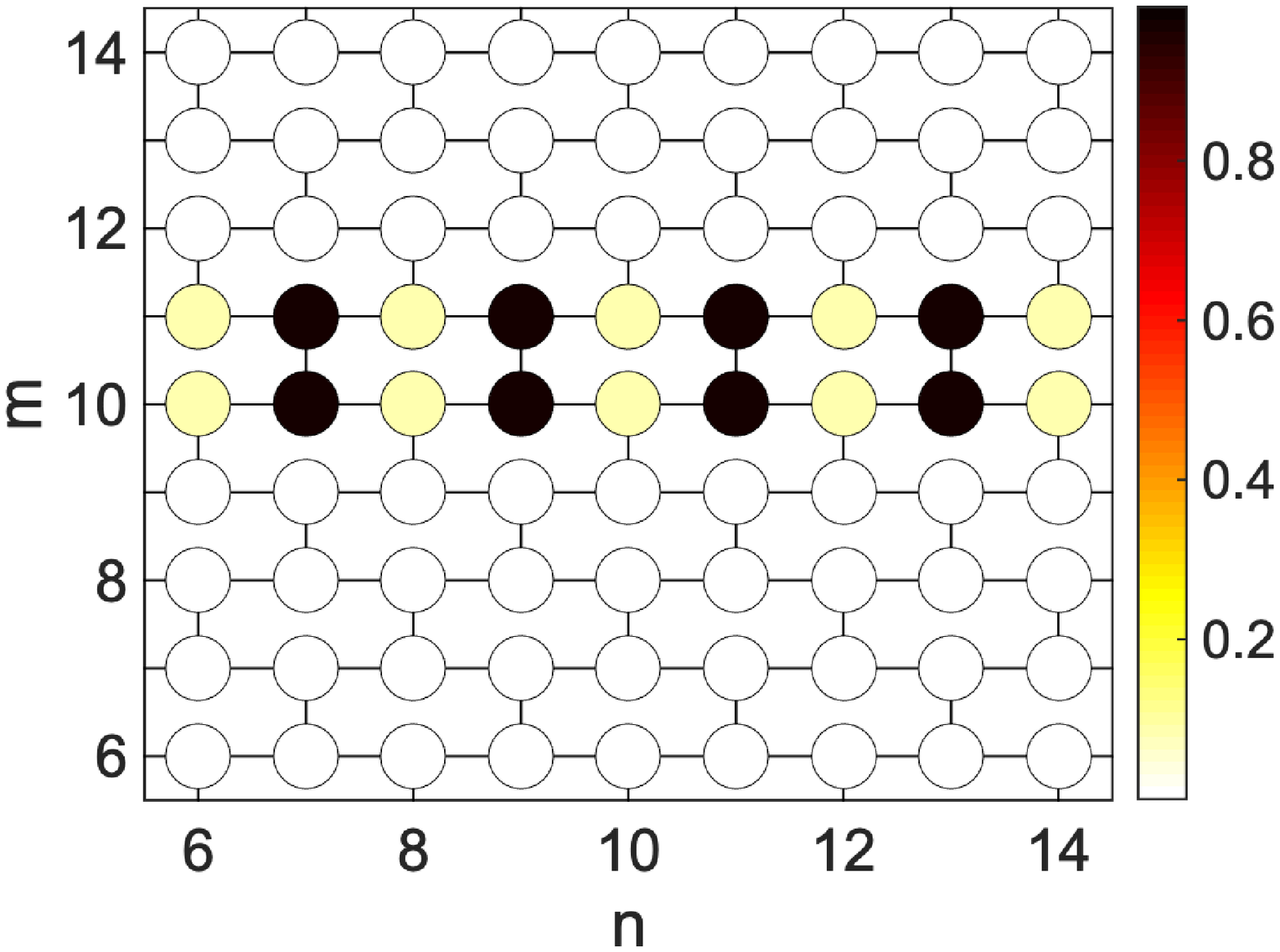}\label{subfig:prof_1D_ch_t2_0_05_c}}
	\subfigure[]{\includegraphics[scale=0.3]{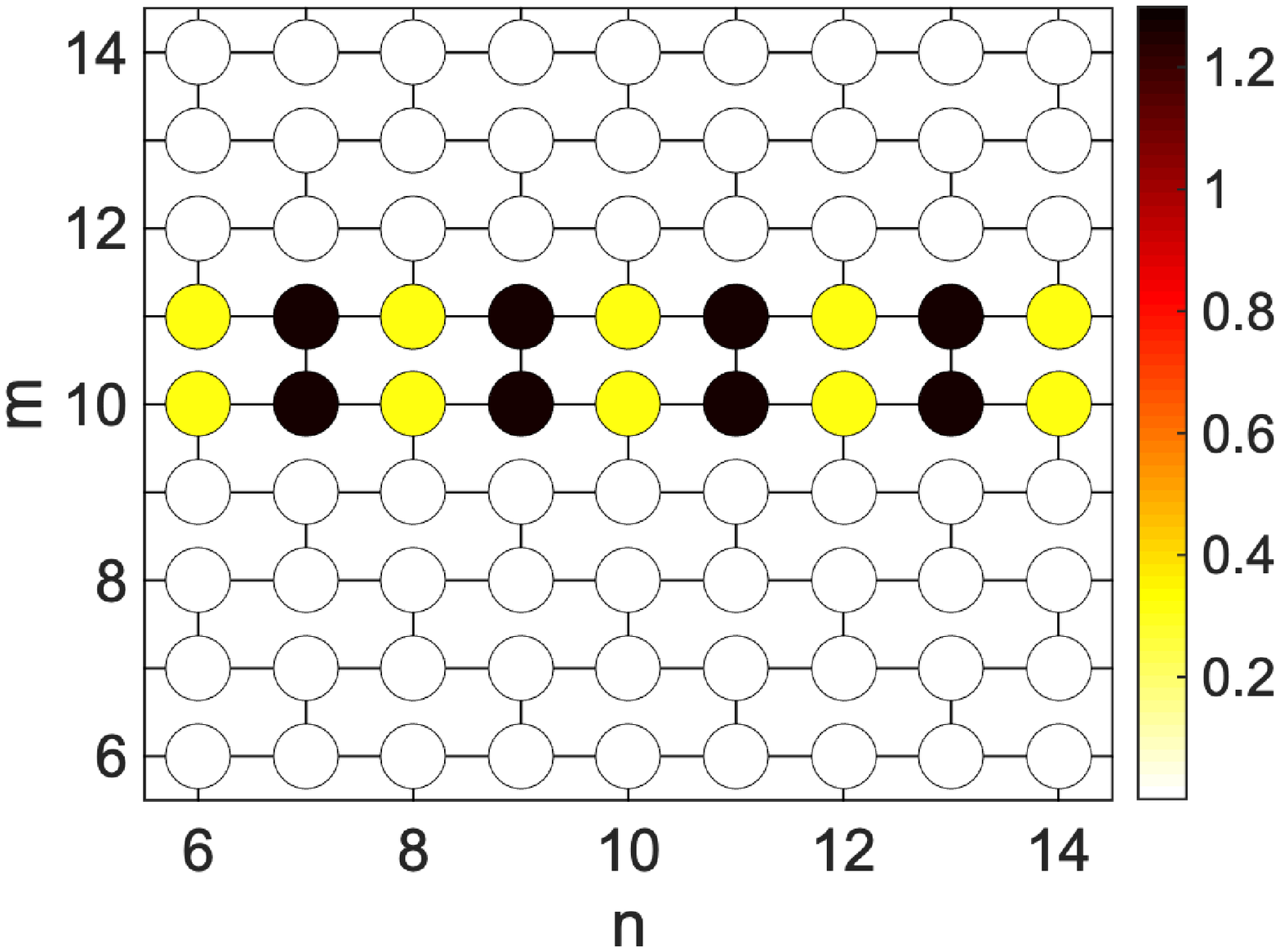}\label{subfig:prof_1D_ch_t2_0_05_d}}
	\caption{{Top-view of planar fronts that are localised in the $m$-direction, but uniform in the $n$-direction, in the discrete system \eqref{eq:dnls_all} with honeycomb lattice. Panels (a), (b) are on-site fronts, while (c) and (d) are intersite ones. Here, $c^{\Yup}=0.05$.
	}}
	\label{fig:prof_1D_ch_t2}
\end{figure*}

\begin{figure*}[t!]
	\centering
	{\includegraphics[scale=0.425]{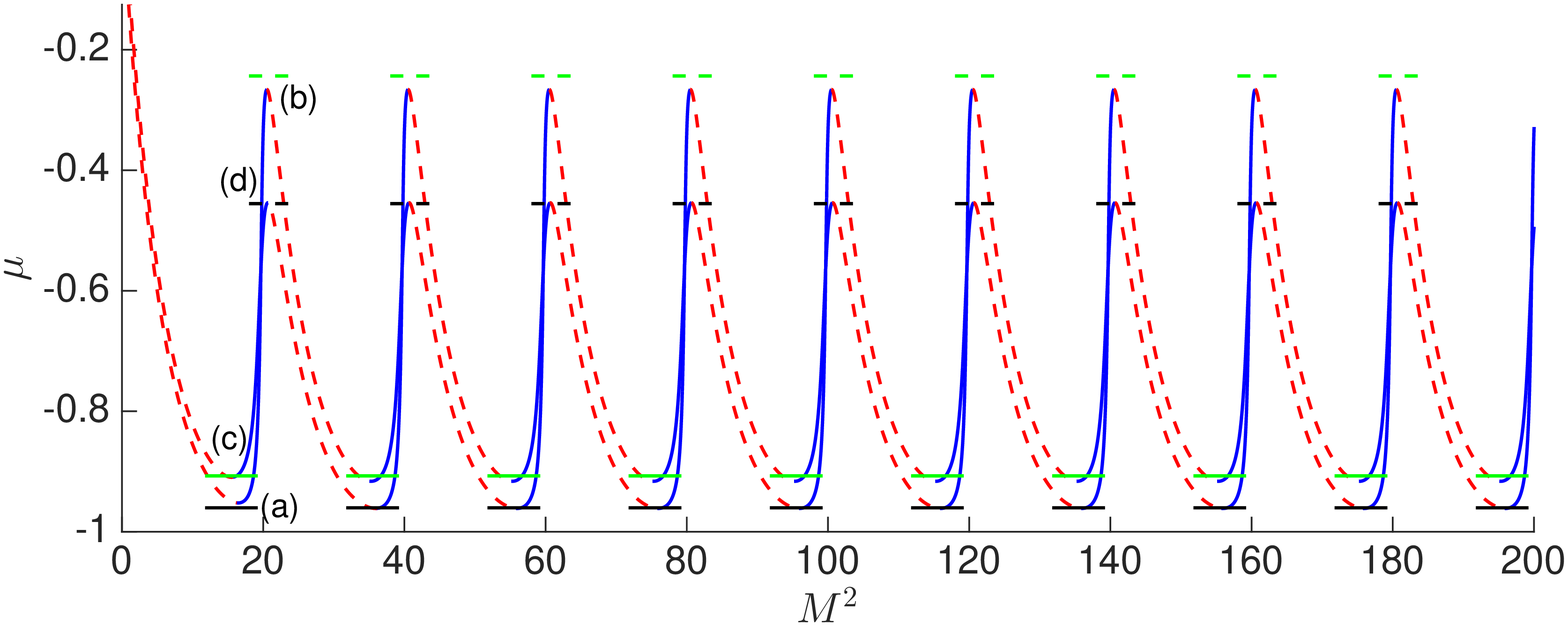}}
	\caption{{Bifurcation diagram of the solutions shown in figure \ref{fig:prof_1D_ch_t2}. Green and black lines show our one-active-cell approximation of type 1 and 2, respectively.}}
	\label{fig:bifur_1D_ch_t2}
\end{figure*}

{It has been reported that the continuous planar Swift-Hohenberg equation has patches as well as planar fronts \cite{Lloyd2008,Coullet2000,Hilali1995,Sakaguchi1996,McCalla2010,Lloyd2009}. The latter type reduces the planar equation into the 1D Swift-Hohenberg equation, that has been discussed in details in \cite{Woods1999,Burke2006,Burke2007,Burke2007a}.}

{In addition to the patches we discussed in Section \ref{sec2d:loc_snake} above, our discrete systems also admit planar fronts on the infinite periodic strip, i.e., fronts in one direction but constant in the transverse direction. We plot in figure\ \ref{fig:prof_1D_loc} several planar fronts of the square lattice, showing solutions that are uniform in the $m$- and localised in the $n$-directions. We have also calculated	bifurcation diagrams of the solutions and plot them in figure \ref{fig:bifur_1D_loc}. Comparing the result with figure \ref{fig:bifur_cp}, it is clear that the planar fronts have a much simpler bifurcation diagram than that of patches. This supports our hypothesis that the complexity of a bifurcation diagram depends on the interfaces of the localised solutions. }

{ We have computed planar fronts of the discrete equation with honeycomb and triangular lattices. We obtained that for the coupling values satisfying $c^{+}=c^{\Yup}=2c^{\varhexstar}$, the bifurcation diagram is independent of the lattice type. This can be explained by the following reduction.} 

{Considering solutions that are localised in the $n$-direction and uniform in the perpendicular $m$-direction, i.e., $u_{n,m\pm1}=u_{n,m},$ the discrete Laplacian becomes $	c^{\square}\Delta^{\square}u_{n,m}=c^{\square}\left(u_{n+1,m}+u_{n-1,m}-2u_{n,m}\right)$ for the square and honeycomb lattices and $c^{\varhexstar}\Delta^{\square}u_{n,m}=2c^{\varhexstar}\left(u_{n+1,m}+u_{n-1,m}-2u_{n,m}\right)$ for the triangular lattices. With this, the two-dimensional discrete Allen-Cahn equation \eqref{eq:dnls_all} can be simplified into the 1D counterpart, regardless of the two-dimensional lattice types, that has been analysed in details in \cite{Taylor2010,Susanto2018}. The one-active site approximation has been developed previously in the case of small coupling in our previous work \cite{Kusdiantara2017,Susanto2018}. When the coupling is large, the snaking boundaries that are in this case exponentially small away from the Maxwell point have also been calculated in \cite{Susanto2011,Susanto2018}. The reader is referred to those papers for the details. 
}	

{Besides planar fronts shown in figure \ref{fig:prof_1D_loc}, we can also consider solutions that are extended in the $n$-direction but localised in the $m$-direction, i.e., $u_{n\pm1,m}=u_{n,m}$. For this case, we still obtain the same 1D counterpart for the square and triangular lattices. However, for the honeycomb lattice, the discrete Laplacian will become
$c^{\Yup}\Delta^{\Yup_\pm} u_{n,m}=c^{\Yup}\left(u_{n,m\pm1}-u_{n,m}\right),$ where $\Delta^{\Yup_+}$ and $\Delta^{\Yup_-}$ correspond to the case when $n+m$ is even or odd, respectively. 
}

{We plot in figure \ref{fig:prof_1D_ch_t2} solution profiles of planar fronts of this type for $c^{\Yup}=0.05$. One can note that the front is uniform in $n$ for onsite solutions (figures \ref{subfig:prof_1D_ch_t2_0_05_a} and \ref{subfig:prof_1D_ch_t2_0_05_b}), but it is rather periodic in $n$ for intersite ones (figures \ref{subfig:prof_1D_ch_t2_0_05_c} and \ref{subfig:prof_1D_ch_t2_0_05_d}). We have also followed their existence by computing their bifurcation diagram. The result is shown in figure \ref{fig:bifur_1D_ch_t2}. The diagram has two types of saddle-node bifurcations that belong to either type 1 or 2 (see figure \ref{fig:active_cell_hc}), that appear both in the site-centred and bond-centred solutions.
}

\section{Conclusions}\label{sec2d:conclusions_2d}
We have considered two-dimensional discrete Allen-Cahn equation with cubic and quintic nonlinearities in the domain of square, honeycomb, and triangular lattices.
We have studied numerically  and analytically the time-independent solutions, i.e., uniform and localised states and their stabilities.

Our numerical results show that the snaking structures from the localised states for all of the lattice structures can have many types of saddle-node bifurcations. Herein, we propose an active-cell approximation to estimate the saddle-node bifurcations. The results show that our assumption gives good agreement for weakly-coupled system, i.e., small coupling strength and fewer exited cells around the interfaces of a localised solution. Moreover, we also showed that our approximation can be used to approximate the critical eigenvalue of localised states. {For the governing equation \eqref{eq:dnls_all}, our method is expected to be valid for $|c^\square|\ll\mu.$ For large coupling $|c^\square|\to\infty$, the pinning region will be exponentially small. In that case, a different approximation is required, which is addressed for future work.} 

{The idea of our work here can also be extended to further higher dimensional equations. For example, pattern formation in 3D continuous systems can display stacked lamellae (parallel 'sheets') and hexagons \cite{thie13} and quasicrystal patterns \cite{subr18}, in addition to body-centred cubic arrangements (which do not fit a hexagonal lattice in plane). In 3D discrete setups, our active cell approximation can still be used to describe lamellar states in a similar way to our work on planar fronts in Section \ref{secadd}, namely by reducing the 'dimensionality' of the problem.}

{Our work can also be extended to spatially continuous systems, where the ``lattice types" can be imposed by introducing spatial heterogeneity through periodic-in-space linear potentials (see, e.g, \cite{Kao2014} for a 1D problem that exhibits similar behaviours with a discrete setup \cite{Kusdiantara2017}).}

{Bifurcation diagrams we presented in the current paper are only principle branches of the fundamental localised solutions. In the continuous planar Swift-Hohenberg equation, it is well established that there are many exotic planar patterns, e.g., localised roll, square and stripe patches, that exhibit snaking and non-snaking behaviour on the same bifurcation branch \cite{Lloyd2008,Avitabile2010}. It is of particular interest to study if similar rich bifurcation structures are also present in the planar discrete Allen-Cahn equation. This is also addressed for future work.}

After we submitted the present work, a new preprint appeared \cite{Bramburger2019} that also considers the existence of homoclinic snaking in discrete systems, but offers a qualitative study with a mathematically rigorous approach. The paper definitely complements our results. 
 
\section{Acknowledgement}
RK gratefully acknowledgement financial from Lembaga Pengelolaan Dana Pendidikan (Indonesia Endowment Fund for Education)(Grant No.- Ref: S-34/LPDP.3/2017). The authors acknowledge the two reviewers for their careful reading and remarks that improved the manuscript. 
%
%

\section*{References}
%
%

\end{document}